\catcode`\@=11

\newif\if@fewtab\@fewtabtrue

{\count255=\time\divide\count255 by 60
\xdef\hourmin{\number\count255}
\multiply\count255 by-60\advance\count255 by\time
\xdef\hourmin{\hourmin:\ifnum\count255<10 0\fi\the\count255}}
\def\ps@draft{\let\@mkboth\@gobbletwo
    \def\@oddhead{}
    \def\@oddfoot
       {\hbox to 7 cm{$\scriptstyle Draft\ version:\ \draftdate$
       \hfil}\hskip -7cm\hfil\rm\thepage \hfil}
    \def\@evenhead{}\let\@evenfoot\@oddfoot}

\def\draftcite#1{\ifnum\draftcontrol=1#1\else{}\fi}

\def\@lbibitem[#1]#2{\item{}\hskip -3cm \hbox to 2cm
{\hfil$\scriptstyle\draftcite{#2}$}\hskip
1cm[\@biblabel{#1}]\if@filesw
     {\def\protect##1{\string ##1\space}\immediate
      \write\@auxout{\string\bibcite{#2}{#1}}}\fi\ignorespaces}

\def\@bibitem#1{\item\hskip -3cm \hbox to 2cm
{\hfil $\scriptstyle\draftcite{#1}$}\hskip 1cm
\if@filesw \immediate\write\@auxout
       {\string\bibcite{#1}{\the\value{\@listctr}}}\fi\ignorespaces}

\def\nsection#1{\section{#1}\setcounter{equation}{0}}

\font\tendl=msbm10  scaled \magstep1
\font\sevendl=msbm7 scaled \magstep1
\font\fivedl=msbm5 scaled \magstep1
\font\tengl=eufm10  scaled \magstep1
\font\sevengl=eufm7 scaled \magstep1
\font\fivegl=eufm5 scaled \magstep1

\newfam\dlfam  
\textfont\dlfam=\tendl \scriptfont\dlfam=\sevendl
\scriptscriptfont\dlfam=\fivedl
\newfam\glfam  
\textfont\glfam=\tengl \scriptfont\glfam=\sevengl
\scriptscriptfont\glfam=\fivegl

\def\draftdate{\number\month/\number\day/\number\year\ \ \ \hourmin }

\global\def\draftcontrol{0}
\catcode`\@=12
\documentclass[aps,superscriptaddress,floatfix,nofootinbib]{revtex4}
\usepackage{epsfig}
\usepackage{float}
\usepackage[centerlast]{subfigure}
\usepackage{bm}
\usepackage{amssymb}
\usepackage{verbatim}
\usepackage{graphicx}
\usepackage{amsmath}

\newcommand\void[1]{}
\newcommand{\be}{\begin{eqnarray}}
\newcommand{\en}{\end{eqnarray}\vs 0.5 cm}

\newcommand{\no}{\noindent}
\newcommand{\vs}{\vskip}

\newcommand{\un}{\underline}

\newcommand{\qq}{\begin{eqnarray}}

\newcommand{\ee}{{\rm e}}

\newcommand{\qqq}{\end{eqnarray}}

\newcommand{\CD}{{\cal D}}

\newcommand{\CF}{{\cal F}}

\newcommand{\CI}{{\cal I}}
\newcommand{\CJ}{{\cal J}}

\newcommand{\CL}{{\cal L}}
\newcommand{\CM}{{\cal M}}

\newcommand{\CP}{{\cal P}}

\newcommand{\CV}{{\cal V}}
\newcommand{\CW}{{\cal W}}
\newcommand{\CX}{{\cal X}}

\newcommand{\s}{\hspace{0.05cm}}
\newcommand{\m}{\hspace{0.025cm}}

\begin{document}
\
\vskip 1.7cm
\begin{center}
{\large{\bf{FLUCTUATION RELATIONS in STOCHASTIC THERMODYNAMICS}}}\m\footnote{Extended version of lectures given at the Mathematics Department of Helsinki
University, November 2012}
\vskip 0.7cm
Krzysztof Gaw\c{e}dzki
\vskip 0.3cm
C.N.R.S., ENS Lyon, Laboratoire de Physique, 
46 Al\'ee d'Italie, F69007 Lyon, France
\end{center}
\date{ }
\vskip 0.3cm


\addtocounter{section}{1}
\vskip 0.8cm

\noindent Fluctuation relations are identities, holding in
non-equilibrium systems, that have attracted a lot of interest
in the last 20 years. This is a series of 4 lectures discussing
various aspects of such relations for stochastic
equations modeling non-equilibrium processes. 
\vskip 0.2cm

\no\hbox to 2.2cm{{\bf{Lecture 1}}:\hfill}Transient 
fluctuation relations for Markov processes
\vskip 0.1cm

- Origin of fluctuation relations

- Jarzynski-Crooks-Hatano-Sasa relations for nonstationary Markov chains 

- Case of continuous time Markov processes  

\vskip 0.2cm

\no\hbox to 2.2cm{{\bf{Lecture 2}}:\hfill}2nd Law of Stochastic 
Thermodynamics
\vskip 0.1cm

- Work, heat and entropy in stochastic thermodynamics

- Fluctuation relations and the 2nd Law of Stochastic Thermodynamics  

- Finite time refinement of the 2nd Law and Landauer Principle
\vskip 0.2cm

\no\hbox to 2.2cm{{\bf{Lecture 3}}:\hfill}Fluctuation-dissipation 
relations 
\vskip 0.1cm

- Jarzynski-Hatano-Sasa relation near stationary state 

- General Fluctuation-Dissipation Theorem

- Green-Kubo formula for diffusions  
\vskip 0.2cm

\no\hbox to 2.2cm{{\bf{Lecture 4}}:\hfill}Large deviations and
stationary fluctuation relations
\vskip 0.1cm

- Gallavotti-Cohen type fluctuation relations    

- Macroscopic fluctuation theory

- A non-trivial example
\vskip 2cm

\addtocounter{section}{-1}
\nsection{{Transient 
fluctuation relations for Markov processes}}
\subsection{A bit of history}

\noindent The history of fluctuation relations may be traced back to 
the late seventies/early eighties papers by Bochkov-Kuzovlev
\cite{BK77,BK79,BK81} that were not remarked at the time.
In the next development, in 1993 Evans-Cohen-Morriss observed 
in \cite{ECM93} a symmetry 
in the distribution of fluctuations of microscopic pressure 
in a thermostatted particle system driven by external shear. 
Attempts to explain this symmetry on the theoretical ground 
led to the formulation of the Evans-Searles transient fluctuation 
relation \cite{ES94} and of the Gallavotti-Cohen stationary 
fluctuation relation \cite{GC95a,GC95b}. On the other hand, 
Jarzynski in 1997 proved in \cite{J97a} a simple equality, 
which appeared to be closely related to the Bochkov-Kuzovlev one, 
see \cite{J07}. Originally formulated  for deterministic non-equilibrium 
evolutions, the fluctuation relations were quickly extended 
to stochastic dynamics in \cite{J97b,K98,LS99}. All those works 
attracted in late nineties and afterwards a widespread interest 
and led to an avalanche of papers. For reviews see 
\cite{ES02,G08,M03,J11,SF12}. 
The present lectures are neither an exhaustive review of the subject
nor follow the thread of history but are designed as a brief informal 
introduction for probablists or mathematical physicists to some
topics involving fluctuation relations that are close to my own 
interests. The lectures should also be accessible to the physics
audience which, nevertheless, was not their primary addresee and might 
find them too dry.

\subsection{Few trivial identities}

\noindent As observed by Maes in \cite{M99}, \,fluctuation relations
in the stochastic setup have their root in tautological identities
comparing two probability measures $\,\CP(dx)\,$
and $\,\CP'(dx^*)\,$ absolutely continuous with respect to each other, where
$\,x\mapsto x^*\,$ is a, possibly trivial, involution. We shall denote
by $\,\mathbb E\,$ and $\,\mathbb E'\,$ \,the expectations with respect to 
$\,\CP(dx)\,$ and $\,\CP'(dx)$, \,respectively. \,If 
$\,\ee^{-\CW(x)}=\frac{\CP'(dx^*)}{\CP(dx)}\,$ is 
the Radon-Nikodym derivative of $\,\CP'(dx^*)\,$ with 
respect to $\,\CP(dx)\,$ then, 
trivially,
\qq
\mathbb E\ \ee^{-\CW(x)}\,=\,1
\label{1}
\qqq
and, \,more generally,
\qq
\mathbb E\ \CF(x)\,\ee^{-\CW(x)}\,=\,\mathbb E'\ \CF(x^*)\,.
\label{2}
\qqq
In particular, taking $\CF(x)=f(\CW(x))$, we obtain
\qq
\mathbb E\ f(\CW(x))\,\,\ee^{-\CW(x)}\,=\,\mathbb E'\ f(-\CW'(x))
\label{3}
\qqq
where $\,\CW'(x)=-\CW(x^*)\,$ so that 
$\,\ee^{-\CW'(x)}=\frac{\CP(dx^*)}{\CP'(dx)}$.
\,Eq.\,(\ref{3}) implies the following relation between 
the probability distribution $\,\pi(d\CW)\,$ of the random 
variable $\,\CW(x)\,$ with respect to $\,\CP(dx)\,$ and 
$\,\pi'(d\CW)\,$ of the random variable $\,\CW'(x)\,$
with respect to $\,\CP'(dx)$:
\qq
\ee^{-\CW}\,\pi(d\CW)\,=\,\pi'(d(-\CW))\,.
\label{4}
\qqq
\vskip 0.2cm

\subsection{Application to Markov chains}

\noindent We shall first apply the above tautological relations to 
discrete-time (nonstationary) {\,\bf Markov chain} $\,(x_n)_{n=0}^{N+1}\,$ 
with space of states $\,{\cal X}$. \,Let us denote by $\,P_n(x,dy)\,$ 
the transition probabilities for the process and by $\,\mu_0(dx)\,$ 
the distribution of $\,x_0$. \,The probability space of the process 
may be taken as $\,{\cal X}_N={\cal X}^{N+2}\,$ with the probability measure
\qq
\CP_{\mu_0}[d\bm x]\,=\,\mu_0(dx_0)\,P_0(x_0,dx_1)\,\cdots\,P_N(x_N,dx_{N+1})\,,
\qqq
where $\,\bm x\equiv(x_0,\dots,x_{N+1})\,$ (we shall denote the trajectorial
or functional dependence by square brackets). The distribution of time $\,n\,$
value $\,x_n\,$ of the process will be denoted $\,\mu_n(dx)$. 
Suppose that 
$\,(x'_n)_{n=0}^{N+1}\,$ is another Markov chain with the same space 
of states $\,{\cal X}\,$ corresponding 
to transition probabilities $\,P'_n(x,dy)\,$ and the initial measure
$\,\mu'_0(dx)$. \,It corresponds to the measure 
\qq
\CP'_{\mu_0'}[d\bm x]\,
=\,\mu'_0(dx_0)\,P'_0(x_0,dx_1)\,\cdots\,P'_N(x_N,dx_{N+1})\,, 
\qqq
Let us now consider an involution $*:{\cal X}\rightarrow{\cal X}$ and 
let us extend it to ${\cal X}_N$ by combining it with the time reflection, 
so that for $\bm x=(x_0,\dots,x_{N+1})$,
\qq
\bm x^*\,=\,(x_{N+1}^*,\dots,x_0^*)\,.
\label{*}
\qqq  
Assuming the relative absolute continuity of measures
$\,\CP_{\mu_0}[d\bm x]\,$ and $\,\CP'_{\mu_0'}[d\bm x^*]\,$ on
$\,{\cal X}_N$,
\,we may apply the scheme described at the beginning of the lecture 
to the case at hand defining
\qq
\ee^{-\CW[\bm x]}\,=\,\frac{\CP'_{\mu_0'}[d\bm x^*]}{\CP_{\mu_0}[d\bm x]}\,=\,\,
\ee^{\hspace{0.03cm}\CW'[\bm x^*]},
\label{comp}
\qqq
where, explicitly,
\qq
\frac{\CP'_{\mu_0'}[d\bm x^*]}{\CP_{\mu_0}[d\bm x]}\,=\,
\frac{\mu'_0(dx^*_{N+1})\,P'_0(x^*_{N+1},dx^*_N)\,\cdots\,P'_N(x^*_1,dx^*_0)}
{\mu_0(dx_0)\,P_0(x_0,dx_1)\,\cdots\,P_N(x_N,dx_{N+1})}\,,
\label{RN}
\qqq
inferring immediately that
\qq
\boxed{\,\mathbb E_{\mu_0}\,\ee^{-\CW[\bm x]}\,=\,1\,},
\label{5}
\qqq
where $\,\mathbb E_{\mu_0}$ stands for the expectation with respect to
$\,\CP_{\mu_0}(d\bm x)\,$ and that
\qq
\boxed{\,\ee^{-\CW}\,\pi(d\CW)\,=\,\pi'(d(-\CW))\,},
\label{6}
\qqq
where $\,\pi(d\CW)\,$ and $\,\pi'(d\CW)\,$ are the probability disytibutions 
of $\,\CW[\bm x]\,$ and $\,\CW'[\bm x]\,$ with respect to the probability 
measure $\,\CP_{\mu_0}[d\bm x]\,$ and $\,\CP'_{\mu_0'}(d\bm x)$, \,respectively. 
The assumed relative absolute continuity of $\,\CP_{\mu_0}[d\bm x]\,$ and 
$\,\CP'_{\mu_0'}[d\bm x^*]\,$ follows if all initial and transition 
probability measures have positive densities with respect to a fixed 
measure $\,\lambda(dx)\,$ on $\,\CX\,$ that we shall take to be the 
counting measure if $\,\CX\,$ is discrete and the Lebesgue measure if 
$\,\CX=\mathbb R^d$.

\subsection{Hatano-Sasa fluctuation relations for Markov chains}

\noindent Suppose that $\,\nu_n(dx)=\ee^{-\varphi_n(x)}\lambda(dx)\,$ 
are probability measures left invariant under transition probabilities 
$\,P_n(x,dy)$:
\qq
\int\limits_X\nu_n(dx)\,P_n(x,dy)\,=\,\nu_n(dy)\,.
\label{leftinv}
\qqq  
They are in general different form the time-$n\,$ distributions $\,\mu_n(dx)\,$
of the process. We shall say that measures $\,\nu_n(dx)\,$ accompany 
the Markov process $\,(x_n)$. \,Take
\qq
P'_n(x,dy)\,=\,\frac{P_{n^*}(y^*,dx^*)}{\nu_{n^*}(dx^*)}\,\nu_{n^*}(dy^*)
\qqq
for $\,n^*\equiv N-n$, \,assuming again that the Radon-Nikodym 
derivatives exist. $\,P'_n(x,dy)\,$ are Markov transition probabilities and, 
\,for $\,\nu'_n(dx)=\nu_{n^*}(dx^*)$, 
\qq
\int\limits_X\nu'_n(dx)\,P'_n(x,dy)\,=\,\int\limits_X
\nu_{n^*}(dx^*)\,\frac{P_{n^*}(y^*,dx^*)}{\nu_{n^*}(dx^*)}\,\nu_{n^*}(dy^*)
\,=\,\nu_{n^*}(dy^*)\,=\,\nu'_n(dy)
\qqq
(the integration is over $x$), so that the measures 
$\,\nu'_n(dx)=\ee^{-\varphi'_n(x)}\lambda(dx^*)\,$ with $\,\varphi'_n(x)=
\varphi_{n^*}(x^*)\,$ are left invariant under the transition probabilities 
$\,P'_n(x,dy)$. \,The Markov process $\,(x'_n)\,$ with such transition 
probabilities has the interpretation of a (particular) time-reversal of the 
original process $\,(x_n)\,$ and $\,\nu'_n(dx)\,$ are its accompanying
measures. \,The Radon-Nikodym derivative (\ref{RN})
takes now the form
\qq
\frac{\CP'_{\mu_0'}[d\bm x^*]}{\CP_{\mu_0}[d\bm x]}
\,=\,\frac{\mu'_0(dx_{N+1}^*)}{\nu_N(dx_{N+1})}
\,\frac{\nu_N(dx_N)}{\nu_{N-1}(dx_N)}
\,\cdots\,\frac{\nu_1(dx_1)}{\nu_{0}(dx_1)}\,
\frac{\nu_0(dx_0)}{\mu_0(dx_0)}
\label{RNHS}
\qqq
and the identities (\ref{5}) and (\ref{6}) hold for any choice
of initial measures $\,\mu_0\,$ and $\mu_0'\,$ with densities 
\qq
\mu_0(dx)=\rho_0(x)\,d\lambda(dx)\,,
\qquad \mu'_0(dx)=\rho'_0(x)\,\lambda(dx^*)\,.
\qqq
Explicitly, 
\qq
\CW[\bm x]\,=\,-\,\ln\rho'_0(x_{N+1}^*)-\varphi_N(x_{N+1})\,+
\sum\limits_{n=1}^N\big(\varphi_n(x_n)-\varphi_{n-1}(x_n)\big)\,+\,\ln\rho_0(x_0)
+\varphi_0(x_0)\,.\label{CWN}
\qqq
In the special case with initial measures $\,\mu_0=\nu_0\,$ and
$\,\mu'_0=\nu'_0$, \,the boundary contributions vanish so that
\qq
\CW[\bm x]\,=\,\sum\limits_{n=1}^N\big(\varphi_n(x_n)-\varphi_{n-1}(x_n)\big)
\qqq
and we obtain the relation 
\qq
\boxed{\,\mathbb E_{\nu_0}\ \ee^{-\sum\limits_{n=1}^N\left(\varphi_n(x_n)-\varphi_{n-1}(x_n)
\right)}\,=\,1\,}
\label{HS}
\qqq
which is the Markov chain version of \,the {\bf\,Hatano-Sasa relation} 
\cite{HS01}. \,In the particular situations when the time
reversed and the original process have the same law, \,relation
(\ref{6}) reduces to the property of the probability density $\,\pi(\CW)$:
\qq
\ee^{-\CW}\,\pi(d\CW)\,=\,\pi(d(-\CW))\,.
\label{7}
\qqq

\subsection{Jarzynski and Crooks relations}
\label{subsec:jarzcroocks}

\noindent Consider the special case when the transition probabilities satisfy
the detailed balance relation with respect to {\,\bf Hamiltonians} $\,H_n(x)\,$ for inverse temperature $\,\beta=\frac{1}{k_BT}\,$ ($k_B$ is the Boltzmann
constant), \,i.e. when
\qq
\lambda(dx)\,P_n(x,dy)\,=\,\lambda(dy)\,P_n(y,dx)
\,\,\ee^{-\beta(H_n(y)-H_n(x))}\,.
\label{detbal}
\qqq
In that case, assuming that the partition functions
\qq
Z_n\,=\,\int\limits_{\cal X}\ee^{-\beta H_n(x)}\lambda(dx)
\qqq
are finite, the Gibbs measures corresponding to Hamiltonians $\,H_n(x)$,
\qq
\nu_n(dx)\,=\,Z_n^{-1}\,\ee^{-\beta H_n(x)}\lambda(dx)
\label{nutH}
\qqq
are left invariant under the transition probabilities $\,P_n(x,dx)\,$
(i.e. are the accompanying measures of the process in the terminology
introduced above) and
\qq
\varphi_n(x)\,=\,\beta\,(H_n(x)-F_n)\,,
\qqq
where $\,F_n=-\beta^{-1}\ln Z_n$ are the Gibbs free energies.
\,The Hatano-Sasa relation (\ref{HS}) reduces in this case to the Markov-chain 
version of \,the {\,\bf Jarzynski equality} \cite{J97a}:
\qq
\boxed{\,{\mathbb E}_{\nu_0}\ \ee^{-\beta\hspace{0.02cm}W[\bm x]}\,=\,
\ee^{-\beta\Delta F}\,}
\label{jarz}
\qqq
for $\,\Delta F=F_N-F_0\,$ and
\qq
W[\bm x]\,=\,\sum\limits_{n=1}^N\big(H_n(x_n)-H_{n-1}(x_n)\big)\,. 
\label{WNj}
\qqq
The quantity $\,W[\bm x]\,$ may be interpreted as \,the {\,\bf work performed 
on the system}. \,To understand why, consider a particle that, 
due to an interaction with thermal 
environment, jumps at discrete times $\,n\,$ between potential wells 
$\,x\in{\cal X}\,$ in the time-varying potential landscape $\,H_n(x)$. 
\,The jump between the potential well $\,x_{n-1}\,$ occupied at time $\,n-1\,$
and the potential well $\,x_n\,$ occupied at time $\,n\,$ is chosen with
the transition probability $\,P_n(x_{n-1},dx_n)\,$ satisfying the detailed
balance with respect to the time $\,n-1\,$ Hamiltonian $\,H_{n-1}(x)$.
\,After that jump, the energy landscape is changed from $\,H_{n-1}(x)\,$ 
to $\,H_{n}(x)$. \,For the occupied well $\,x_n$, \,this requires work 
equal to $\,H_n(x_n)-H_n(x_n)\,$ and $\,W[\bm x]\,$ is the sum of
such works along the particle trajectory. \,Note that while free energy 
$\,F_0\,$ pertains to the Gibbs distribution $\,\nu_0\,$ of 
the initial value $\,x_0\,$ of the process, $\,F_N\,$ corresponds
to the Gibbs measure $\,\nu_N\,$ that, in general, is different from 
the distribution $\,\mu_N\,$ of $\,x_N\,$ (or $\,\mu_{N+1}\,$ of $\,x_{N+1}$).
\vskip 0.1cm

The identity (\ref{jarz}) implies by the Jensen inequality that
\qq  
\boxed{\,\mathbb E_{\nu_0}\,W[\bm x]\,\geq\,\Delta F\,} 
\label{2workN}
\qqq
i.e. that the average work is bounded below by the change of 
free energy between the initial and final times. \,The Jarzynski equality 
(\ref{jarz}) contains, however, more information. For example, it implies that 
the probability of observing the trajectories with $\,W[\bm x]
\leq\Delta F-a\,$ for positive $a$ is exponentially small. Indeed,
\qq
&&\ee^{\beta a}\,\,\mathbb E_{\nu_0}\,1_{\{W[\bm x]\leq\Delta F-a\}}\,=\,
\ee^{\beta\Delta F}\,\mathbb E_{\nu_0}\ \ee^{-\beta(\Delta F-a)}\,
1_{\{W[\bm x]\leq\Delta F-a\}}\cr\cr
&&\leq\,\ee^{\beta\Delta F}\,
\mathbb E_{\nu_0}\,\,\ee^{-\beta W[\bm x]}\,
1_{\{W[\bm x]\leq\Delta F-a\}}\,\leq\,\ee^{\beta\Delta F}\,
\mathbb E_{\nu_0}\,\,\ee^{-\beta W[\bm x]}\,\leq\,1\,.
\qqq
\vskip 0.1cm

In the case with detailed balance, \,the time reversed transition 
probabilities are
\qq
P'_n(x,dy)\,=\,P_{n^*}(x^*,dy^*)\,.
\qqq
They preserve the Gibbs measures
\qq
\nu'_n(dx)\,=\,{Z_n'}^{-1}\,\ee^{-\beta H'_n(x)}\,\lambda(dx^*)
\qqq
for $\,H'_n(x)=H_{n^*}(x^*)\,$ and $Z_n'=Z_{n^*}$. \,If
\qq
W'[\bm x]\,=\,\sum\limits_{n=1}^N\big(H'_n(x_n)-H'_{n-1}(x_n)\big)\,
=\,-W[\bm x^*]\,, 
\qqq
and $\,\pi(dW)=p(W)\lambda(dW)\,$ ($\pi'(dW)=p'(W)\lambda(dW)$) \,is 
the probability distribution of $\,W[\bm x]\,$ ($W'[\bm x]$) \,with respect 
to $\,\CP_{\nu_0}[d\bm x]\,$ ($\CP'_{\nu'_0}[d\bm x]$)
\,then relation (\ref{6}) implies the Markov chain version
of \,the {\,\bf Crooks relation} \cite{C99} 
\qq
\boxed{\,\ee^{-\beta\hspace{0.01cm}W}\,p(W)\,=\,\ee^{-\beta\hspace{0.01cm}
\Delta F}\,p'(-W)}\,.
\label{cr}
\qqq
On the right hand side, $\,p'(-W)\,$ may be replaced by $p(-W)$ 
for time-reversible processes with $\,P'_n(x,dx)=P_n(x,dx)$. 
\vskip 0.2cm

The utility of Eq.\,(\ref{jarz}) and (\ref{cr}) is that it permits
to extract the free energy difference between initial and final
Gibbs states in a nonstationary Markov chain with instantaneous
detailed balance from the statistics of the work $\,W[\bm x]\,$ performed 
on the system. Usually in thermodynamics, the free energy difference 
is equal to the {\,\bf work\,}
performed in a quasi-stationary process between the initial and final Gibbs
state and that requires very long times. Here, however, there is no assumption 
that the process has to be close to stationary and, although the initial
state $\mu_0$ was assumed to be equal to the Gibbs one $\,\nu_0$, 
\,the final state $\,\mu_N\,$ (the distribution of $\,x_N$) \,is, 
\,as a rule, different from $\,\nu_N$, \,as was already mentioned. 
\,Hence the interest of the above 
Jarzynski and Crooks identities for numerical calculations or for
experimental determination of free 
energy differences for a mesoscopic system in different Gibbs states, 
as, for example, in DNA/RNA-hairpin stretching experiments and simulations, 
\,see FIG.\,1. \,In particular, \,Eq.\,(\ref{cr}) shows that 
$\Delta F$ may be found as the value of $W$ for 
which $p(W)=p'(-W)$.

\begin{figure}[H]
\begin{center}
\leavevmode
\hspace*{-0.3cm}
        \includegraphics[width=10cm,height=7cm]{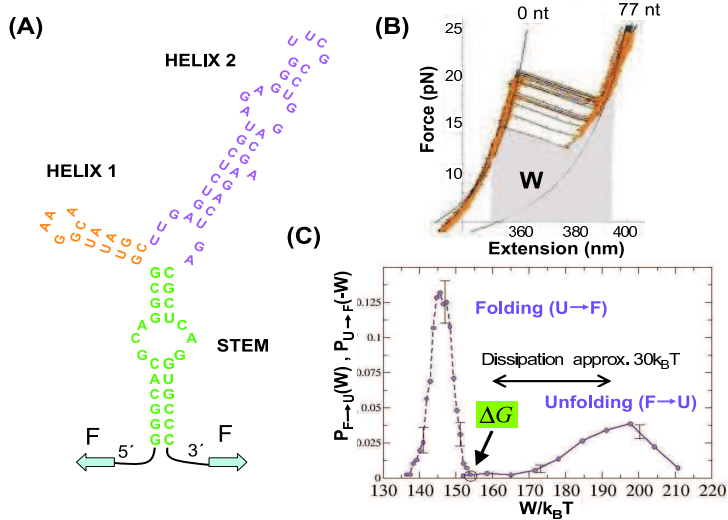}\\
\vskip -0.2cm
        \caption{Work statistics in RNA stretching, from  \cite{R06}}
\end{center}
\end{figure}

\subsection{Continuous time limit}
 
\noindent Similar considerations apply to nonstationary continuous-time
{\,\bf Markov processes}. \,On the formal level, \,such processes may be 
obtained by taking the limit of discrete-time Markov processes when 
their time-step $\,\epsilon\,$ tends to zero with the total time interval
$\,N\epsilon=\tau\,$ kept constant, provided that the transition probabilities
for $\,n\,$ such that $\,n\hspace{0.01cm}\epsilon=t\,$ is fixed have
the behavior
\qq
P_n(x,dy)\,=\,\epsilon\hspace{0.01cm}w_t(x,dy)\,+\,\Big(1-\epsilon\int\limits_X
w_t(x,dz)\Big)\,\delta_x(dy)\ +\ o(\epsilon)\,.
\qqq
Here $\,\delta_x(dy)\,$ denotes the Dirac measure supported at $\{x\}$.
\,The limiting continuous-time Markov process $\,(x_t)\,$ for $\,t\in[0,\tau]\,$
has the initial distribution $\,\mu_0(dx)$. \,Quantities $\,w(x,dy)$,
\,that are defined modulo signed measures concentrated on the diagonal,
may be distributional but are positive measures away from the diagonal 
and give there the transition rates of the continuous time Markov 
process. The backward generators of the limiting process defined by 
the identity
\qq
\frac{d}{dt}\,\hspace{0.02cm}\mathbb E_{\mu_0}\,\hspace{0.02cm}f(x_t)\ 
=\ \mathbb E_{\mu_0}\,\hspace{0.02cm}(L_tf)(x_t)
\qqq
are given by the formula
\qq
(L_tf)(x)\,=\,\int\limits_{\cal X}\big(f(y)-f(x)\big)\,w_t(x,dy)\,.
\qqq
The transition probabilities $\,P_{s,t}(x,dy)\,$ of the process $\,(x_t)\,$
for $\,s\leq t\,$ evolve in time according to the equations
\qq
\partial_sP_{s,t}(x,dy)\,=\,-\,L_s(x)\,P_{s,t}(x,dy)\,,\qquad
\partial_tP_{s,t}(x,dy)\,=\,L^*_t(y)\,P_{s,t}(x,dy)\,,
\qqq
where $L_t^*$ is the adjoint operator acting on measures.
\,At equal times, $\,P_{t,t}(x,dy)=\delta_x(dy)$. \,In the operator
notation,
\qq
P_{s,t}\,=\,\overrightarrow{\exp}\Big[\int\limits_s^t L_{\sigma}\,
d\sigma\Big]
\qqq
with the time-ordered exponential, where we write the 1-dimensional 
Lebesgue measure $\,\lambda(d\sigma)\,$ as $\,d\sigma$.
\,The time $t$ probability distributions $\,\mu_t(dx)\,$ of the process 
defined by
\qq
\mathbb E_{\mu_0}\,f(x_t)\,=\,\int\limits_X f(y)\,\mu_t(dy)
\qqq
are given by the relation
\qq
\mu_t(dy)\,=\,\int\limits_X\mu_0(dx)\, P_{0,t}(x,dy)\,.
\qqq
They evolve according to the equation
\qq
\partial_t\mu_t\,=\,L_t^*\mu_t\,.
\label{muev}
\qqq
On the other hand, \,measures $\nu_t(dx)\,$ satisfying the relation
\qq
L^*_t\nu_t\,=\,0\,.
\qqq
will be said to accompany the process. They would be invariant under
time evolution if the transition rates where fixed at the subsequent times
to their times-$t$ values.
\vskip 0.1cm

If state space $\,\CX\,$ is discrete, \,a continuous time Markov 
process jumps from $\,x(t)=x\,$ to $\,x(t+dt)=y\not=x\,$ with probability
$\,w(x,dy)dt\,$ and otherwise stays at $x$. 
\,If $\,\CX=\mathbb R^d$, \,the process is a diffusion with a drift 
or a jump process or a combination of both.

\vskip 0.4cm

\noindent{\bf Examples 1}. \ A general diffusion process.
\vskip 0.2cm

\noindent Consider a stochastic differential equation in $\,\mathbb R^d\,$ 
\qq
dx\,=\,X_{0t}(x)\hspace{0.03cm}dt\,+\,X_{\alpha t}(x)\circ dW_\alpha(t)\,,
\label{gendif}
\qqq
written with the {\,\bf Stratonovich convention\,} indicated by 
symbol $\circ$, \,with arbitrary time-dependent 
vector fields $\,X_{0t},\,X_{\alpha t},\ \alpha=1,\dots,A\,$ 
and independent standard one-dimensional {\,\bf Wiener processes} 
$\,W_\alpha(t)\,$ (we employ the summation convention so $\alpha$ is summed 
over). \,\,Eq.\,(\ref{gendif}),
together with the initial distribution, defines a continuous-time
Markov process with transition rates
\qq
w_t(x,dy)\,=\,\big(X_{0t}(x)\cdot\nabla_x+\frac{_1}{^2}(X_{\alpha t}(x)
\cdot\nabla_x)^2\big)\hspace{0.02cm}\delta(x-y)\,\lambda(dy)
\qqq
and the backward generator
\qq
L_t\,=\,X_{0t}\cdot\nabla\,+\,\frac{_1}{^2}\,(X_{\alpha t}\cdot\nabla)^2\,.
\label{L0}
\qqq
The densities of the time $t$ measures $\,\mu_t(dx)=\rho_t(x)\,
\lambda(dx)\,$ of the process evolve according to the {\bf Fokker-Planck 
equation}
\qq
\partial_t\rho_t(x)\,=\,L_t^\dagger\rho_t(x)\,=\,
-\nabla\cdot j_t(x)\,,
\label{FP}
\qqq
where $L_t^\dagger$ is the formal adjoint of $L_t$ with respect to the
Lebesgue measure $\lambda(dx)$ and
\qq
j_t(x)\,=\,\Big(\rho_t\,X_{0t}\,-\,\frac{_1}{^2}\,
\big(\nabla\cdot(\rho_t X_{\alpha t})\big)\,X_{\alpha t}\Big)(x)\,
\,=\,\big(\rho_t\,\widehat X_{0t}\,-\,\CD_t\nabla\rho_t\big)(x)\,\equiv\,
j_{\rho_t}(x)
\label{jt0}
\qqq
is the {\,\bf probability current}, \,where
\qq
\widehat X_{0t}\,=\,X_{0t}-\frac{_1}{^2}(\nabla\cdot 
X_{\alpha t})X_{\alpha t}\,,
\qquad\CD_t\,=\,\frac{_1}{^2}X_{\alpha t}\otimes X_{\alpha t}\,.
\label{hatcd}
\qqq
If, \,following \cite{N67}, \,we introduce
the {\,\bf current velocity} $\,v_t(x)\,$ by the relation 
\qq
j_t\,=\,\rho_t\hspace{0.01cm}v_t
\qqq
then the Fokker-Planck equation (\ref{FP}) may be rewritten
as the advection equation
\qq
\partial_t\rho_t\,+\,\nabla\cdot(\rho_tv_t)\,=\,0\,,
\label{adv0}
\qqq
which will be used a lot below. Note, nevertheless, that
\qq
v_t\,=\,\widehat X_{0t}-\,
\CD_t\nabla\ln{\rho_t}
\label{currv0}
\qqq
depends on $\rho_t$. \,The current velocity may be interpreted as the
mean velocity of the process conditioned to pass at time $\,t\,$
through point $\,x\,$:
\qq
v_t(x)\,=\,\frac{{\mathbb E}_{\mu_0}\,\delta(x-x_t)\circ\frac{dx_t}{dt}}
{\rho_t(x)}\,=\,\frac{\lim\limits_{\epsilon\to0}\,{\mathbb E}_{\mu_0}\,\delta(x-x_t)
\,\frac{x_{t+\epsilon}-x_{t-\epsilon}}{2\epsilon}}
{\rho_t(x)}\,.
\qqq

\vskip 0.4cm

\noindent{\bf Examples 2}. \,Langevin process.
\vskip 0.2cm

\noindent A particular diffusion process in $\mathbb R^d$ is given 
by the {\,\bf Langevin equation} 
\qq
dx\,=\,\CM\big(-(\nabla H_t)(x)+f_t(x)\big)\,dt\,+\,(2\CD)^{1/2}dW(t)
\label{Lang}
\qqq
where $\CM$ (the {\,\bf mobility}) \,is a matrix with non-negative 
symmetric part, and $\CD$ (the {\bf diffusivity}) \,is a non-negative matrix.
$\,H_t(x)\,$ is the time-dependent Hamiltonian, $\,f_t(x)\,$ is 
a non-conservative force and $\,W(t)\,$ the standard $\,d$-dimensional Wiener 
process. For simplicity, we took matrices $\,\CM\,$ and $\,\CD\,$ as 
independent of $\,t\,$ and $\,x$. \,The Markov process corresponding 
to Eq.\,(\ref{Lang}) has the transition rates
\qq
w(x,dy)\,=\,\big(-\CM(\nabla H_t)(x)+\CM f_t(x)+\CD\nabla_x\big)
\cdot\nabla_x\delta(x-y)\,\lambda(dy)
\qqq
and the backward generator 
\qq
L_t\,=\,\big(-\CM(\nabla H_t)+\CM f_t(x)+\CD\nabla\big)\cdot\nabla\,.
\label{Lgen}
\qqq
The probability current takes in this case the form
\qq
j_t(x)\,=\,\big(-\CM(\nabla H_t)(x)+\CM f_t(x)-\CD\nabla\big)\hspace{0.02cm}
\rho_t(x)
\label{jt}
\qqq
and the current velocity is
\qq
v_t(x)\,=\,-\CM(\nabla H_t)(x)+\CM f_t(x)-\CD\nabla\ln\rho_t(x)\,.
\label{currv}
\qqq
One says that $\CM$ and $\CD$ satisfy the {\,\bf Einstein relation\,} if 
\qq
\CM+\CM^t\,=\,2\beta\,\CD\,.
\qqq
If this is the case and the non-conservative force $f_t$ vanishes then
the Gibbs measures 
\qq
\nu_t(dx)\,=\,Z_t^{-1}\,\ee^{-\beta H_t(x)}\,\lambda(dx)
\qqq
accompany the Markov process described by Eq.\,(\ref{Lang}).
\vskip 0.4cm

\noindent{\bf Examples 3}. \ (Einstein-Smoluchowski) Brownian motion.
\vskip 0.2cm

\noindent In the even dimensional case set $x=(q,p)$, where  
$q$ is the position and $p$ the momentum. Let
\qq
&&\CM\,=\,\Big(\begin{matrix}0&\ -I\cr I&\ M^{-1}\end{matrix}\Big)\,,
\hspace{2.73cm}
\CD\,=\,\Big(\begin{matrix}0&\ 0\cr 0&\ M^{-2}D\end{matrix}\Big)\,,\cr
&&H_t(q,p)\,=\,\frac{_1}{^2}p\cdot m^{-1}p+U_t(x)\,,\qquad 
\ \,f_t(q,p)=(\phi_t(x),0)\,,
\qqq
where $M,D,m$ are half-dimension positive matrices, the first two
commuting. Then the Langevin stochastic equation takes the form
\qq
dq\,=\,m^{-1}p\,dt\,,\qquad dp\,=\,\Big(-M^{-1}m^{-1}p\,
-\,(\nabla U)(q)\,+\,\phi_t(q)\Big)\,dt\,+\,M^{-1}(2D)^{1/2}
\,dW(t)\,.\quad
\qqq
This is the {\,\bf underdamped\,} Langevin equation with the Hamiltonian
dynamics modified by the addition of friction and random forces. 
\,In this example, 
one uses the involution $\,(q,p)^*=(q,-p)\,$ for the time reversal. 
\,The Einstein relation reads here 
\qq
M\,=\,\beta\hspace{0.02cm}D
\label{E1}
\qqq
aligning the inverse friction coefficient $\,M\,$ with the diffusivity $\,D\,$
(physically, both friction and diffusion come from the same source: the 
interaction with molecules of the thermal environment). The case with 
$\,U_t=0\,$ and $\,\phi_t=0\,$ 
describes the Einstein-Smoluchowski Brownian motion. \,In the 
limit $\,m\to0$, \,the underdamped Langevin
equation reduces to the {\,\bf overdamped\,} one for $\,q(t)\,$ which reads:
\qq
dq\,=\,M(-\nabla U_t(q)+\phi_t)\,dt\,+\,(2D)^{1/2}dW(t)\,.
\label{Langov}
\qqq
This has again the form of Eq.\,(\ref{Lang}) with $\,x=q=x^*$, 
$\,\CM=M=\CM^t\,$ and $\,\CD=D$. 
\vskip 0.4cm

\noindent{\bf Examples 4}. \ L\'evy process.
\vskip 0.2cm

\noindent The (possibly nonstationary) {\,\bf L\'evy jump process\,} 
in $\,\mathbb R^d\,$ corresponds to the transition probability rates
\qq
w_t(x,dy)\,=\,w_t(d(y-x))
\qqq 
where $w_t(dy)$ is a positive measure.
\vskip 0.4cm

For a continuous-time Markov process on time interval $\,[0,\tau]$, 
\,the time-reversed process defined  by analogy to the one 
for the discrete-time case has the transition rates
\qq
w'_t(x,dy)\,=\,\frac{w_{t^*}(y^*,dx^*)}{\nu_{t^*}(dx^*)}\,
\nu_{t^*}(dy^*)\,,
\qqq
where $\,t^*\equiv\tau-t\,$ and an initial measure $\,\mu_0'$. 
\,By a formal limiting argument, we may infer that 
the fluctuation relations for the functional $\,\CW[\bm x]\,$ defined by
relation (\ref{comp}) carry over to the case of continuous-time Markov 
processes. In particular, writing 
\qq
&&\mu_0(dx)\,=\,\rho_0(x)\,\lambda(dx)\,,\hspace{0.88cm}\mu_0'(dx)\,=\,\rho'_0(x)\,
\lambda(dx^*)\,,\\ \cr
&&\nu_t(dx)\,=\,\ee^{-\varphi_t(x)}\,\lambda(dx)\,,\qquad
\nu'_t(dx)\,=\,\ee^{-\varphi_t'(x)}\,\lambda(dx^*)
\qqq
for $\,\varphi'_t(x)=\varphi_{t^*}(x^*)\,$ and
\qq
\CW[\bm x]\,=\,-\,\ln{\rho'_0(x_{\tau}^*)}-\varphi_{\tau}(x_{\tau})\,+
\int\limits_0^\tau\partial_t\varphi_t(x(t))\,dt\,
+\,\ln{\rho_0(x_0)}+\varphi_0(x_0)\,=\,-\CW'[\bm x^*]\,,\label{CWT}
\qqq
where $\,(\bm x^*)_t=x_{t^*}^*\,$ for $\,\bm x=(x_t)$, \,identity (\ref{comp})
still holds resulting in relations (\ref{5}) and (\ref{6}).
\,In the special case when $\,\mu_0=\nu_0\,$ and $\,\mu'_0=\nu'_0$, 
\,expression (\ref{CWT}) reduces to
\qq
\CW[\bm x]\,=\,\int\limits_0^\tau(\partial_t\varphi_t)(x_t)\,dt
\qqq
and we obtain the Hatano-Sasa equality \cite{HS01}
\qq
\boxed{\,\mathbb E_{\nu_0}\ \ee^{-\int\limits_0^\tau(\partial_t\varphi_t)(x_t)
\,dt}\,=\,1\,}\,,
\label{HSc}
\qqq
see Eq.\,(\ref{HS}), \,and, if the direct and the reversed process
have the same law, also the identity (\ref{7}).
\vskip 0.2cm

The detailed balance condition for the continuous-time
Markov process is defined similarly as for the discrete time and takes the form
\qq
\lambda(dx)\,w_t(x,dy)\,=\,\lambda(dy)\,w_t(y,dx)\,\ee^{-\beta(H_t(y)-H_t(x))}\,,
\label{DBc}
\qqq
compare to Eq.\,(\ref{detbal}).
It implies again that the Gibbs measures
\qq
\nu_t(dx)\,=\,Z_t^{-1}\,\ee^{-\beta H_t(x)}\,\lambda(dx)\,.
\label{nutHc}
\qqq
accompany the process. The transition rates for the time-reversed 
process take then the form
\qq
w'_t(x,dy)\,=\,w_{t^*}(x^*,dy^*)
\qqq
and one obtains (the continuum time versions of) the Jarzynski equality 
(\ref{jarz}) and of the Crooks identity (\ref{cr}) 
for
\qq
W[\bm x]\,=\,\int\limits_0^\tau(\partial_tH_t)(x_t)\,dt
\label{workc}
\qqq
and $\,W'(\bm x)\,$ given by the similar formula with $\,H_t(x)\,$
replaced by $\,H'_t(x)=H_{t^*}(x^*)$.  
\,Inequality (\ref{2workN}) also carries over to the continuous time 
case.

\subsection{Fluctuation relations for general diffusion processes}
\label{subsec:genfr}

\noindent The examples of fluctuation relations discussed above constitute 
a tip on an iceberg of more general relations of similar type.
In particular, in \cite{CG08}, we established a family of fluctuation 
relations for general diffusion processes solving stochastic equations 
(\ref{gendif}) on the time interval $\,[0,\tau]$. \,Upon an arbitrary 
division of the drift field
\qq
X_{0t}(x)\,=\,X^+_{0t}(x)+X^-_{0t}(x)\,,
\qqq
we defined the time-reversed diffusion process as the one corresponding 
to the drift and diffusion vector fields
\qq
X'_{0t}\,=\,(X^+_{0\hspace{0.01cm}t^*})^*-(X^-_{0\hspace{0.01cm}t^*})^*\,,
\qquad X'_{\alpha t}\,=\,(X_{\alpha\hspace{0.01cm}t^*})^*
\label{r12}\,,
\qqq
where $\,X^*$ is the push forward of the vector field $\,X\,$ by
an involution $\,x\mapsto x^*$, \,i.e.
\qq
(X^*)^i(x^*)\,=\,\frac{\partial{x^*}^i}{\partial x^j}\,X^j(x)
\qqq
in coordinates.
In other words, $\,X^+_0\,$ was chosen to transform by the vector 
and $\,X^-_0\,$ by the pseudo-vector rule under the involution $\,x\mapsto 
x^*$. \,The choice
of the rule for $\,X_\alpha\,$ is immaterial. We showed, combining 
the Girsanov and Feynman-Kac formulas, that in this case the functionals
$\,\CW[\bm x]\,$ and $\CW'[\bm x]\,$ defined by relation (\ref{comp})
have the form
\qq
\CW[\bm x]\,=\,-\,\ln\rho_0'(x_\tau^*)\,
+\,\int\limits_0^\tau\hspace{-0.1cm}
\CJ_t[\bm x]\,dt\,+\,
\ln\rho_0(x_0)\,=\,-\CW'[\bm x^*]\,,
\label{gCWT}
\qqq
where
\qq
\CJ_t[\bm x]\,=\,\widehat X^{+}_{0t}(x_t)
\cdot\CD_t^{-1}(x_t)\Big(\circ\frac{dx_t}{dt}-
X^{-}_{0t}(x_t)\Big)\,-\,\nabla\cdot X^{-}_{0t}(x_t)
\label{gCQ}
\qqq
in the notations of (\ref{hatcd}). \,For the sake of illustration, 
let us consider four particular time reversals. 

\subsubsection{\bf Case (a)}

\noindent First, we shall show that
the case studied before is, indeed, a particular instance of such
a general scheme. Upon taking 
\qq
\widehat X^+_{0t}(x)\,=\,-\,\CD_t(x)\,\nabla\varphi_t(x)
\qqq
for $\,\ee^{-\varphi_t}\,$ such that $\,L_t^\dagger\ee^{-\varphi_t}=0$,
\,we obtain
\qq
\CJ_t[\bm x]\,=\,-\,\nabla\varphi_t(x_t)\cdot\Big(\circ\frac{dx_t}{dt}-
X^{-}_{0t}(x_t)\Big)\,-\,\nabla\cdot X^{-}_{0t}(x_t)\,=\,
-\,\nabla\varphi_t(x_t)\cdot\circ\frac{dx_t}{dt}\,.
\label{CQ1}
\qqq
The last equality holds because, with the use of (\ref{hatcd}),
\qq
&&0\ =\ \ee^{\varphi_t}\hspace{0.03cm}
L_t^\dagger\ee^{-\varphi_t}\,=\,\ee^{\varphi_t}
\hspace{0.03cm}\partial_i\Big(-X_{0t}^i\hspace{0.03cm}\ee^{-\varphi_t}\,+\,
\frac{_1}{^2}\,X_{\alpha t}^i\hspace{0.02cm}\partial_j
\big(X_{\alpha t}^j\ee^{-\varphi_t}\big)\Big)\cr
&&=\,\ee^{\varphi_t}\hspace{0.03cm}\nabla\cdot\Big(-\widehat X_{0t}
\hspace{0.03cm}\ee^{-\varphi_t}\,+\,\widehat X_{0t}^{+}\hspace{0.02cm}
\ee^{-\varphi_t}\Big)
\,=\,-\,\ee^{\varphi_t}\hspace{0.03cm}\nabla\cdot\Big(X_{0t}^{-}
\hspace{0.03cm}\ee^{-\varphi_t}\Big)
\,=\,(\nabla\varphi_t)\cdot X^{-}_{0t}\,-\,\nabla\cdot X^{-}_{0t}\,.\qquad
\qqq
Now, \,since 
\qq
-\int\limits_0^\tau\nabla\varphi_t(x_t)\cdot\circ\hspace{0.02cm}dx_t\,=\,
-\varphi_\tau(x_\tau)+\varphi_0(x_0)\,
+\int\limits_0^\tau(\partial_t\varphi_t)(x_t)\,dt\,,
\qqq
functional (\ref{gCWT}) reduces in this case to expression
(\ref{CWT}). 

\subsubsection{\bf Case (b)}

\noindent In another important example that will be used below, let us take 
$\,X_{0t}=X_0+Y_{t}\,$ with $\,X_0\,$ and $\,X_\alpha\,$ time independent. 
\,Let $\,d\nu(dx)=\ee^{-\varphi(x)}\lambda(dx)\,$ be the invariant measure 
for $Y_{t}=0$. \,Taking 
\qq
\widehat X^+_{0t}\,=\,-\,\CD(x)\,\nabla\varphi(x)
\label{Xplus}
\qqq
and proceeding as before, we obtain
\qq
\CJ_t[\bm x]\,=\,-\nabla\varphi(x_t)\cdot\Big(\circ\frac{dx_t}{dt}
-Y_{t}(x_t)\Big)\,-\,\nabla\cdot Y_t(x_t)\,.
\label{CQ2}
\qqq  

\subsubsection{\bf Case (c)}
\label{subsubsec:revprot}

\noindent Taking $\,X^+_{0t}=X_{0t}\,$ so that $\,X^-_{0t}=0\,$ corresponds
to 
\qq
\CJ_t[\bm x]\,=\,\widehat X_{0t}(x_t)
\cdot\CD_t^{-1}(x_t)\circ\frac{dx_t}{dt}\,.
\qqq
Such time reversal is usually called the reversed protocol. 

\subsubsection{\bf Case (d)}

\noindent Finally, suppose that we split the drift vector field taking
\qq
\widehat X^{+}_{0t}(x)\,=\,0\,.
\qqq
In this case,
\qq
\CJ_t[\bm x]\,=\,-\,\nabla\cdot X^{-}_{0t}(x_t)\,.
\qqq 
The latter time reversal makes sense also in the limit of deterministic 
dynamical processes with $\,X_{\alpha t}=0\,$ when $\,X_{0t}^-=X_{0t}\,$ 
so that $\,\CJ_t[\bm x]\,$ becomes the phase-space contraction rate. It gives 
rise in that case to the Evans-Searles transient fluctuation 
relation \cite{ES94,ES02}.
\vskip 0.6cm

\noindent{\bf Example 5.} \,Consider the underdamped Langevin dynamics
of an {\,\bf anharmonic chain}, \,with phase-space points 
$\,x=(q_0,p_0,\dots,q_L,p_L)$, $\,q_i,p_i\in\mathbb R^d$, \,which is governed 
by the stochastic equations
\qq
dq_i\,=\,m^{-1}p_i\,dt\,,\qquad dp_i\,=\,\Big(\hspace{-0.1cm}-M_i^{-1} 
m^{-1}p_i-\nabla_{q_i}\hspace{-0.05cm}
U(\un{q})\Big)+(2\beta_i^{-1}M_i^{-1})^{1/2}dW_i(t)
\label{chain}
\qqq
with scalars $\,m,M_i,\beta_i>0\,$ and with potential
\qq
U(\un q)\,=\,\frac{k}{2}\sum\limits_{i=1}^{L}
(q_i-q_{i-1})^2
\,+\,\sum\limits_{i=0}^L\big(\frac{r}{2}q_i^2+\frac{g}{4}(q_i^2)^2
\big)
\qqq 
for $k,r,\hspace{0.02cm}g>0$. \,The backward generator of the corresponding 
Markov process
is
\qq
L\,=\,\sum\limits_{i=0}^L\Big(m^{-1}p_i\cdot
\nabla_{{q}_i}\,-\,(\nabla_{{q}_i}\hspace{-0.05cm}U)(\un q)\cdot
\nabla_{{p}_i}\,-\,M_i^{-1}m^{-1}p_i\cdot\nabla_{{p}_i}
\,+\,\beta^{-1}_iM_i^{-1}\nabla_{{p}_i}^2\Big).
\label{gen}
\qqq
If $\,\beta_i\equiv\beta\,$ then the dynamics 
(\ref{chain}) has the Gibbs state $\,\nu(dx)=Z^{-1}\,
\ee^{-\beta H(x)}\lambda(dx)\,$ as an invariant measure, \,where
\qq
H(x)\,=\,\sum\limits_{i=0}^L\frac{p_i^2}{2m}\,+\,U(\un q)
\qqq
is the Hamiltonian of the chain. We shall be mostly interested in the case
when the dynamics inside the chain is Hamiltonian, that is when
$\,M_i^{-1}=0\,$ for $\,i\not=0,L$. \,In that situation, 
the coefficients $\,\beta_i\,$ for $\,i\not=0,L\,$ disappear from 
the stochastic equations (\ref{chain}). \,Dividing the drift $\,X_0\,$
into the vector and pseudo-vector parts so that
\qq
X_0^+(x)\,=\,\big(0,-M_i^{-1}m^{-1}p_i\big)\,,\qquad 
X_0^-(x)\,=\,\big(m^{-1}p_i,
-\nabla_{q_i}\hspace{-0.05cm}U(\un q)\big)
\label{division}
\qqq
one infers from Eq.\,(\ref{gCQ}) that
\qq
\CJ_t[\bm x]\,=\,-\sum\limits_{i=0}^L
\beta_i\hspace{0.01cm}
m^{-1}p_{it}\cdot\big(\circ\frac{dp_{it}}{dt}+\nabla_{q_i}\hspace{-0.05cm}
U(\un q_t)\big)
\qqq
for such a choice. Note that the friction coefficients $\,M_i^{-1}\,$ do 
not enter into the latter expression which also does not depend on the
choice of interpolating $\,\beta_i$, $\,i\not=0,L$, \,if $\,M_i^{-1}=0\,$
for $\,i\not=0,L$. \,A simple calculation shows that
\qq
\CJ_t[\bm x]\,=\,\frac{d}{dt}\hspace{0.03cm}\ln\rho_0(x_t)\,+\,
\sum\limits_{i=1}^{L}(\beta_{i}-\beta_{i-1})\,j_{(i-1,i)}(x_t)\,,
\qqq
where
\qq
\mu_0(dx)&=&\rho_0(x)\,\lambda(dx)\cr
&=&Z_0^{-1}\exp\Big[\sum\limits_{i=0}^L\beta_i\big(\frac{p_i^2}{2m}
+\frac{r}{2}q_i^2+\frac{g}{4N}(q_i^2)^2\big)\,+\,
\sum\limits_{i=1}^{L}\frac{\beta_{i-1}+\beta_{i}}{2}\frac{k}{2}
(q_{i-1}-q_{i})^2\Big]\lambda(dx)\quad
\qqq
is the a {\bf\,local equilibrium\,} measure and
\qq
j_{(i-1,i)}(x)\,=\,\frac{k}{{2m}}(p_{i-1}+p_{i})\cdot(q_{i-1}-q_{i})
\label{hfl}
\qqq
is the \,{\bf energy} \,(or \,{\bf heat}) \,{\bf flux} \,from site 
$i-1$ to site $i$. \,Taking $\,\mu_0'=\mu_0$, \,we obtain 
from (\ref{gCWT}) the quantity
\qq
\CW[\bm x]\,=\,\sum\limits_{i=1}^L(\beta_i-\beta_{i-1})\int\limits_0^\tau
j_{(i-i,i)}(x_t)\,dt
\label{CWbetai}
\qqq
for which the transient fluctuation relations (\ref{5}) and (\ref{6}) hold
with all the implications, e.g. the inequality
\qq
\mathbb E_{\mu_0}\,\CW[\bm x]\,\geq\,0\,.
\label{CWnu}
\qqq
If the dynamics inside the chain is Hamiltonian then different choices
of $\,\beta_i\,$ for $\,i\not=0,L\,$ lead to different fluctuation relations
for the same system. E.g. for a linear interpolation between $\,\beta_0\,$ 
and $\,\beta_L$,
\qq
\CW[\bm x]\,=\,(\beta_L-\beta_0)\,
\frac{1}{L}\sum\limits_{i=1}^L\int\limits_0^\tau
j_{(i-i,i)}(x_t)\,dt\,,
\label{CWa}
\qqq
whereas for the piecewise constant interpolation with a jump between 
sites $i-1$ and $i$,
\qq
\CW[\bm x]\,=\,(\beta_L-\beta_0)\,\int\limits_0^\tau
j_{(i-i,i)}(x_t)\,dt\,.
\label{CWb}
\qqq
Different choices correspond to different local equilibrium measures $\,\mu_0(dx)\,$
none of which is a stationary state if $\,\beta_0\not=\beta_L\,$ because
\qq
L^\dagger\rho_0\,=\,\sum\limits_{i=1}^L(\beta_i-\beta_{i-1})
j_{(i-1,i)}\,\not\equiv0\,.
\qqq 
It was shown in refs. \cite{EPR99a,EPR99b} that in the stationary 
state with an invariant measure $\,\nu(dx)\,$ (that in fact has not been 
proven to exist for the version of the model that we consider, 
see however \cite{EPR99a}),
\qq
\mathbb E_\nu\,\CW[\bm x]\,\geq\,0
\label{CWmu}
\qqq
with the equality (for $\,\tau>0\,$) if and only if 
$\,\beta_0=\beta_N$. \,The expectation on the right hand side of inequality 
(\ref{CWmu}), unlike the one in (\ref{CWnu}), is independent on the choice 
of interpolating $\,\beta_i\,$ (why?). Taking $\,\CW[\bm x]\,$ in the 
form (\ref{CWb}), the latter result shows that, in the (putative) stationary 
state, the heat flows in average from the hot to the cold end of the chain, 
a reassuring result.

\nsection{Stochastic Thermodynamics}

\noindent Although of tautological origin, the fluctuation relations
considered in the previous section have important consequences
that permit to make contact with the thermodynamical concepts
in simple nonequilibrium situations relevant for the modelisation 
of dynamics of mesoscopic systems, like colloids, polymers, 
or bio-molecules, in contact with heat bath(s). Such considerations
make up an actively developing field known under the name
of Stochastic Thermodynamics, see \cite{S12} for a recent review.

\subsection{Entropy and entropy production}

\noindent Let us return to systems described by discrete-time Markov 
processes $\,(x_n)_{n=0}^{N+1}$. \,Assuming that the time-$n\,$ 
distributions $\,\mu_n(dx)\,$ of such a Markov process have the form 
$\,\rho_n(x)\,\lambda(dx)\,$ with positive densities relative 
to the reference measure $\,\lambda(dx)$, \,we shall define 
the time-$n\,$ fluctuating entropy of the system by the Boltzmann-type 
expression
\qq
S^{sys}_n[\bm x]\,=\,-k_B\,\ln{\rho_n(x_n)}\,.
\qqq
Note that its expectation value 
\qq
{\mathbb E}_{\mu_0}\,S^{sys}_n[\bm x]\,=\,-\,k_B\int\ln{\rho_n(x)}\,\rho_n(x)
\,\lambda(dx)\,\equiv\,S[\mu_n]
\qqq
is the Gibbs-Shannon entropy of measure $\,\mu_n$. \,The change 
of the fluctuating entropy of the system during the process 
is consequently given by the expression
\qq
\Delta S^{sys}[\bm x]\,\equiv\,S^{sys}_{N+1}[\bm x]\,-\,S^{sys}_{0}[\bm x]\,=
\,-\,k_B\,\ln{\rho_{N+1}(x_{N+1})}\,+\,k_B\,\ln{\rho_0(x_0)}\,.
\qqq
whose average is
\qq
{\mathbb E}_{\mu_0}\,\Delta S^{sys}[\bm x]\ =\ 
S[\mu_{N+1}]-S[\mu_0]\,\equiv\,\Delta S[\mu]\,.
\label{DGS}
\qqq
Similarly, for a continuous time 
Markov process $\,(x_t)_{0\leq t\leq\tau}\,$ with time-$t$ distributions
$\,\mu_t(dx)=\rho_t(x)\,\lambda(dx)$, \,we define the fluctuating entropy
of the system as
\qq
S^{sys}_t[\bm x]\,=\,-\,k_B\,\ln{\rho_t(x_t)}\,
\label{flent}
\qqq
with the change during the process
\qq
\Delta S^{sys}[\bm x]\,\equiv\,S^{sys}_\tau[\bm x]\,-\,S^{sys}_0[\bm x]
\,=\,-\,k_B\,\ln{\rho_\tau(x_\tau)}\,+\,k_B\,\ln{\rho_0(x_0)}
\qqq
and the averages
\qq
{\mathbb E}_{\mu_0}\,S^{sys}_t[\bm x]\,=\,S[\mu]\,,\qquad
{\mathbb E}_{\mu_0}\,\Delta S^{sys}[\bm x]\,=\,S[\mu_\tau]-S[\mu_0]\,\equiv\,
\Delta S[\mu]\,.
\label{DGSc}
\qqq
\vskip 0.2cm

Nonequilibrium processes change entropy of both the system and the
environment. The total entropy production in the units of the Boltzmann 
constant $\,k_B\,$ will be identified with the functional $\,\CW[\bm x]\,$ 
defined by Eq.\,(\ref{comp}) that compares the path measures of the direct
and the time-reversed processes,
\qq
\Delta S^{tot}[\bm x]\,=\,k_B\,\CW[\bm x]\,,
\qqq
provided that the initial distribution of the time-reversed process
is fixed by the final distribution of the direct process by the 
relations
\qq
\mu'_0(dx)\,=\,\begin{cases}\mu_{N+1}(dx^*)\cr
\mu_\tau(dx^*)\end{cases}
\label{tied}
\qqq
in the discrete- or continuous-time cases, respectively. 
Note that the total entropy change $\,\Delta S^{tot}\,$ defined this way 
is a fluctuating quantity depending on 
the trajectory $\,\bm x\,$ of the Markov process. 
Fluctuation relation (\ref{5}) for $\,\mu'_0\,$ fixed according to
(\ref{tied}) may now be rewritten as the identity
\qq
\boxed{\,{\mathbb E}_{\mu_0}\ \ee^{-k_B^{-1}\Delta S^{tot}[\bm x]}\ =\ 1}\,.
\label{SS}
\qqq
Similarly,  upon the introduction of the same notions for the 
time-reversed process, Eq.\,(\ref{6}) may be rewritten as the relation
\qq
\boxed{\,\ee^{-k_BS}\,p^{tot}(S)\,=\,p'^{tot}(-S)}\,,
\label{SSp}
\qqq 
where $\,p^{tot}(S)\,$ ($p'^{tot}(S)$) \,is the probability density
function of the total entropy production in the direct (time-reversed) process.
Note the similarity of relation (\ref{SS}) to the Hatano-Sasa one (\ref{HS}).
The latter, however, was obtained for a different choice of
the initial distributions, namely, for $\,\mu_0(dx)=\nu_0(dx)\,$ 
and $\,\mu'_0(dx)=\nu'_0(dx)$.  
\vskip 0.1cm

We may separate the total entropy production into two contributions: 
\qq
\Delta S^{tot}[\bm x]\,=\,\Delta S^{sys}[\bm x]\,+\,\Delta S^{env}[\bm x]
\qqq
where $\,\Delta S^{env}[\bm x]\,$
is the (fluctuating) entropy production in the environment. For $\,\CW[\bm x]\,$
given by Eqs.\,(\ref{CWN}) or (\ref{CWT}), we obtain:
\qq
\Delta S^{env}[\bm x]\,=\,\begin{cases}-k_B\,\varphi_N(x_{N+1})\,
+\,k_B\sum\limits_{n=1}^N
\big(\varphi_n(x_n)-\varphi_{n-1}(x_n)\big)\,+\,k_B\,\varphi_0(x_0)\,,\cr
-k_B\,\varphi_\tau(x_\tau)\,+\,k_B\int\limits_0^\tau
(\partial_t\varphi)_t(x_t)\,dt\,+\,k_B\,\varphi_0(x_0)\,,
\end{cases}
\label{Stot}
\qqq
for discrete or continuous time, respectively, 
whereas for the diffusion processes employing the more general 
time-reversal schemes discussed in Sec.\,\ref{subsec:genfr},
\qq
\Delta S^{env}[\bm x]\,=\,k_B\int\limits_0^\tau{\cal J}_t[\bm x]\,dt
\qqq
The entropy production in the environment depends in the latter case on 
the choice of the time-reversal but the fluctuation relations (\ref{SS}) 
and (\ref{SSp}) hold for all choices.

\subsection{$1^{\rm st}$ Law of Stochastic Thermodynamics}

\noindent In order to motivate the introduction of the above thermodynamical
concepts in the general situation, let us consider the special case when 
the transition probabilities
satisfy the detailed balance conditions (\ref{detbal}) or (\ref{DBc}) and
and the accompanying measures have the Gibbsian form 
(\ref{nutH}) or (\ref{nutHc}).
The total entropy production given by Eqs.\,(\ref{Stot})
may be rewritten in that case in  the form 
\qq
\Delta S^{env}[\bm x]\,=\,\frac{W[\bm x]\,-\,\Delta H[\bm x]}{T}
\label{DSen}
\qqq
where $\,W[\bm x]\,$ is the work given by Eqs. (\ref{WNj}) or (\ref{workc}),
\qq
\Delta H[\bm x]\,\,=\,\begin{cases}H_N(x_{N+1})\cr H_\tau(x_\tau)\end{cases}
\hspace{-0.45cm}\Bigg\}\,-\,H_0(x_0)
\qqq 
is the change of energy along trajectory $\,\bm x$, \,and $\,T\,$ is 
the temperature. 
\,The $\,{\bf 1}^{\rm\bf st}$ {\bf Law of Thermodynamics\,}
expressing the conservation of energy asserts that work minus dissipated
heat is equal to the change of internal energy of the system.
We shall impose it on the level of fluctuating quantities defining 
heat dissipated along trajectory $\,\bm x\,$ as
\qq
Q[\bm x]\,=\,W[\bm x]\,-\,\Delta H[\bm x]\,.
\label{dissh}
\qqq
With such a definition, expression for the entropy production
in environment (\ref{DSen}) becomes the Clausius-type relation
\qq
\Delta S^{env}[\bm x]\,=\,\frac{Q[\bm x]}{T}
\label{Claus}
\qqq
for the change of the entropy in isothermal quasi-stationary 
processes. The use of such a relation is well justified for an environment 
with a very fast relaxation leading to a Markovian evolution of the system
interacting with it.

\subsection{$2^{\rm nd}$ Law of Stochastic Thermodynamics}

\noindent From the fluctuation relation (\ref{SS}) it follows 
by the Jensen inequality that
\qq
\boxed{\,{\mathbb E}_{\mu_0}\,\Delta S^{tot}[\bm x]\ \geq\ 0\,.}
\label{2ndLaw}
\qqq
This is the ${\bf2}^{\rm\bf nd}$ {\bf Law of Stochastic Thermodynamics\,} 
formulated
as the Clausius inequality. In the case with detailed balance,
where 
\qq
\Delta S^{tot}[\bm x]\,=\,\Delta S^{sys}[\bm x]\,+\,\frac{Q[\bm x]}{T}\,=\,
\Delta S^{sys}[\bm x]\,+\,\frac{W[\bm x]-\Delta H[\bm x]}{T}\,,
\label{expfl}
\qqq
the Clausius inequality (\ref{2ndLaw}) may be rewritten as the lower bound
on the mean dissipated heat:
\qq
\boxed{\,{\mathbb E}_{\mu_0}\,Q[\bm x]\ \geq\ -T\hspace{0.07cm}
{\mathbb E}_{\mu_0}\,\Delta S^{sys}[\bm x]\,=\,-T\,\Delta S[\mu]}\,,
\label{2ndLawp}
\qqq
see Eqs.\,(\ref{DGS}) and (\ref{DGSc}), \,or as the bound
\qq
{\mathbb E}_{\mu_0}\,W[\bm x]\ \geq\ -\,{\mathbb E}_{\mu_0}\,\big(\Delta H[\bm x]
\,-\,T\Delta S^{sys}[\bm x]\big)\ =\ \Delta F\,-\,T\,S[\mu_0\Vert\nu_0]\,+\,
T\begin{cases}S[\mu_{N+1}\Vert\nu_N]\,,\cr
S[\mu_\tau\Vert\nu_\tau]\,,\end{cases}
\label{thisineq}
\qqq
where for two probability measures $\,\mu(dx)\,$ and $\,\nu(dx)\,$ absolutely
continuous with respect to each other, $\,S[\mu\Vert\nu]\,$ denotes
their relative entropy defined by the formula
\qq
S[\mu\Vert\nu]\,=\,\int\ln\Big(\frac{\mu(dx)}{\nu(dx)}\Big)\,\mu(dx)\,.
\qqq
The relative entropy vanishes for $\,\mu=\nu\,$ and is positive in all
other cases. Taking as the initial measure of the
process $\,\mu_0\,$ the accompanying time-zero Gibbs measure $\,\nu_0$, 
\,we infer from inequality (\ref{thisineq}) that the $2^{\rm nd}$ Law of 
Stochastic Thermodynamics (\ref{2ndLaw}) is in this case stronger 
than the bound (\ref{2workN}) established before, a fact not always 
appreciated in the literature on Stochastic Thermodynamics where the 
latter bound is often assimilated with the $2^{\rm nd}$ Law.
\vskip 0.1cm

For the diffusion processes employing the general time-reversal, as
discussed in Sec.\,\ref{subsec:genfr}, the direct calculation 
\cite{CG08} gives 
\qq 
{\mathbb E}_{\mu_0}\,\Delta S^{tot}[\bm x]\,=\,k_B\int\limits_0^\tau dt
\int\limits_{\cal X}\big[\big(\widehat X^+_{0t}-{\cal D}_t\nabla\rho_t\big)(x)
\cdot(\rho_t\CD_t)^{-1}\big(\widehat X^+_{0t}-{\cal D}_t\nabla\rho_t\big)\big](x)
\,\lambda(dx)\,,
\qqq 
where $\,\rho_t(x)\,$ denote, as before, the densities of the time-$t\,$ 
distributions $\,\mu_t(dx)\,$ of the process w.r.t. the reference
measure $\,\lambda(dx)$, see Sec.\,\ref{subsec:means} below for
the derivation of the above relation in a special case. \,The right hand
side is an explicitly non-negative expression that for the time 
reversal (c) of Sec.\,\ref{subsubsec:revprot} reduces to the identity
\qq
{\mathbb E}_{\mu_0}\,\Delta S^{tot}[\bm x]\,=\,k_B\int\limits_0^\tau dt
\int\limits_{\cal X}\big[v_t\cdot\CD_t^{-1}v_t\big](x)\,
\rho_t(x)\,\lambda(dx)\,,
\label{spcas} 
\qqq 
where $\,v_t(x)\,$ is the current velocity given by (\ref{currv0}).

\subsection{Work heat and entropy production in overdamped Langevin dynamics}
\label{subsec:means}

\noindent To illustrate further the above discussion, let us
consider a system described by an overdamped Langevin equation 
(\ref{Langov}) in $\mathbb R^d$
with positive mobility and diffusivity matrices $M$ and $D$
related by the Einstein relation (\ref{E1}).
Stochastic equation (\ref{Langov}), together with the initial
distribution $\,\mu_0(dq)$, \,defines a nonstationary 
continuous-time Markov process $\,(q_t)_{0\leq t\leq\tau}\,$ with 
transition rates
\qq
w_t(q,dq')\,=\,\big(-M(\nabla U_t)(q)+D\nabla_q\big)
\cdot\nabla_q\hspace{0.01cm}\delta(q-q')\,\lambda(dq')
\qqq
satisfying the detailed balance relations
\qq
\lambda(dq)\,w_t(q,dq')\,=\,\lambda(dq')\,w_t(q',dq)
\,\,\ee^{-\beta\big(U_t(q')-U(q)\big)}
\qqq
(check it!). Consequently, the Gibbs measures
\qq
\nu_t(dq)\,=\,Z_t^{-1}\,\ee^{-\beta\hspace{0.01cm}U_t(q)}\lambda(dq)
\qqq
accompany the process, i.e. $\,L_t^*\nu_t=0\,$
for the backward generators
\qq
L_t=\big(-M(\nabla U_t)+D\nabla\big)\cdot\nabla\,.
\label{Ltq}
\qqq
On the other hand, the time-$t\,$ probability distributions 
$\,\mu_t(dq)=\rho_t(q)\lambda(dq)\,$ 
evolve according to the advection equation (\ref{adv0}) 
with the current velocity
\qq
v_t(q)\,=\,-\,M(\nabla U_t)(q)-D\nabla\ln\rho_t(q)\,.
\label{vt}
\qqq
Definitions (\ref{workc}) and (\ref{dissh}) give here the expressions 
\qq
W[\bm q]\,=\,\int\limits_0^\tau(\partial_t U_t)(q_t)\,dt
\label{W}
\qqq
for the {\,\bf work\,} performed on the system
during time interval $[0,\tau]\,$ and 
\qq
Q[\bm q]\,=\,W[\bm q]-\Delta U[\bm q]\,=\,-\int\limits_0^\tau(\nabla U_t)
(q_t)\circ dq(t)
\label{Q}
\qqq
for the {\,\bf heat\,} dissipated
in the same time interval. The assignment of the names
may seem somewhat arbitrary and counter-intuitive (it is
the right-hand-side of (\ref{Q}) that looks as the work of 
the gradient force). To understand it better, let us consider
an example of a simple system: a bead connected by a spring to
the wall whose position $\,y_t\,$ may be manipulated externally,
see FIG.\,2.

\begin{figure}[th]
\leavevmode
\begin{center}
\vskip -0.2cm
\hspace*{-0.3cm}
\includegraphics[width=6cm,height=2.5cm]{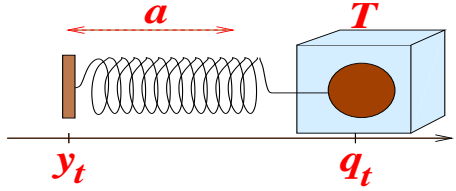}\\
\caption{A simple system undergoing an overdamped
Langevin dynamics}
\end{center}
\end{figure}
 
\noindent The position $\, q_t\,$ of the bead satisfies the simple overdamped
Langevin equation (\ref{Langov}) with the time-dependent potential 
$\,U_t(q)=\frac{k}{2}(q-a-y_t)^2\,$ if the inertia of the bead can be 
neglected. In this case,
\qq
W[\bm q]\,=\,-k\int\limits_0^\tau(q_t-a-y_t)\dot{y}_t\hspace{0.03cm}dt
\label{workL}
\qqq
is the work performed when externally manipulating the wall and
\qq
Q[\bm q]\,=\,-k\int\limits_0^\tau(q_t-a-y_t)\circ dq_t
\label{heatL}
\qqq
is the work done on the bead by the force exerted by the spring 
that, in the overdamped regime, is entirely dissipated as heat due to 
friction. For more on the rational behind the definition
(\ref{workL}) of work, see \cite{J07}. The mean values of the work and heat
are given by the formulae
\qq
{\mathbb E}_{\mu_0}\,W[\bm q]&=&\int\limits_0^\tau dt
\int(\partial_tU)(q)\,\rho_t(q)\,\lambda(dq)\,,\cr
{\mathbb E}_{\mu_0}\,Q[\bm q]&=&{\mathbb E}_{\mu_0}\,
\big(W[\bm q]-\Delta U[\bm q]\big)\,=\,\int\limits_0^\tau dt
\bigg[\int(\partial_t U_t)(q)\,\rho_t(q)\,\lambda(dq)
-\frac{d}{dt}\int U_t(q)\,\rho_t(q)\,\lambda(dq)\bigg]\cr
&=&-\int\limits_0^\tau dt\int U_t(q)\,\partial_t\rho_t(q)\,\lambda(dq)
\,=\,\int\limits_0^\tau dt\int U_t(q)\,\nabla\cdot(\rho_t v_t)(q)\,
\lambda(dq)\cr
&=&-\int\limits_0^\tau dt
\int(\nabla U_t)(q)\cdot(\rho_tv_t)(q)\,\lambda(dq)\,,
\qqq
where we used the advection equation (\ref{adv0}) assuming that there
are no boundary contributions from the integration by parts.
\vskip 0.1cm

The fluctuating entropy of the system is represented for the 
overdamped Langevin equation (\ref{Langov}) by the expression
\qq
S^{sys}_t[\bm q]\,=\,-k_B\,\ln{\rho_t(q_t)}\,,
\qqq  
see (\ref{flent}), and the fluctuating total entropy production by
\qq
\Delta S^{tot}[\bm q]\,=\,\Delta S^{sys}[\bm q]+\frac{Q[\bm q]}{T}\,.
\label{summ}
\qqq
For the averages, \hspace{0.02cm}this gives:
\qq
&&{\mathbb E}_{\mu_0}\,\Delta S^{sys}[\bm q]\,=\,
\Delta S[\mu]\,
=\,-\,k_B\int\limits_0^\tau dt\,\frac{_d}{^{dt}}\hspace{-0.1cm}
\int\ln\rho_t(q)\,
\rho_t(q)\,\lambda(dq)\cr
&&=k_B\int\limits_0^\tau\hspace{-0.08cm}dt\hspace{-0.08cm}
\int\big[\nabla\cdot(\rho_tv_t)(q)\,
+\,\ln\rho_t(q)\,\nabla\cdot(\rho_tv_t)(q)\big]\,\lambda(dq)\cr
&&=\,-k_B\int\limits_0^\tau\hspace{-0.07cm}dt\hspace{-0.07cm}
\int\hspace{-0.07cm}(\nabla\ln\rho_t)(q)\,(\rho_tv_t)(q)\,\lambda(dq)\,,
\qquad\quad
\label{tos1}
\qqq
and from (\ref{summ}), using relation (\ref{vt}), we obtain:
\qq
{\mathbb E}_{\mu_0}\,\Delta S^{tot}[\bm q]&=&k_B\int\limits_0^\tau
\big[-\beta(\nabla U_t)-(\nabla\ln{\rho_t})\big](q)\cdot(\rho_t v_t)(q)\,
\lambda(dq)\cr
&=&k_B\int\limits_0^\tau
dt\int v_t(q)\cdot D^{-1}v_t(q)\,\rho_t(q)\,\lambda(dq)\,,
\label{BB}
\qqq
which is the special case of identity (\ref{spcas}).

\subsection{Landauer Principle}

\noindent The $2^{\rm nd}$ Law (\ref{2ndLaw}), rewritten for the non-stationary 
Markov dynamics satisfying the detailed balance as inequality 
(\ref{2ndLawp}), bounds the dissipated heat in terms of the change of 
the Gibbs-Shannon entropy between initial and final statistical states 
of the system. In this form, it is closely related to the principle
formulated by Landauer in 1961 \cite{L61}, see also \cite{B82}, 
stating that the erasure of one bit of information during a computation 
process conducted in thermal environment requires a release of heat 
equal to at least $\,k_BT\hspace{0.01cm}\ln{\hspace{-0.03cm}2}\,$ 
(in average).
As an example, consider a bi-stable system which may be in two distinct 
states and which undergoes a process that at final time leaves it always in, 
say, the second of those states, with a loss of memory of the initial state.
Such a device may be realized in the context of Stochastic Thermodynamics 
by an appropriately designed overdamped Langevin process 
$\,(q_t)_{0\leq t\leq\tau}\,$ that starts from 
the Gibbs state corresponding to a potential $\,R_0(q)=-k_BT\ln\rho_0(q)\,$ 
with two symmetric wells separated by a high barrier and ends in a Gibbs 
state corresponding to a potential $\,R_{\tau}(q)=-k_BT\ln\rho_{\tau}(q)\,$ 
with only one of those wells, see FIG.\,3.

\begin{figure}[H]
\begin{center}
\leavevmode
\vskip -0.3cm
\hspace*{-0.1cm}
        \includegraphics[width=8.5cm,height=4.8cm]{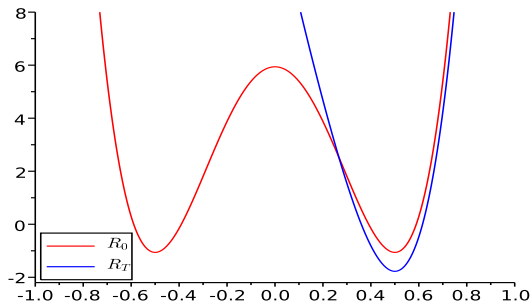}
\vskip -0.2cm
        \caption{Initial and final Gibbs potentials in a memory erasure 
process}
\end{center}
\end{figure}

\noindent The change of the Gibbs-Shannon
entropy of the system in such a process is approximately 
\qq
\Delta S[\mu]\,\approx\,-\,\ln{\hspace{-0.05cm}1}
+\,2\hspace{0.03cm}(\ln{\hspace{-0.04cm}\frac{_1}{^2}})\hspace{0.03cm}
\frac{_1}{^2}\hspace{0.03cm}\,
=\,-\hspace{0.03cm}\ln{\hspace{-0.04cm}2}\,,
\label{SLand}
\qqq
with the better and better approximation the deeper the wells or the lower 
the temperature. Landauer's lower bound for average heat release follows now
from inequality (\ref{2ndLawp}) reading
\qq
{\mathbb E}_{\mu_0}\,Q[\bm q]\,\geq\,-T\Delta S[\mu]
\label{Q>S}
\qqq
in the case at hand.

\subsection{Finite time Thermodynamics for overdamped Langevin processes}

\noindent It is well known that the $2^{\rm nd}$ Law bound can be
saturated for quasi-stationary processes that move infinitely
slowly so that at intermediate times the instantaneous 
measures $\mu_t$ are (almost) equal to the accompanying 
measures $\nu_t$.  
\,Suppose however, that we cannot afford to go too slowly. Indeed, in 
computational devices, we are interested in fast dynamics that arrives 
at the final state quickly but produces as little heat as possible. We 
are therefore naturally led to two questions:
\vskip -0.55cm
\ 

\begin{itemize}
\item What is the lower bound for the total entropy production or the 
average heat release in a process that interpolates between given 
states in a time interval of fixed duration?
\item What is the dynamical protocol that leads to such a minimal
total entropy production or heat release?
\end{itemize}
\ 

\vskip -0.6cm
\noindent These questions make sense in a variety of setups. 
They are among the core 
ones of the so called Finite-Time Thermodynamics \cite{A11} 
that was developed during last decades mostly with an eye on technological 
applications. In the context of Stochastic Thermodynamics, they were first 
asked and solved for Gaussian overdamped Langevin processes in 
\cite{SS07}. Here we shall study them in the framework 
of general overdamped Langevin equations (\ref{Langov}) following 
\cite{AMM11,AGM12}. 
\vskip 0.1cm

\subsection{Benamou-Brenier minimization and optimal mass transport}
\label{subsec:BB}

\noindent We would then like to find the minimum of the right hand side of 
Eq.\,(\ref{BB}) over
all control potentials $\,U_t\,$ that lead to the overdamped Langevin 
evolution (\ref{Langov}) from the fixed initial density 
$\,\rho_0(q)\,$ to the fixed final one $\,\rho_{\tau}(q)\,$ in the fixed 
time interval $\,[0,\tau]$. \,In \cite{BB97,BB99}, \,Benamou-Brenier have
solved a closely related problem of minimization of the functional
\qq
\int\limits_0^\tau dt\int v_t(x)^2\,\rho_t(x)\,\lambda(dx)
\qqq
over velocity fields $\,v_t(q)\,$ constrained by the advection equation
(\ref{adv0}), with the densities $\,\rho_t\,$ fixed at the initial 
and final times. Introducing the positive matrix $\,D^{-1}\,$ 
into the above functional does not pose a problem (this may be done by
a linear change of variables). Note, however, that the current velocities
$\,v_t\,$ that appear on the right hand side of (\ref{BB}) are of
the special gradient form, see Eq.\,(\ref{vt}). Let us first ignore 
the latter restriction at the price that adapting the argument 
of \cite{BB97,BB99} we may end up with a non-optimal bound 
(this will prove not to be the case {\it a posteriori}). Introducing 
for velocities $\,v_t\,$ the Lagrangian flow 
$\,q_t(\bm x)\,$  satisfying the ODE
\qq
\partial_t q_t(q_0)\,=\,v_t(q_t(q_0))\,,\qquad q_t(q_0)|_{t=0}\,=\,q_0\,,
\qqq
and writing the solution of the advection equation in the form
\qq
\rho_t(q)\,=\,\int\delta\big(q-q_t(q_0)\big)\,\rho_0(q_0)\,\lambda(dq_0)\,,
\label{dens}
\qqq
one may transform the functional on the right hand side of \m Eq.\,(\ref{BB})
as follows:
\qq
\int\limits_0^\tau dt\int
v_t(q)\cdot D^{-1}v_t(q)\,\rho_t(q)\,\lambda(dq)&=&
\int\limits_0^\tau dt\int\lambda(dq)\int v_t(q)\cdot D^{-1}v_t(q)\,\,
\delta(q-q_t(q_0)\,\lambda(dq_0)\cr
&=&\int\limits_0^\tau dt\int(\partial_tq_t)(q_0)\cdot D^{-1}
(\partial_tq_t)(q_0)\,\,\rho_0(q_0)\,\lambda(dq_0)\cr
&=&\int\Big(\int\limits_0^\tau (\partial_tq_t)(q_0)\cdot D^{-1}
(\partial_tq_t)(q_0)\,dt\Big)\,\rho_0(q_0)\,\lambda(dq_0)\,.\qquad
\qqq
In the first step, one minimizes the right hand side over the curves 
$\,t\mapsto q_t(q_0)\,$ with fixed endpoint $\,q_{\tau}(q_0)$.
The minima are realized on straight lines leading to the
functional
\qq
\frac{1}{\tau}\int\int(q_{\tau}(q_0)-q_0)\cdot D^{-1}
(q_{\tau}(q_0)-q_0)\,\,\rho_0(q_0)\,\lambda(dq_0)\ \equiv\ \frac{1}{\tau}\,
K[q_\tau(\cdot)]
\label{cost}
\qqq
that, multiplied by $\,\tau\m$ is, by definition, the quadratic 
cost function $\,K[q_\tau(\cdot)]\,$ of the map 
$\,q_0\mapsto q_{\tau}(q_0)$.
\,In the second step, one is left with the celebrated {\,\bf optimal 
mass transport\,} problem of Monge-Kantorovich \cite{V03} consisting 
of the minimization
of quadratic cost $\,K[q_\tau(\cdot)]\,$ over the diffeomorphisms 
$\,q_0\mapsto q_{\tau}(q_0)\,$ under the constraint that
\qq
\int\limits  
\int\delta\big(q-q_{\tau}(q_0)\big)\,\rho_0(q_0)\,\lambda(dq_0)\,=\,\rho_{\tau}(q)\,,
\qqq
or, \,equivalently, that
\qq
\rho_{\tau}(q_{\tau}(q_0))\,\frac{\partial(q_{\tau}(q_0))}{\partial(q_0)}\,=\,\rho_0(q_0)\,,
\qqq
i.e. demanding that the map $\,q_0\mapsto q_{\tau}(q_0)\,$ transports the density
$\,\rho_0$ to $\rho_{\tau}$. \,One of the results of the optimal mass transport
theory states that if the normalized densities $\,\rho_0\,$ and 
$\,\rho_{\tau}\,$ are positive, smooth and have the $2^{\rm nd}$ moments 
then the minimal cost is attained on a unique
diffeomorphism that is a gradient of a smooth convex function
\qq
q_{\tau}(q_0)\,=\,D\nabla\m\Psi(q_0)\,.
\label{grad}
\qqq
The corresponding minimizing velocity $\,v_t\,$ with the linear Lagrangian 
flow $\,q^{lin}_t(q_0)\,=\,tq_0+(1-t)q_{\tau}(q_0)\,$ satisfies the {\,\bf
inviscid Burgers equation}
\qq
\partial_t v\,+\,v_t\cdot\nabla v_t\,=\,0
\qqq
(which just states that the Lagrangian trajectories have no acceleration).
Even more importantly for us, as a consequence of Eq.\,(\ref{grad}),
the minimizing velocity $\,v_t\,$ is also of a gradient type:
\qq
v_t\,=\,D\nabla \psi_t\,=\,0\,,
\label{vtgr}
\qqq
where the function
\qq
\psi_t(q)\,=\,\frac{1}{2t}(q-q_0)\cdot D^{-1}(q-q_0)-\frac{1}{2\tau}
q_0\cdot D^{-1}q_0+\frac{1}{\tau}\Psi(q_0)\qquad{\rm for}
\qquad q=q^{lin}_t(q_0)
\label{psit}
\qqq
satisfies the {\,\bf Hamilton-Jacobi equation}
\qq
\partial_t\psi_t\,+\,\frac{_1}{^2}(\nabla\psi_t)\cdot D(\nabla\psi_t)\,
=\,0\,.
\qqq  
The corresponding interpolating densities $\,\rho_t\,$ are given 
by Eq.\,(\ref{dens}):
\qq
\rho_t(q)\,=\,\int\delta\big(q-q^{lin}_t(q_0)\big)\,\rho_0(q_0)\,
\lambda(dq_0)\,.
\label{dens1}
\qqq
Relation (\ref{vtgr}) means that, although the Benamou-Brenier
minimization was over general velocities not necessarily
of the gradient type, \,the minimizer $\,v_t\,$ is a current velocity 
for the overdamped Langevin process with control potential $\,U_t\,$ 
such that 
\qq
\nabla U_t\,=\,-\beta^{-1}\nabla\big(\psi_t+\ln\rho_t(q)\big),
\label{optpr}
\qqq
which fixes $\,U_t\,$ up to a time-dependent constant. In particular
the Benamou-Brenier minimizer also minimizes the right hand side
of Eq.\,(\ref{BB}) over the current velocities of the Langevin
processes (\ref{Langov}).

\subsection{Finite-time refinement of the $2^{\rm nd}$ Law and of the Landauer 
bound}
\label{subsec:FTT}

\noindent We obtain this way
\vskip 0.4cm

\noindent{\bf Theorem \,(Finite-time refinement of the $\bf2^{\bf\rm nd}$ Law).}
\ The mean total entropy production in an overdamped Langevin evolution
during time $\tau$ between the states with probability densities $\rho_0$ 
and $\rho_{\tau}$ satisfies the bound 
\qq
\boxed{\,\mathbb E_{\mu_0}\,\Delta S^{tot}[\bm q]\ \geq\ 
\frac{k_B}{\tau}\,K_{min}[\rho_0,\rho_{\tau}]\,}
\label{2ndLawft}
\qqq
where $\,K_{min}[\rho_0,\rho_{\tau}]\,$ is the minimal quadratic cost
of transport of density $\rho_0$ to $\rho_{\tau}$, see (\ref{cost}).
The above bound is saturated by the optimal protocol
with the control potential satisfying Eq\,.(\ref{optpr}), where 
$\,\psi_t\,$ and $\,\rho_t\,$ are given by Eqs. (\ref{psit}) and 
(\ref{dens}), respectively, in terms of the linear interpolation 
$\,q^{lin}_t(q_0)\,$ between $q_0$ and its image $q_{\tau}(q_0)$
under the optimal transport map.
\vskip 0.4cm

The above result has a geometric interpretation. 
The minimal quadratic
cost $\,K_{min}(\rho_0,\rho_{\tau})\,$ is, by definition, the square of
the {\,\bf Wasserstein distance\,} between the measures 
$\mu_0$ and $\mu_{\tau}$
that, formally, corresponds to the Riemannian metric on the 
space of probability densities \cite{JKO98}, with the square 
of the tangent vectors given by
\qq
\Vert\partial_t\rho\Vert^2_W\,=\,\int\big[(\partial_t\rho)
(-\nabla\cdot\rho D\hspace{0.02cm}
\nabla)^{-1}(\partial_t\rho)\big](q)\,\lambda(dq)\,.
\label{Wa}
\qqq
The Fokker-Planck equation for the (\ref{FP}) corresponds to the
gradient flow in metric (\ref{Wa}) for the functional
\qq
\beta\CF_t[\rho]\,=\,\int\big[\beta U_t+\ln\rho](q)\,\rho(q)\,\lambda(dq)
\qqq  
equal, up to the factor $\,\beta$, \,to the free energy $\,\CF_t[\rho]$, 
\,and one has
\qq
\mathbb E_{\mu_0}\,\Delta S^{tot}[\bm q]\,=\,
k_B\int\limits_0^\tau\Vert\partial_t\rho_t\Vert^2_W\,dt\,\geq\,
\frac{k_B}{\tau}\,d_W(\mu_0,\mu_{\tau})^2
\qqq
with the optimal protocol giving the (shortest) geodesic between $\rho_0$
and $\rho_{\tau}$.
\vskip 0.4cm

\noindent{\bf Corollary.} \ Under the same assumptions,
the mean heat release satisfies the bound
\qq
\mathbb E_{\mu_0}\,Q_{\tau}[\bm q]\,\geq\,=\,-
T\Delta S[\mu]\,+\,\frac{k_B}{\tau}
\,K_{min}[\rho_0,\rho_{\tau}]\,,
\label{Q>Sft}
\qqq
with the inequality saturated by the same optimal protocol.
\vskip 0.4cm

The latter inequality providing a finite-time refinement of the estimate 
(\ref{Q>S}), implies also a finite-time refinement
of the Landauer bound in the situation where the change of
the mean system entropy is given by Eq.\,(\ref{SLand}). Such
a refinement may be relevant in future computer designs \cite{S11}
(the present day computers still dissipate much more heat
than the minimum allowed by the thermodynamical considerations).
A recent experiment \cite{B12} with a colloidal particle manipulated
by laser tweezers measured the heat released in a process of
memory erasure interpolating at room temperature between two states 
with Gibbs potential from FIG.\,2, with the results plotted in 
FIG.\,3.

\begin{figure}[H]
\begin{center}
\leavevmode
\vskip -0.3cm
\hspace*{-0.3cm}
        \includegraphics[width=4.7cm,height=8.5cm,angle=-90]{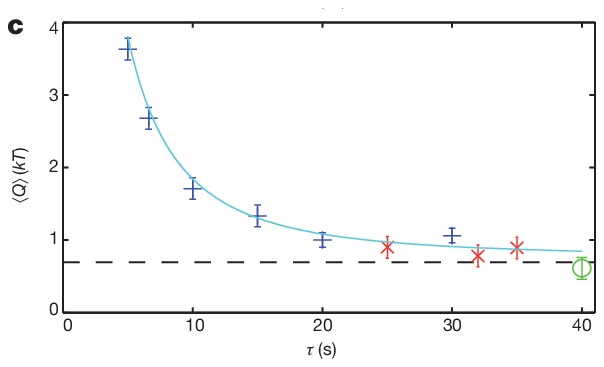}\\
        \caption{Mean heat release as a function of time in the memory 
                 erasure experiment of \cite{B12}}
\end{center}
\end{figure}

\noindent The control potential used in the experiment did not follow 
the optimal protocol and, as a result, the heat release in the $10s$ 
run exceeded $2.5$ times the Landauer bound instead of the optimal $40\%$. 
    
\begin{figure}[H]
\begin{center}
\leavevmode
\vskip -0.2cm
\hspace*{0.7cm}
{%
      \begin{minipage}{0.45\textwidth}\hspace*{-0.4cm}
        \includegraphics[width=7cm,height=5cm,angle=0]{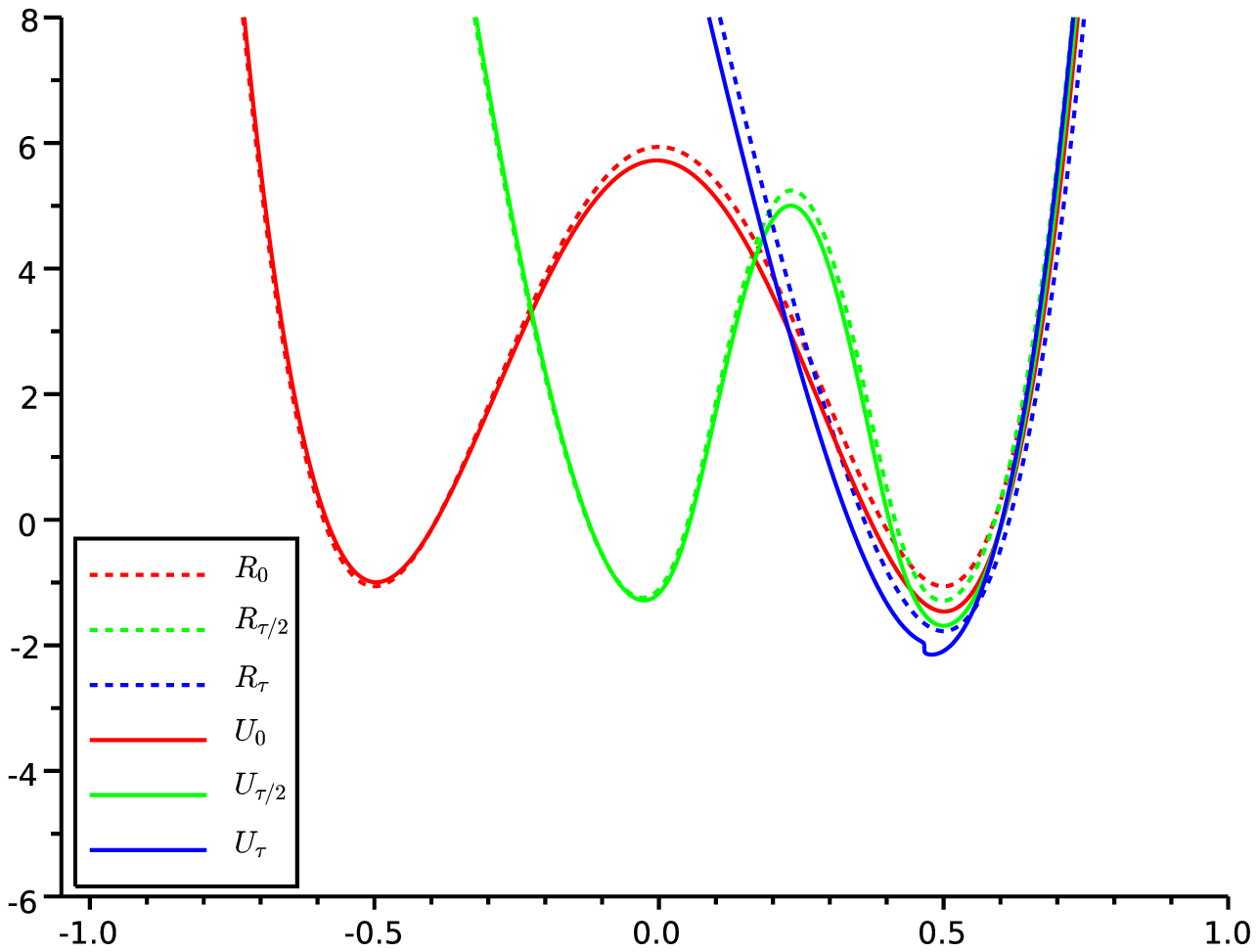}\\
        \vspace{-0.73cm} \strut
        \end{minipage}}
    \hspace*{-0.1cm}
{%
      \begin{minipage}{0.45\textwidth}\vspace{0.17cm}\hspace*{-0.3cm}
        \includegraphics[width=7cm,height=5cm]{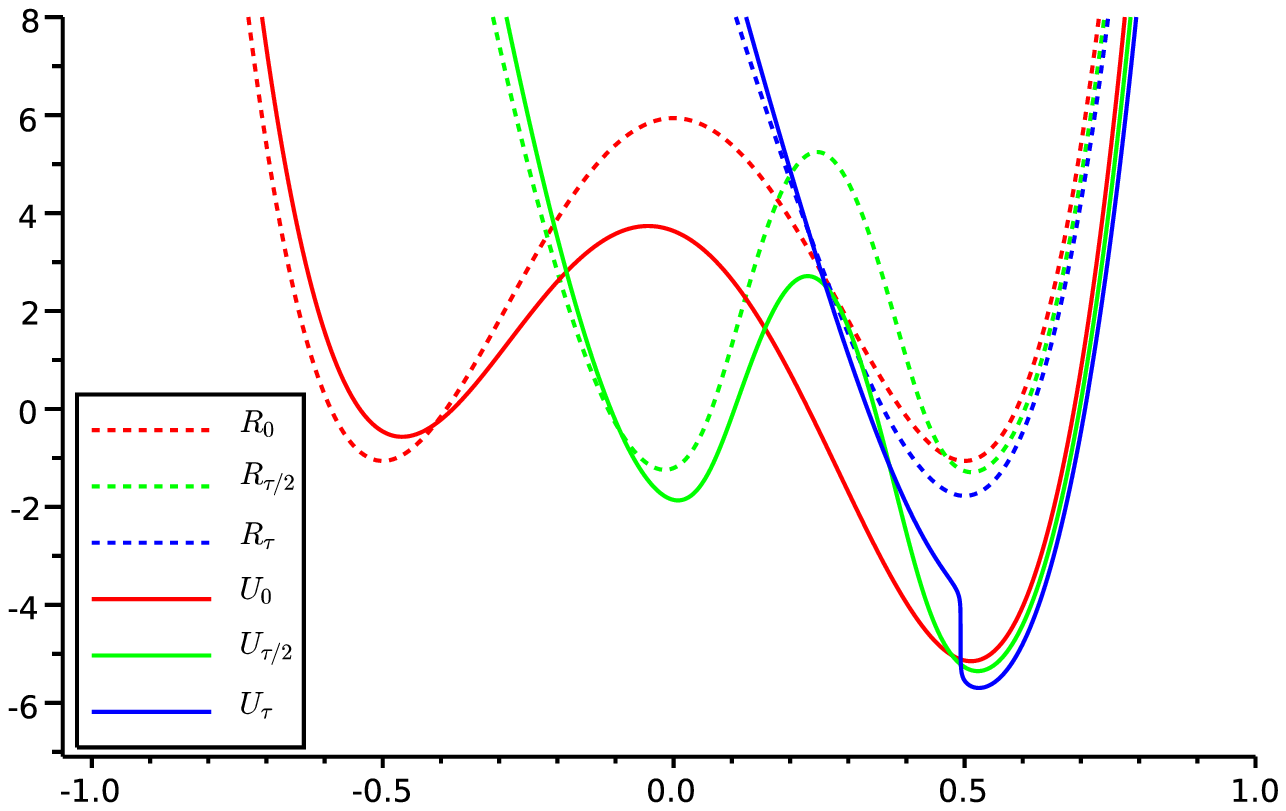}\\
        \vspace{-0.7cm} \strut
        \end{minipage}}
    \hspace*{0.63cm}
\hspace*{0.1cm}
\end{center}
\vskip -0.3cm
\caption{Gibbs and control potentials $R_t$ and $U_t$ for optimal $10s$ 
(left) and $1s$ (right) runs}
\end{figure}

\begin{figure}[H]
\begin{center}
\leavevmode
\vskip -0.4cm
\hspace*{0.7cm}
{%
      \begin{minipage}{0.45\textwidth}\hspace*{-0.4cm}
        \includegraphics[width=7cm,height=5cm,angle=0]{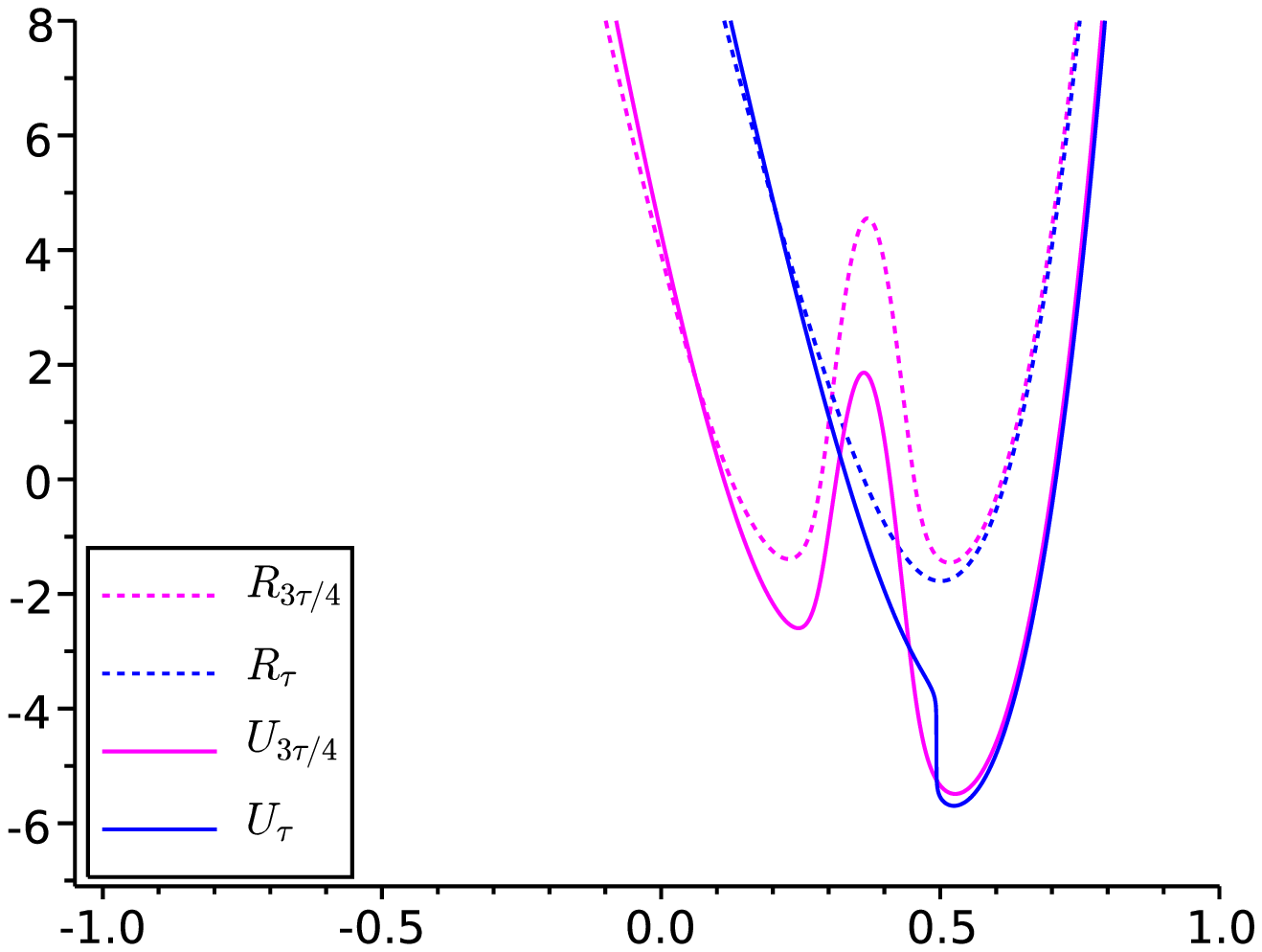}\\
        \vspace{-0.73cm} \strut
        \end{minipage}}
    \hspace*{-0.1cm}
{%
      \begin{minipage}{0.45\textwidth}\vspace{0.17cm}\hspace*{-0.3cm}
        \includegraphics[width=7cm,height=5cm]{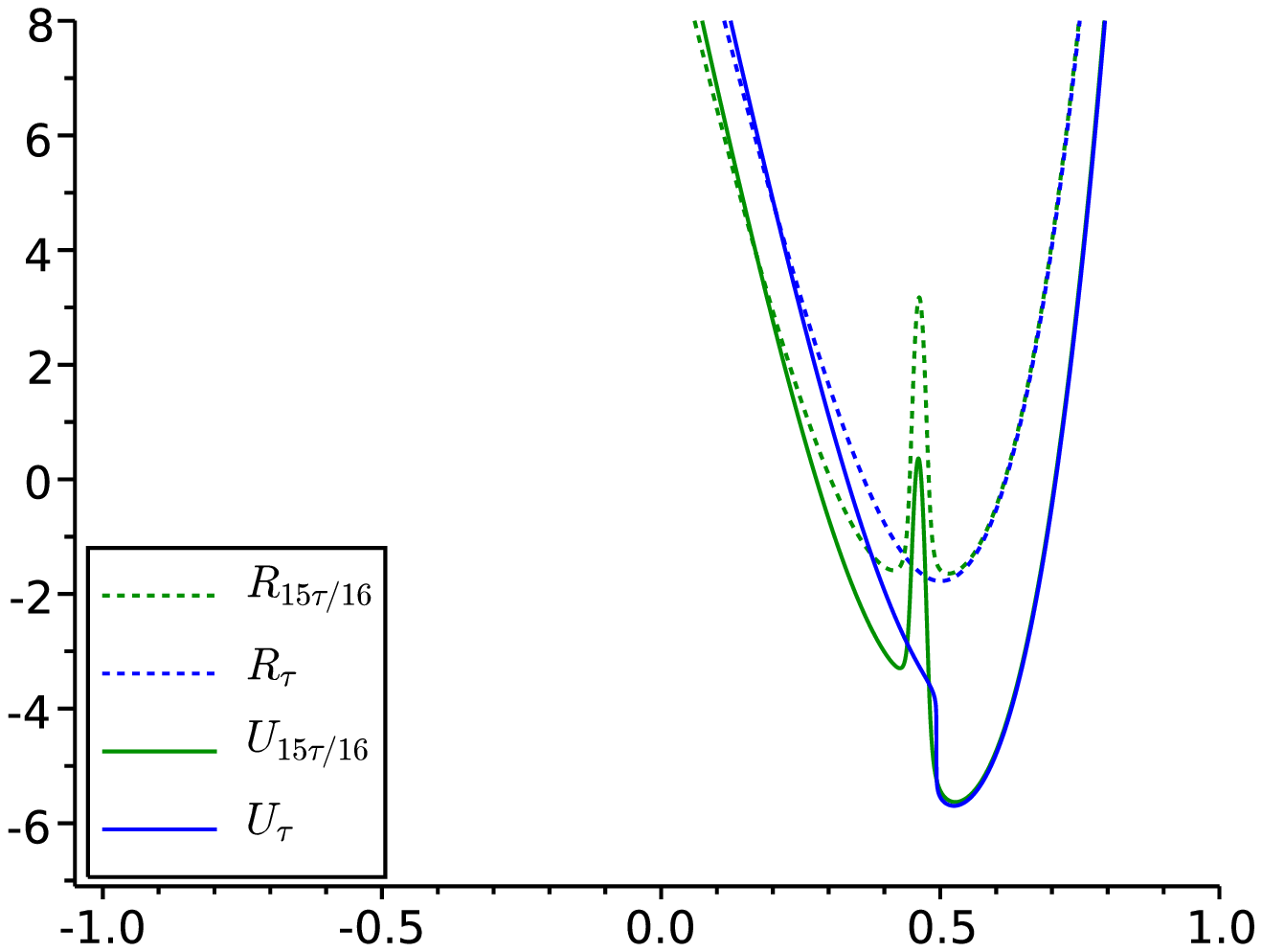}\\
        \vspace{-0.7cm} \strut
        \end{minipage}}
    \hspace*{0.63cm}
\hspace*{0.1cm}
\end{center}
\vskip -0.3cm
\caption{Potentials $R_t$ and $U_t$ for times
$\frac{3}{4}\tau$ and $\tau$ (left) and for $\frac{15}{16}\tau$
and $\,\tau\,$ (right) for $1s$ run}
\end{figure}
\vskip 0.1cm

\noindent The optimal protocols for $10s$ and $1s$ runs (the latter releasing 
almost 4 times more heat than the Landauer bound) are illustrated 
on FIG.\,4 at initial, half- and final time. For $10s$, the Gibbs 
potentials $\,R_t=-\beta^{-1}\ln\rho_t\,$ are very close to control 
potentials $\,U_t\,$ so that the optimal protocol is almost quasi-stationary. 
For $1s$, the $\,R_t\,$ differ considerable from $\,U_t$, \,also 
at the initial and final times, showing a more intricate structure
in late times, with the persistence of the barrier separating the two wells,
see FIG.\,5. The initial and final jumps of the 
potential in the optimal protocol were first discovered in
the case with quadratic potentials by an explicit calculation in
\cite{SS07}. \,The optimal transport map $\,q_0\mapsto q_\tau(q_0)\,$ 
leading to the optimal protocol has the form, in the problem in question, 
of a kink on the shifted identity map and the corresponding current
velocities $v_t$ build  in time an (almost) shock, see FIG.\,6. 

\begin{figure}[H]
\begin{center}
\leavevmode
\vskip -0.4cm
\hspace*{0.7cm}
{%
      \begin{minipage}{0.45\textwidth}\hspace*{-0.4cm}
        \includegraphics[width=6.1cm,height=4.7cm,angle=0]{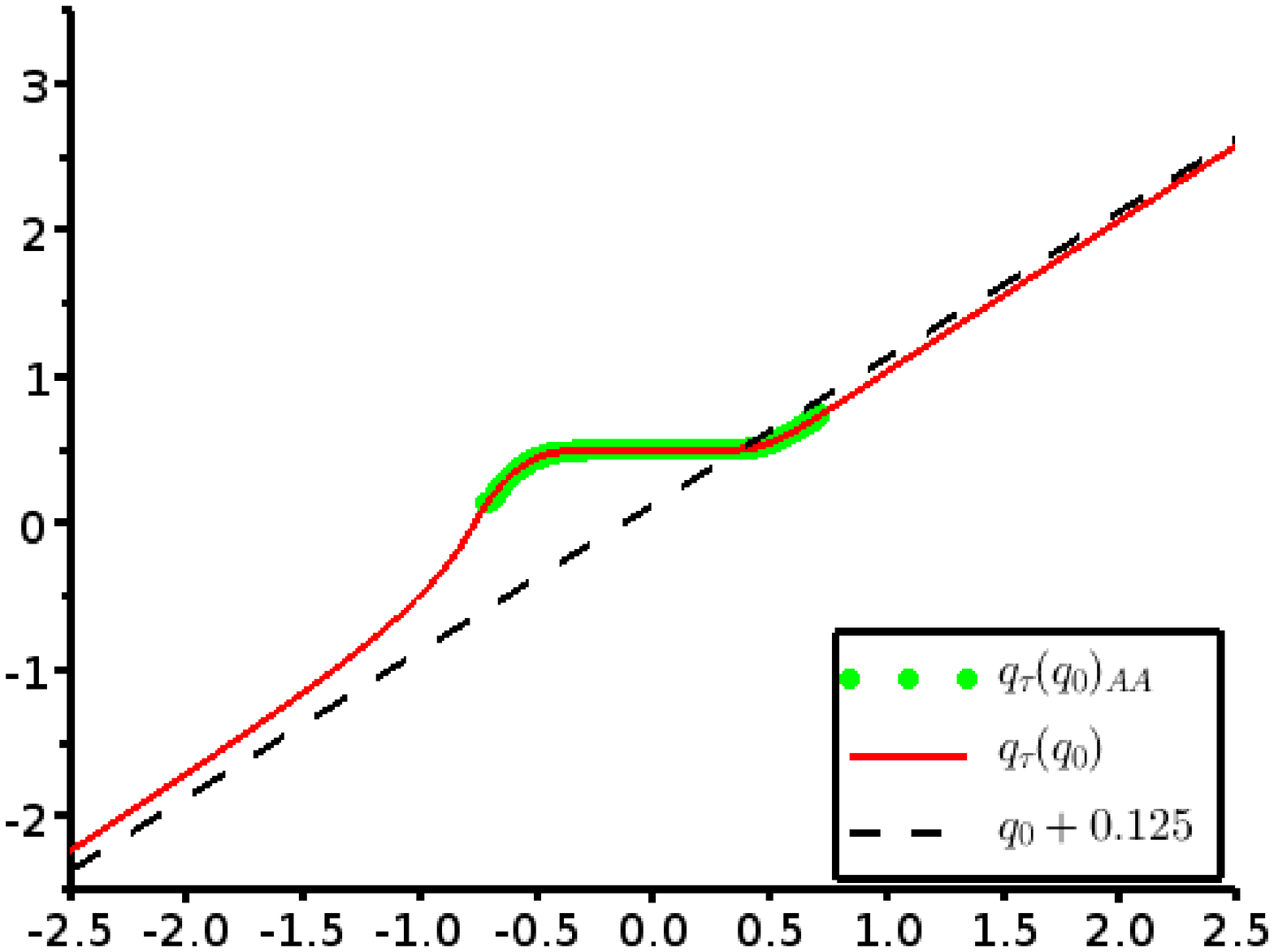}\\
        \vspace{-0.73cm} \strut
        \end{minipage}}
    \hspace*{-0.1cm}
{%
      \begin{minipage}{0.45\textwidth}\vspace{0.17cm}\hspace*{-0.3cm}
        \includegraphics[width=7cm,height=4.6cm]{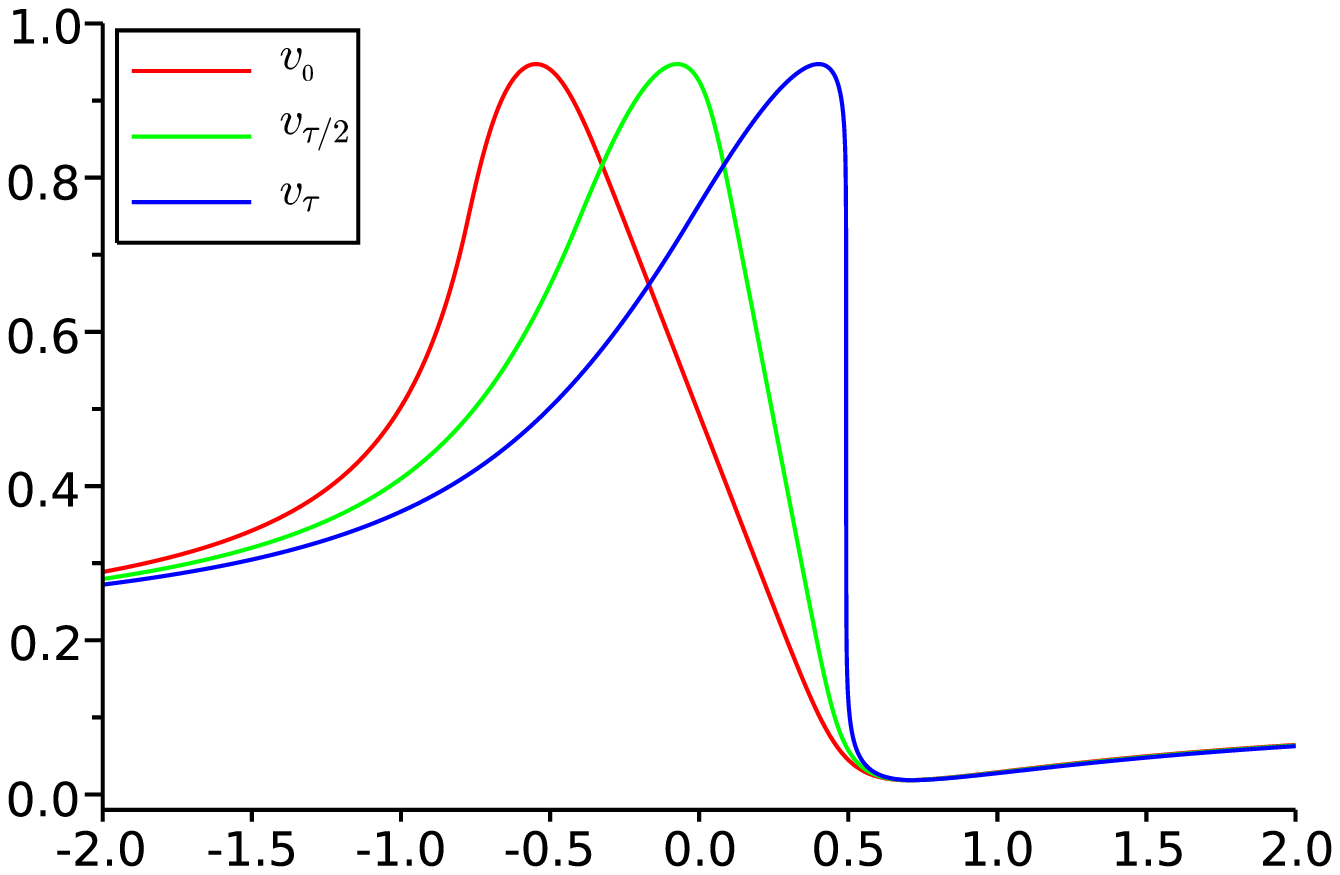}\\
        \vspace{-0.7cm} \strut
        \end{minipage}}
    \hspace*{0.63cm}
\hspace*{0.1cm}
\end{center}
\vskip -0.3cm
\caption{Optimal map $q_0\mapsto q_\tau(q_0)$ (left) and the corresponding 
current velocities (right)}
\end{figure}
\vskip 0.2cm

\noindent In one dimensional problem as above, the optimal map
$\,q_0\mapsto q_\tau(q_0)\s$ may be found numerically 
by sorting the points distributed with densities $\rho_0$ and 
$\rho_\tau$  in the increasing order. In more than one 
dimension, finding such maps requires a more sophisticated
Auction Algorithm, see \cite{BF03}.
\vskip 0.2cm

Above, we dealt with nonstationary overdamped Langevin evolution
without non-conservative forces. Adding such forces $\,\phi_t\,$ as
in Eq.\,(\ref{Langov}) but keeping the Einstein relation
(\ref{E1}) modifies the expression (\ref{Q}) for the 
dissipated heat to
\qq
Q_{\tau}[\bm q]\,=\,-\int\limits_0^\tau\big(-\nabla U_t+\phi_t\big)(q_t)
\circ dq(t)\,.
\qqq
The Fokker-Planck equation takes still the form of the advection equation
(\ref{adv0}) but the expression for the current velocity is modified 
by the addition of $\,\phi_t\,$ to
\qq
v_t(q)\,=\,M\big(-\nabla U_t+\phi_t\big)(q)-D\nabla\ln\rho_t(q)
\label{vtf}
\qqq
and is no more of the gradient type. The mean total entropy production 
is still given by Eq.\,(\ref{BB}). Since the Benamou-Brenier 
minimization held for arbitrary velocities, the bounds (\ref{2ndLawft})
and (\ref{Q>Sft}) hold in this case as well, but are saturated by 
the protocol discussed above without a non-conservative force.
\vskip 0.2cm

\subsection{Generalization to arbitrary diffusions}

\noindent Much of the discussion of Secs.\,\ref{subsec:BB} and 
\ref{subsec:FTT} extends to the general diffusion processes
studied in Sec.\,\ref{subsec:genfr} with a diffusion
matrix $\,\CD_t(x)\,$ time-independent and everywhere non-degenerate, 
provided that we define the total entropy production employing the
reversed protocol of Sec.\,\ref{subsubsec:revprot}. The average
total entropy production is then given by expression (\ref{spcas})
with $\,\CD_t(x)=\CD(x)$. \,The right hand side of Eq.\,(\ref{spcas})  
may be bounded by the Benamou-Brenier argument of Sec.\,\ref{subsec:BB} 
resulting in the finite time improvement
\qq
{\mathbb E}_{\mu_0}\,\Delta^{tot} S[\bm x]\ \geq\ \frac{k_B}{\tau} 
K_{min}[\rho_0,\rho_\tau]
\qqq
of the $2^{\rm nd}$ Law. \,Now
$\,K_{min}[\rho_0,\rho_\tau]\,$ minimizes over the deterministic
maps $\,x_0\mapsto x_\tau(x_0)\,$ that transport 
measure $\,\mu_0(dx)=\rho_0(x)\,\lambda(dx)\,$
to $\,\mu_\tau(dx)=\rho_\tau(x)\,\lambda(dx)\,$ the quadratic cost
$\,K[x_\tau(\cdot)]\,$ given by the formula
\qq
K[x_\tau(\cdot)]\,=\,\int\limits_{\CX}d_{\CD^{-1}}(x_0,x_\tau(x_0))^3
\,\rho_0(x_0)\,\lambda(dx_0)\,,
\qqq
where $\,d_{\CD^{-1}}(x,y)\,$ stands for the distance function in the 
Riemannian metric $\,g=dx\cdot\CD^{-1}(x)\,dx$. \,In-depth information 
about optimal 
mass transport in the context of Riemannian geometry may be found in 
ref.\,\cite{VO4}. 
\vskip 0.2cm

More discussion of optimization problems in Stochastic Thermodynamics 
is contained in refs.\,\cite{GSS08,EK10,AMM12,MMP12}. One of the problems 
still open is the extension of the above analysis to general
underdamped Langevin evolutions which are diffusion processes with
degenerate diffusion matrix. For such processes one has to use 
a different time reversal to define the total entropy production,
see Example 5 in Sec.\,\ref{subsec:genfr}. For general underdamped 
Langevin processes, the passage in the latter quantity to the overdamped 
limit contains some surprises, see \cite{H80,SSD82,CBE12}.

\nsection{Fluctuation-dissipation relations}

\noindent This is a lecture devoted to relations between the fluctuation
relations and the laws governing the linear response to perturbations around
stationary states that, historically, were among the first results 
about nonequilibrium dynamics.

\subsection{Hatano-Sasa fluctuation relation and the general 
Fluctuation-Dissipation Theorem}

\noindent Recall that the Hatano-Sasa transient fluctuation relation
(\ref{HSc}) holds for arbitrary continuous time nonstationary Markov process 
with backward generators $\,L_t\,$ and the family of accompanying measures 
$\,\nu_t(x)=\ee^{-\varphi_t(x)}\lambda(dx)\,$ such that $\,L_t^*\nu_t=0$.
\,Following \cite{PJP09}, see also \cite{H78}, let us consider 
a family of stationary 
transition rates $\,w_{\epsilon}(x,dy)\,$ parametrized by 
$\,\epsilon=(\epsilon^a)\,$ in a neighborhood of $\,\epsilon=0$,
\,that correspond to backward generators $\,L_\epsilon\,$ with a family
of invariant measures $\,\nu_\epsilon(dx)\,=\,\ee^{-\varphi_\epsilon(x)}
\lambda(dx)\,$ such that
\qq
L_\epsilon^*\nu_\epsilon\,=\,0\,.
\label{inco}
\qqq
For each time-dependent protocol 
$\,t\mapsto\epsilon_t\,$ such that $\,\epsilon_t=0\,$
for $t\leq 0$, we shall consider for $t\geq 0\,$ the nonstationary Markov 
process with backward generators $\,L_{\epsilon_t}\,$ that starts from 
the measure $\,\nu\equiv\nu_0$. \,Note that measures $\,\nu_{\epsilon_t}\,$
accompany such a process. For each of protocols $\,\bm\epsilon=
(\epsilon_t)$, \,we have the Hatano-Sasa relation
\qq
\mathbb E^{\bm\epsilon}_{\nu}\,\,\ee^{-\int\limits_0^\tau\frac{d\epsilon^a_t}
{dt}\,\partial_{\epsilon^a}\varphi_{\epsilon_t}(x_t)\,dt}\,=\,1
\qqq
which, \,expanded up to the second order in $\bm\epsilon$, \,gives the identity:
\qq
&&-\int\limits_0^\tau\frac{d\epsilon^a_t}
{dt}\,\mathbb E_{\nu}\,\partial_{\epsilon^a}\varphi(x_t)\,dt\,-\,
\int\limits_0^\tau\frac{d\epsilon^a_t}
{dt}\,\epsilon^b_t\,\mathbb E_{\nu}\,
\partial_{\epsilon^b}\partial_{\epsilon^a}
\varphi(x_t)\,dt\cr
&&-\,\int\limits_0^\tau\frac{d\epsilon^a_t}
{dt}\,dt\int\limits_0^\tau\epsilon^b_s\,\frac{\delta}{\delta\epsilon_s^b}
\Big|_{\bm\epsilon=\bm 0}\,\mathbb E_{\nu}^{\bm\epsilon}\,
\partial_{\epsilon^a}\varphi(x_t)\,\,ds\cr
&&+\,\frac{_1}{^2}
\int\limits_0^\tau \frac{d\epsilon^a_t}
{dt}\,dt\int\limits_0^\tau \frac{d\epsilon^b_s}
{dt}\,\mathbb E_{\nu}\,\partial_{\epsilon^b}\varphi(x_s)
\,\partial_{\epsilon^a}\varphi(x_t)\,\,ds\ =\ 0\,,
\label{2so}
\qqq
where $\,\varphi\equiv\varphi_0\,$ and $\,\mathbb E_\nu\equiv
\mathbb E_\nu^{\bm 0}$. \,Expanding similarly the normalization condition
\qq
\int\ee^{-\varphi_\epsilon(x)}\,\lambda(dx)\,=\,1
\qqq
holding for all $\,\epsilon\,$ to the second order,
we obtain:
\qq
\int\Big(-\epsilon^a\,\partial_{\epsilon^a}\varphi(x)\,-\,
\frac{_1}{^2}\epsilon^a\epsilon^b\,\partial_{\epsilon^b}
\partial_{\epsilon^a}\varphi(x)\,+\,
\frac{_1}{^2}\epsilon^a\epsilon^b\,\partial_{\epsilon^b}\varphi(x)
\,\partial_{\epsilon^a}\varphi(x)\Big)\,\ee^{-\varphi(x)}\,
\lambda(dx)\,=\,0\,.
\qqq
The last identity implies that
\qq
\mathbb E_{\nu}\,\partial_{\epsilon^a}\varphi(x_t)\,=\,
\int
\partial_{\epsilon^a}\varphi(x_t)\,\,\ee^{-\varphi(x)}\,
\lambda(dx)\,=\,0
\qqq
and that
\qq
\mathbb E_{\nu}\,
\partial_{\epsilon^b}\partial_{\epsilon^a}
\varphi(x_t)&=&\int\partial_{\epsilon^b}\partial_{\epsilon^a}
\varphi(x)\,\,\ee^{-\varphi(x)}\,\lambda(dx)\,=\,
\int\partial_{\epsilon^b}\varphi(x)\,\,\partial_{\epsilon^a}\varphi(x)\,
\,\ee^{-\varphi(x)}\,\lambda(dx)\cr
&=&\mathbb E_{\nu}
\,\partial_{\epsilon^b}\varphi(x_t)\,\,
\partial_{\epsilon^a}\varphi(x_t)\,,
\qqq
where the right hand side is $t$-independent.
Substituting these identities to Eq.\,(\ref{2so}), we obtain:
\qq
&&-\,\frac{_1}{^2}
\int\limits_0^\tau\frac{d\epsilon^a_t}
{dt}\,dt\int\limits_0^\tau\frac{d\epsilon^b_s}{ds}\,ds
\,\,\mathbb 
E_{\nu}\,
\partial_{\epsilon^b}\varphi(x_t)\,\,\partial_{\epsilon^a}
\varphi(x_t)\,-\,\int\limits_0^\tau\frac{d\epsilon^a_t}
{dt}\,dt\int\limits_0^\tau\epsilon^b_s\,\frac{\delta}{\delta\epsilon_s^b}
\Big|_{\bm\epsilon=\bm 0}\,\mathbb E_{\nu}^{\bm\epsilon}\,
\partial_{\epsilon^a}\varphi(x_t)\,\,ds\cr
&&+\,\frac{_1}{^2}
\int\limits_0^\tau \frac{d\epsilon^a_t}
{dt}\,dt\int\limits_0^\tau \frac{d\epsilon^b_s}
{dt}\,\mathbb E_{\nu}\,\partial_{\epsilon^b}\varphi(x_s)
\,\partial_{\epsilon^a}\varphi(x_t)\,\,ds\cr
&&=\ -\,\frac{_1}{^2}
\int\limits_0^\tau\frac{d\epsilon^a_t}
{dt}\,dt\int\limits_0^\tau\frac{d\epsilon^b_s}{ds}\,\,\mathbb 
E_{\nu}\,
\partial_{\epsilon^a}\varphi(x_t)\,\,\partial_{\epsilon^b}
\varphi(x_t)\,\,ds\cr
&&-\,\int\limits_0^\tau\frac{d\epsilon^a_t}
{dt}\,dt\int\limits_0^\tau\frac{d\epsilon^b_s}
{ds}\,ds\int\limits_s^\tau
\frac{\delta}{\delta\epsilon_\sigma^b}
\Big|_{\bm\epsilon=\bm 0}\,\mathbb E_{\nu}^{\bm\epsilon}\,
\partial_{\epsilon^a}\varphi(x_t)\,\,d\sigma\cr
&&+\,\frac{_1}{^2}
\int\limits_0^\tau \frac{d\epsilon^a_t}
{dt}\,dt\int\limits_0^\tau \frac{d\epsilon^b_s}
{dt}\,\mathbb E_{\nu}\,\partial_{\epsilon^b}\varphi(x_s)
\,\partial_{\epsilon^a}\varphi(x_t)\,\,ds\ =\ 0\,.
\label{2so2}
\qqq
Upon stripping of the arbitrary functions $\,\frac{d\epsilon_t}{dt}$, \,the 
last equation is equivalent to the identity 
\qq
&&\mathbb E_{\nu}\,\partial_{\epsilon^b}\varphi(x_s)
\,\partial_{\epsilon^a}\varphi(x_t)\,-\, 
\mathbb E_{\nu}\,\partial_{\epsilon^b}\varphi(x_t)
\,\partial_{\epsilon^a}\varphi(x_t)\cr
&&=\,\int\limits_s^\tau\frac{\delta}{\delta\epsilon_\sigma^b}
\Big|_{\bm\epsilon=\bm 0}\,\mathbb E_{\nu}^{\bm\epsilon}\,
\partial_{\epsilon^a}\varphi(x_t)\,d\sigma\,+\,
\,\int\limits_t^\tau\frac{\delta}{\delta\epsilon_\sigma^a}
\Big|_{\bm\epsilon=\bm 0}\,\mathbb E_{\nu}^{\bm\epsilon}\,
\partial_{\epsilon^b}\varphi(x_s)\,d\sigma
\label{FR0}
\qqq
Because, by causality,
\qq
\frac{\delta}{\delta\epsilon_\sigma^a}
\Big|_{\bm\epsilon=\bm 0}\,\mathbb E_{\nu}^{\bm\epsilon}\,
\partial_{\epsilon^b}\varphi(x_s)\,=\,0
\qquad
{\rm for}\qquad s\leq\sigma\,,
\qqq
Eq.\,(\ref{FR0}) reduces upon taking $\,s\leq t\,$  to the relation
\qq
&&\mathbb E_\nu\,\partial_{\epsilon^b}\varphi(x_s)
\,\partial_{\epsilon^a}\varphi(x_t)\,-\, 
\mathbb E_\nu\,\partial_{\epsilon^b}\varphi(x_t)
\,\partial_{\epsilon^a}\varphi(x_t)\,=\,\int\limits_s^t
\frac{\delta}{\delta\epsilon_\sigma^b}
\Big|_{\bm\epsilon=\bm 0}\,\mathbb E_{\nu}^{\bm\epsilon}\,
\partial_{\epsilon^a}\varphi(x_t)\,d\sigma\,,\qquad
\label{FDTi}
\qqq
or, \,in the differential form, \,to the identity
\qq
\boxed{\,\partial_s\,\mathbb E_\nu\,\partial_{\epsilon^b}\varphi(x_s)
\,\partial_{\epsilon^a}\varphi(x_t)\,=\,-\,
\frac{\delta}{\delta\epsilon_s^b}
\Big|_{\bm\epsilon=\bm 0}\,\mathbb E_{\nu}^{\bm\epsilon}\,
\partial_{\epsilon^a}\varphi(x_t)\,}\,.
\label{FDTd}
\qqq
This a one of general forms of the {\,\bf Fluctuation-Dissipation
Theorem} (FDT) \cite{H78,PJP09}. The left hand side is the time derivative 
of \,the {\,\bf dynamical 
correlation function\,} of observables $\,\partial_{\epsilon^a}\varphi(x)\,$
in the stationary state, whereas the quantity on the right hand side 
is the {\,\bf response function\,} measuring the change of the dynamical 
one point function of $\,\partial_{\epsilon^a}\varphi(x)\,$ under a small 
perturbation of $\,\epsilon\,$ concentrated around an earlier time. 
Note that such a perturbation makes the dynamics nonstationary. \,The entry 
$\,\partial_{\epsilon^a}\varphi(x_t)\,$ plays a passive role in
the identity (\ref{FDTd}) and could be replaced by an arbitrary
function $\,O_a(x_t)$. \,On the other hand, for a general 
stationary dynamical correlation function with $\,s\leq t\,$ one has
\qq
&&\partial_t\,\mathbb E_{\nu}\,O_b(s)\,O_a(t)\,=\,
\int\nu(dx)\,O_b(x)\int \partial_tP_{s,t}(x,dy)\,O_a(y)\,=\,
\mathbb E_{\nu}\,O_b(s)\,(L\hspace{0.02cm}O_a)(t)\,,\cr
&&\partial_s\,\mathbb E_{\nu}\,O_b(s)\,O_a(t)\,=\,
\int\nu(dx)\,O_b(x)\int\partial_sP_{s,t}(x,dy)\,O_a(y)\,=\,
-\,\mathbb E_{\nu}\,(L'O_b)(s)\,(L\hspace{0.02cm}O_a)(t)\,,\qquad\quad
\qqq
where 
\qq
L'\,=\,\ee^{\varphi}\hspace{0.03cm}L^\dagger\,\ee^{-\varphi}
\qqq
is the adjoint of $\,L\,$ with respect to the invariant measure $\,\nu(dx)$.
\,We infer that the FDT (\ref{FDTd}) may be rewritten in the form
\qq
\,\mathbb E_\nu\,L'\big(\partial_{\epsilon^b}\varphi(x_s)\big)
\,\,O_a(x_t)\,=\,
\frac{\delta}{\delta\epsilon_s^b}
\Big|_{\bm\epsilon=\bm 0}\,\mathbb E_{\nu}^{\bm\epsilon}\,
O_a(x_t)\,.
\label{FDTn}
\qqq 

\subsection{Other forms of the Fluctuation-Dissipation Theorem}

\noindent We shall follow in this section the approach to the FDT developed
in \cite{CG}. Let us consider a particular family of transition rates 
of the form
\qq 
w_\epsilon(x,dy)\,=\,\ee^{-\frac{\beta}{2}\epsilon^a O_a(x)}\hspace{0.03cm}
w(x,dy)\,\ee^{\frac{\beta}{2}\epsilon^a(O_a(y))}
\label{pertw}
\qqq
for a family of functions $\,O_a(x)$,
\,corresponding to the perturbed backward generators
\qq
L_{\epsilon}&=&\ee^{-\frac{\beta}{2}\epsilon^a O_a(x)}L\,
\ee^{\frac{\beta}{2}\epsilon^a O_a(x)}\,-\,\ee^{-\frac{\beta}{2}\epsilon^a 
O_a(x)}\big(L\hspace{0.03cm}\ee^{\frac{\beta}{2}\epsilon^a O_a(x)}\big)\cr
&=&L\,+\,\frac{_\beta}{^2}\epsilon^a\Big([L,O_a]-(L\hspace{0.02cm}O_a)\Big)\ 
+\ o(\epsilon)\,.
\qqq
Above, $\,\beta\,$ is introduced just for dimensional reason, see however 
below. The invariance
condition (\ref{inco}) for the measures $\,\nu_\epsilon(dx)\,$ gives
now to the $1^{\rm st}$ order in $\,\epsilon\,$ the condition
\qq
-\,\epsilon^a\,L^\dagger
\big(\ee^{-\varphi}\,\partial_{\epsilon^a}\varphi\big)
\,+\,\frac{_\beta}{^2}\epsilon^a\Big([L,O_a]-(L\hspace{0.02cm}
O_a)\Big)^\dagger
\big(\ee^{-\varphi}\big)\,=\,0\,,
\qqq
i.e.
\qq
L'\big(\partial_{\epsilon^a}\varphi\big)\,=\,-\,\frac{_\beta}{^2}
\big(L'O_a\,+\,L\hspace{0.02cm}O_a\big)\,.
\label{tbu}
\qqq
Plugging this expression into Eq.\,(\ref{FDTn}), we may rewrite it
in the form
\qq
\boxed{\,\partial_s\,
\mathbb E_\nu\,O_b(x_s)\,O_a(x_t)
\,-\,\mathbb E_\nu(L\hspace{0.02cm}O_b)(x_s)\,O_a(x_t)\,
=\,\frac{2}{\beta}\,\frac{\delta}{\delta\epsilon_s^b}
\Big|_{\bm\epsilon=\bm 0}\,\mathbb E_{\nu}^{\bm\epsilon}\,O_a(x_t)\,}\,.
\label{FDTf}
\qqq
This is another form of the general FDT 
\cite{CKP94,LCZ05,BMW09}. It does not require the knowledge of
the invariant states, but involves explicitly the generator
of the stationary process.
\vskip 0.2cm

We may also rewrite the right hand side of Eq.\,(\ref{tbu}) as
$\,\,-\,\beta L'O_a\,-\,\frac{_\beta}{^2}(L-L')O_a$. \,This results
in yet another form of the general FDT:
\qq
\partial_s\,
\mathbb E_\nu\,O_b(x_s)\,O_a(x_t)
\,-\,\mathbb E_\nu\,\big(\frac{_1}{^2}(L-L')O_b\big)(x_s)
\,O_a(x_t)\Big)\,=\,\frac{1}{\beta}\,\frac{\delta}{\delta\epsilon_s^b}
\Big|_{\bm\epsilon=\bm 0}\,\mathbb E_{\nu}^{\bm\epsilon}\,O_a(x_t)\,.
\qqq
The latter form is useful for the Langevin process where 
\qq
L\,=\,\big(-\CM(\nabla H)+\CM f+\CD\nabla\big)\cdot\nabla
\label{Ls}
\qqq
and were
\qq
L'&=&\ee^{\varphi}\,L^\dagger\,
\ee^{-\varphi}\,=\,\big(\CM(\nabla H)-\CM f\big)\cdot\nabla\,
+\,2\,\ee^{\varphi}\,(\nabla\cdot\ee^{-\varphi})\CD\nabla
\,+\,\nabla\cdot\CD\nabla
\qqq
since $\,L^\dagger(\ee^{-\varphi})=0$. \,Hence in this case,
\qq
\frac{_1}{^2}(L-L')\,=\,\big(-\CM(\nabla H)+\CM f-\ee^{\varphi}
\CD(\nabla\ee^{-\varphi})\big)\cdot\nabla\,=\,v\cdot\nabla\,,
\qqq
where $\,v(x)\,$ is the current velocity, see Eq.\,(\ref{currv}),
in the stationary state. Using the the latter relation, we obtain
the general FDT for the Langevin
dynamics \cite{CFG08}:
\qq
\boxed{\,\partial_s\,
\mathbb E_\nu\,O_b(x_s)\,O_a(x_t)
\,-\,\mathbb E_\nu\,\big(v\cdot\nabla O_b\big)(x_s)
\,O_a(x_t)\Big)\,=\,\frac{1}{\beta}\,\frac{\delta}{\delta\epsilon_s^b}
\Big|_{\bm\epsilon=\bm 0}\,\mathbb E_{\nu}^{\bm\epsilon}\,O_a(x_t)\,}\,.
\label{FDTv}
\qqq
Note that for the stationary Langevin process with $\,\CM=\CM^t=\beta\CD$, 
the perturbation (\ref{pertw}) corresponds to the change of
the Hamiltonian $\,H(x)\to H(x)-\epsilon^a O_a(x)$.
\vskip 0.2cm

In the situation with detailed balance
for the stationary process with $\,\epsilon=0$, \,see Eq.\,(\ref{DBc}),
\,generators $\,L\,$ and $\,L'\,$ coincide and 
$\,\varphi_\epsilon(x)=\beta\big(H(x)\,-\,\epsilon^aO_a(x)-\,F_\epsilon\big)$,
\,where $\,F_\epsilon\,$ is a constant.
As a result, all three forms of the FDT reduce to the identity
\qq
\boxed{\,\partial_s C^{ab}(t-s)\,\equiv\,\partial_s\,
\mathbb E_\nu\,O_b(x_s)\,O_a(x_t)
\,=\,\frac{1}{\beta}\,\frac{\delta}{\delta\epsilon_s^b}
\Big|_{\bm\epsilon=\bm 0}\,\mathbb E_{\nu}^{\bm\epsilon}\,O_a(x_t)\,\equiv\,
\frac{1}{\beta}\,R^{ab}(t-s)\,}
\label{FDTe}
\qqq
which is the classic equilibrium FDT \cite{K57}
relating the equilibrium dynamical correlation function $\,C^{ab}(t-s)\,$
to the response function $\,R^{ab}(t-s)\,$ of the equilibrium state to
small perturbations. 
\vskip 0.4cm

\noindent{\bf Example 6.} \,For the Einstein-Smoluchowski Brownian
motion of Example 3 with scalar mass matrix $\,m$, $\,O_a(q,p)=p^a\,$ 
and $\,O_b(q,p)=q^b$, \,the stationary form of the dynamical correlation 
function is
\qq
C^{ab}(t-s)\,=\,m\hspace{0.03cm} D\,\delta^{ab}\,\ee^{-\frac{t-s}{mM}}
\qqq
(there is no complete stationary state in that case since the expectation
value of $\,q_t^2\,$ diverges linearly in time), \,and the response
function takes the form
\qq
R^{ab}(t-s)\,=\,\delta^{ab}\,\ee^{-\frac{t-s}{mM}}\,.   
\qqq
Relation (\ref{FDTe}) reduces then to the Einstein relation
(\ref{E1}), \,a prototype of the equilibrium FDT.
\vskip 0.4cm 

The possible usage of the FDT (\ref{FDTe}) is for extracting the response
function, more difficult to measure, from the stationary dynamical
correlation function, more easily accessible, or for inferring the temperature 
$\beta^{-1}$ of a system in thermal equilibrium if both the response
function and the dynamical correlation function are accessible. Although
near a nonequilibrium stationary states (NESS), the FDT does not
have such a simple form, the ratio of the dynamical correlation
to the response function is often used for systems out of equilibrium,
in particular in glassy systems \cite{C11}, to define their effective 
temperatures. 
\vskip 0.1cm

It was observed in \cite{CFG08}), see also \cite{SS06}, 
that the form (\ref{FDTv}) of the FDT for Langevin systems implies 
that one recovers the equilibrium form of the FDT in Lagrangian frame
of the current velocity. It was shown subsequently in \cite{CG09} that 
any nonequilibrium Langevin diffusion rewritten in the Lagrangian 
frame of its current velocity recovers the detailed balance property.

\subsection{Relation of the response function to dissipation}

\noindent The original name of the Fluctuation-Dissipation Theorem
for the identity (\ref{FDTe}) comes from the fact that 
the dynamical 2-time correlation function describes the correlation 
between the fluctuations of the random variables $\,O(x_t)$ whereas 
the response function is related to dissipation of energy or heat.
To understand the latter connection, let us consider the case 
of a periodic perturbation of the stationary equilibrium dynamics 
by taking $\,\epsilon_t=\epsilon_0\sin(\omega t)\,$
so that $\,H_t(x)\,=\,H(x)-\epsilon_0\sin(\omega t)\,O(x)$. 
For such a system, the average expectation of the work 
\qq
W_\tau[\bm x]\,=\,\int\limits_0^\tau(\partial_tH_t)(x_t)\,dt
\,=\,-\,\epsilon_0\hspace{0.03cm}\omega
\int\limits_0^\tau\cos(\omega t)\,O(x_t)\,dt
\qqq
is per unit time equal to
\qq
&&\lim\limits_{\tau\to\infty}\ 
\frac{1}{\tau}\,\mathbb E_{\nu}^{\bm\epsilon}\,W_\tau[\bm x]\,=\,
-\lim\limits_{\tau\to\infty}\ 
\frac{\epsilon_0\hspace{0.02cm}\omega}{\tau}\int\limits_0^\tau\,\cos(\omega t)
\,\mathbb E_{\nu}^{\bm\epsilon}\,O(x_t)\,dt\cr
&=&-\lim\limits_{\tau\to\infty}\ 
\frac{\epsilon_0\hspace{0.02cm}\omega}{\tau}\int\limits_0^\tau\,\cos(\omega t)
\,\Big(\mathbb E_{\nu}^{0}\,O(x_t)\,+\,\epsilon_0\int\limits_0^t\sin(\omega s)\,
\frac{\delta}{\delta\epsilon_s}\,\mathbb E_\nu\,O(x_t)\,ds\,
+\,o(\epsilon_0)\Big)\,dt\,.\qquad
\qqq
The stationary contribution vanishes in the long time limit, so that,
denoting 
\qq
\frac{\delta}{\delta\epsilon_s}\,\mathbb E_\nu\,O(x_t)\,\equiv\,
R(t-s)\,,
\qqq
we obtain
\qq
\lim\limits_{\tau\to\infty}\ 
\frac{1}{\tau}\,\mathbb E_{\nu}^{\bm\epsilon}\,W_\tau[\bm x]&=&
-\lim\limits_{\tau\to\infty}\ 
\frac{\epsilon_0^2\omega}{\tau}\int\limits_0^\tau dt
\int\limits_0^t\,\cos(\omega t)\,\sin(\omega s)\,R(t-s)\,ds\ 
+\ o(\epsilon_0^2)\cr
&=&-\lim\limits_{\tau\to\infty}\ 
\frac{\epsilon_0^2\omega}{\tau}\int\limits_0^\tau\sin(\omega s)\,ds
\int\limits_0^{\tau-s}\,\cos(\omega(s+\sigma))\,R(\sigma)\,
d\sigma\ +\ o(\epsilon_0^2)\cr
&=&-\lim\limits_{\tau\to\infty}\ 
\frac{\epsilon_0^2\omega}{\tau}\Big(\int\limits_0^\tau\sin(\omega s)\cos(\omega s)\,ds
\int\limits_0^{\tau-s}\,\cos(\omega\sigma)\,R(\sigma)\,
d\sigma\cr
&&\hspace{1.5cm}-\,\int\limits_0^\tau\sin^2(\omega s)\,ds
\int\limits_0^{\tau-s}\,\sin(\omega\sigma)\,R(\sigma)\,
d\sigma\Big)\ +\ o(\epsilon_0^2)\cr
&=&\Big(\lim\limits_{\tau\to\infty}\ \frac{\epsilon_0^2\omega}{\tau}
\int\limits_0^\tau\sin^2(\omega s)\,ds\Big)\,\int\limits_0^\infty
\sin(\omega\sigma)\,R(\sigma)\,d\sigma\ +\ o(\epsilon_0^2)\cr
&=&\frac{\epsilon_0^2\omega}{2}\,{\rm Im}\hspace{0.03cm}\widehat R(\omega)\ +\ 
o(\epsilon_0^2)\,,
\qqq   
assuming the integrable decay of $\,R(\sigma)\,$ when $\,\sigma\to\infty$,
\,where 
\qq
\widehat R(\omega)\,=\,\int\limits_0^\infty\ee^{i\omega\sigma}\,
R(\sigma)\,d\sigma
\qqq
is the Fourier transform of the response function (recall that the latter
vanishes for negative arguments). In experiments,
often the dissipation rate is directly measured, giving access to
the imaginary part of the Fourier-space response function 
$\,\widehat R(\omega)$. The real part of $\,\widehat R(\omega)\,$ may then
be obtained from the imaginary part by the Kramers-Kronig dispersion
relation
\qq
{\rm Re}\hspace{0.03cm}\widehat R(\omega)\,=\,\frac{_1}{\pi}\,\CP
\int\frac{{\rm Im}\hspace{0.03cm}\widehat R(\omega')}{\omega'-\omega}
\,\lambda(d\omega')  
\qqq
holding for Fourier transforms of function vanishing at negative times.

\subsection{A simple one-dimensional example}
\label{subsec:circledif}

\noindent Let us consider an overdamped Langevin dynamics of a particle 
moving on a circle that is described by by the stochastic equation
\qq
d\theta\,=\,M\big(F-\partial_\theta U(\theta)\big)\hspace{0.02cm}dt\,
+\,\sqrt{2D}\,dW(t)\,,
\label{poc}
\qqq
where $\,\theta\,$ is the angle modulo $\,2\pi\,$ parametrizing the position
of the particle, $\,M\,$ is the mobility and $\,D=\beta^{-1}M\,$ is 
the diffusivity of the particle. \,Periodic function $\,U(\theta)\,$ gives the potential
and $\,F\,$ is a constant part of the force (any non-conservative force
may be separated into a constant plus a potential part in that situation).
Eq.\,(\ref{poc}) models the dynamics of a colloidal particle
of radius $1\mu m$ kept on a circular trajectory of radius $4.12\mu m$
by a laser tweezer in an experiment performed at ENS Lyon \cite{GPC09}
in which case
\qq
&&MU(\theta)\,=\,H_0\sin(\theta)\qquad{\rm for}\qquad H_0=0.87\,rad\,s^{-1},
\label{mu}\\  
&&MF\,=\,0.85\,rad\,s^{-1}\,,\hspace{1.65cm}D\,=\,1.26\times 10^{-2}\,
rad^2\,s^{-1}\,.\label{mf}
\qqq 
The diffusion (\ref{poc}) has the backward generator
\qq
L\,=\,M\big(F-(\partial_\theta U)\big)\hspace{0.02cm}
\partial_\theta+D\hspace{0.02cm}\partial_\theta^2
\label{L1}
\qqq
and a NESS with the invariant measure $\,\nu(\theta)=\rho(\theta)
\hspace{0.02cm}\lambda(d\theta)\,$ for 
\qq
\rho(\theta)\,=\,Z^{-1}\,\ee^{\,\beta(F\theta-U(\theta))}\hspace{-0.1cm}
\int\limits_\theta^{\theta+2\pi}\hspace{-0.1cm}
\ee^{-\beta(F\vartheta-U(\vartheta))}\,\lambda(d\vartheta)
\ \equiv\ \ee^{-\varphi(\theta)}
\label{invm}
\qqq
that corresponds to the constant probability current
\qq
j\,=\,\Big(M(F-(\partial_\theta U)(\theta))-D\hspace{0.02cm}
\partial_\theta\Big)\hspace{0.02cm}\ee^{-\varphi(\theta)}\,=\,
D\hspace{0.02cm}Z^{-1}\Big(1-\ee^{-2\pi\beta F}\Big)
\label{j0}
\qqq
and to the current velocity
\qq
v(\theta)\,=\,j\,\rho(\theta)^{-1}\,=\,
\frac{D(1-\ee^{-2\pi\beta F})}{\ee^{\,\beta(F\theta-U(\theta))}\hspace{-0.1cm}
\int\limits_\theta^{\theta+2\pi}\hspace{-0.1cm}
\ee^{-\beta(F\vartheta-U(\vartheta))}\,\lambda(d\vartheta)}\,.
\label{v0}
\qqq
The generator adjoint to $\,L\,$ with respect to the invariant measure 
$\,\nu\,$ is
\qq
L'\,=\,\ee^{\hspace{0.02cm}\varphi}L^\dagger\ee^{-\varphi}\,=\,
\big(M(F-(\partial_\theta U))-2v\big)\hspace{0.02cm}\partial_\theta
+D\hspace{0.02cm}\partial_\theta^2.
\label{L'1}
\qqq
It is the backward generator of the time-reversed process 
$\,(\theta'_t)\,$ defined with the rule of case ${\bf(a)}$ of 
Sec.\,\ref{subsec:genfr} (with $\theta^*=\theta$) and satisfying 
the overdamped Langevin equation
\qq
d\theta'\,=\,\big(M(F-(\partial_{\theta}U)(\theta'))-2v(\theta')\big)
\hspace{0.02cm}dt\,+\,\sqrt{2D}\,dW(t)\,.
\label{invF}
\qqq
Three different forms (\ref{FDTv}), (\ref{FDTf}) and (\ref{FDTd}) 
of the FDT for this system  were checked by comparing with the
experimental measurements of the correlation and response functions,
with similar results confirming the theoretical predictions for
times up to few seconds \cite{GPC09,GPC11}), see FIG.\,7 for the
case of relation (\ref{FDTv}). For the system in question, the anomalous
term on the fluctuation side of the FDT ($B(t)$ on FIG.\,7) dominates
the equilibrium term ($C(0)-C(t)$ on FIG.\,7) so that the
equilibrium form (\ref{FDTe}) of the FDT is grossly violated
and has to be replaced by one of its nonequilibrium versions.  

\begin{figure}[H]
\begin{center}
\leavevmode
\vskip -0.2cm
\hspace*{0.7cm}
{%
      \begin{minipage}{0.45\textwidth}\hspace*{-0.4cm}
        \includegraphics[width=6.1cm,height=4.7cm,angle=0]{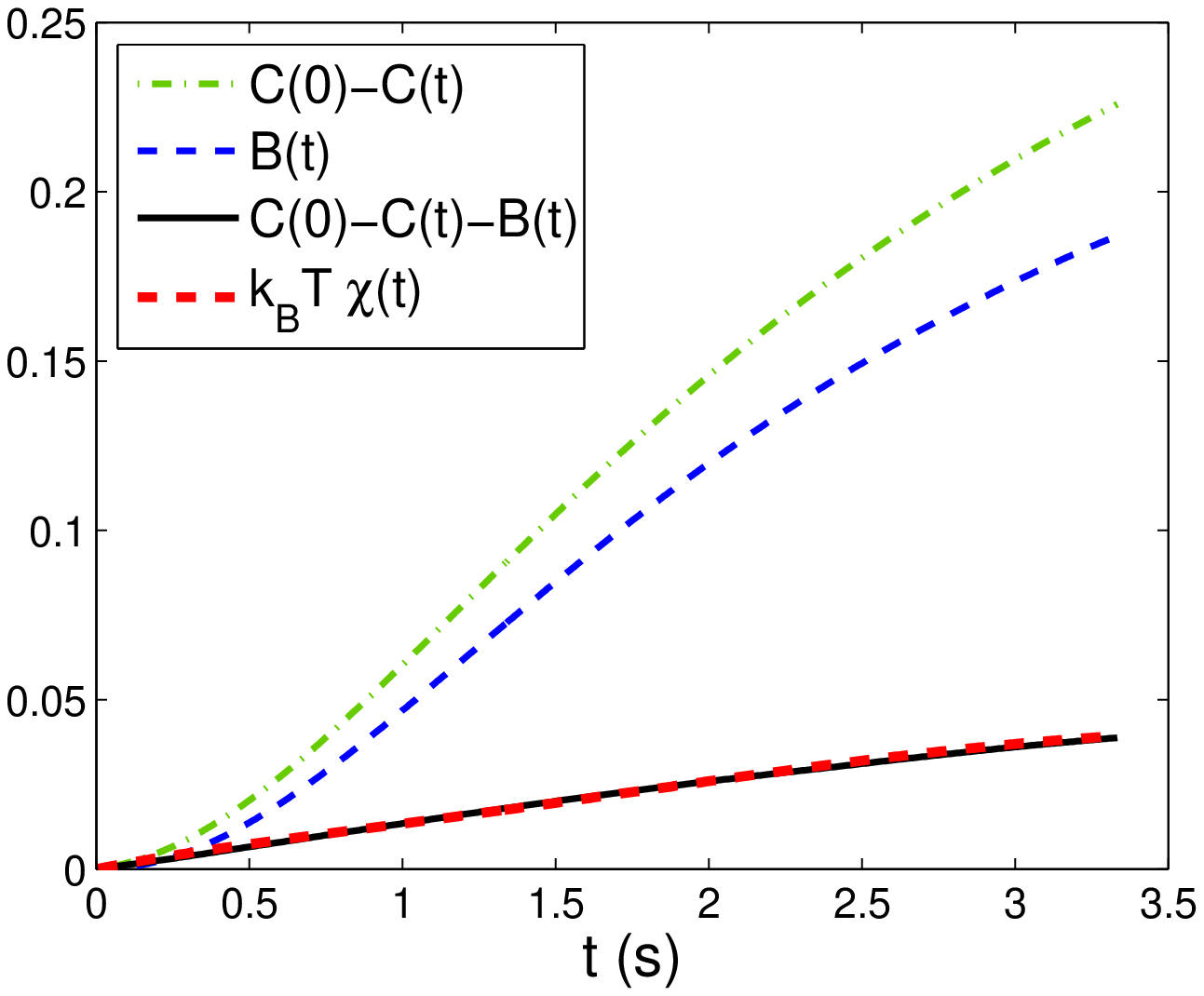}\\
        \vspace{-0.73cm} \strut
        \end{minipage}}
    \hspace*{-0.1cm}
{%
      \begin{minipage}{0.45\textwidth}\vspace{0.17cm}\hspace*{-0.3cm}
        \includegraphics[width=7cm,height=4.6cm]{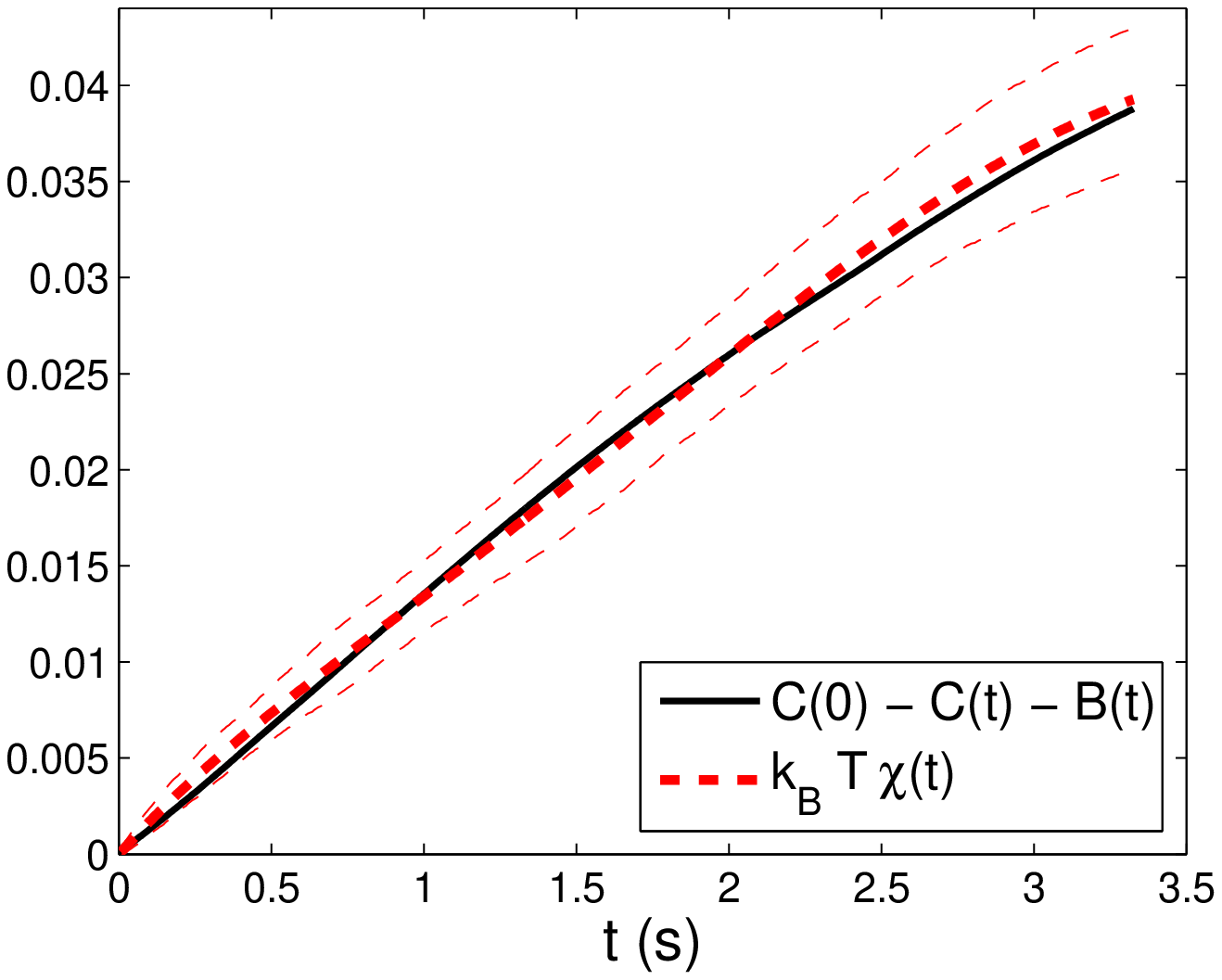}\\
        \vspace{-0.7cm} \strut
        \end{minipage}}
    \hspace*{0.63cm}
\hspace*{0.1cm}
\end{center}
\vskip -0.5cm
\caption{Experimental verification of the time-integrated form of
the FDT (\ref{FDTv}), different terms (left), error brackets (right),
from \cite{GPC09}}
\end{figure}

\subsection{Green-Kubo formula for diffusions}

\noindent Let us consider a diffusion process given by the stochastic
equation (\ref{gendif}) with $\,X_{0t}=X_0+\epsilon^a_tY_{a}\,$
and $\,X_0,Y_a$, \,and $\,X_\alpha\,$ time independent. Let $\,\nu(dx)=
\ee^{-\varphi(x)}\lambda(dx)\,$ be the invariant measure when 
$\,\bm\epsilon=0$. \,As was discussed
in Lecture 1 in Sec.\,\ref{subsec:genfr}, \,a fluctuation relation
\qq
\mathbb E^{\bm\epsilon}_{\mu_0}\,\ee^{-\CW_\tau[\bm x]}\,=\,1
\label{Jarz2}
\qqq
holds in this case for functional $\,\CW_\tau\,$ given by Eqs.\,(\ref{gCWT})
and (\ref{CQ2}) with $\,Y_{t}(x)=\epsilon^a_tY_a$. \,In
particular, the choice $\,\rho_0(x)\,=\ee^{-\varphi(x)}=\rho'_0(x^*)\,$
results in the expression
\qq
\CW_\tau[\bm x]\,=\,\int\limits_0^\tau\epsilon^a_t\,\big((\partial_i\varphi)\,
Y_a^i\,-\,\partial_iY_a^i\big)(x_t)\,dt\,\equiv\,\int\limits_0^\tau
\epsilon^a_t\,\CJ_a(x_t)\,dt\,.
\qqq
Expanding Eq.\,(\ref{Jarz2}) to the second order in $\,\bm\epsilon\,$
around $\,\bm\epsilon=0$, \,we 
obtain the relation
\qq
-\int\limits_0^\tau\epsilon^a_t\,\,\mathbb E_\nu\,\CJ_a(x_t)\,dt\,
-\int\limits_0^\tau\epsilon^a_t\,dt\int\limits_0^t\,\epsilon^b_s
\,\,\frac{\delta}{\delta\epsilon^b_s}\Big|_{\bm\epsilon=\bm 0}\,
\mathbb E^{\bm\epsilon}_\nu\,
\CJ_a(x_t)\,ds\hspace{0.9cm}\cr
+\,\frac{_1}{^2}\,\int\limits_0^\tau\epsilon^a_t\,dt
\int\limits_0^\tau\epsilon^b_s\,\,\mathbb E_\nu\,\CJ_b(x_s)\,\CJ_a(x_t)
\,ds\,\,=\,0
\qqq
or, \,stripping it from the arbitrary functions $\,\bm\epsilon\,$ and taking
$\,s\leq t$,
\qq
&&\mathbb E_\nu\,\CJ_a(x_t)\,=\,0\,,\\
&&\mathbb E_\nu\,\CJ_b(x_s)\,\CJ_a(x_t)\,=\,
\frac{\delta}{\delta\epsilon^b_s}\Big|_{\bm\epsilon=\bm 0}\,
\mathbb E^{\bm\epsilon}_\nu\,\CJ_a(x_t)\,.
\qqq
The first of those equations states that the stationary expectation of 
$\,\CJ_a\,$ vanishes. The integration of the second equation over 
$\,s\,$ from zero to $\,t\,$ gives
\qq
\int\limits_0^t\mathbb E_\nu\,\CJ_b(x_s)\,\CJ_a(x_t)\,ds\,=\,
\partial_{\epsilon^b}|_{\epsilon=0}\,\mathbb E_{\nu}^\epsilon\,\CJ_a(x_t)\,,
\label{togk}
\qqq
where on the right hand side $\,\epsilon=(\epsilon^a)\,$ is taken
time independent. \,Assuming that for $\,t\to\infty\,$ the expectation
\qq
\mathbb E_{\nu}^\epsilon\,\CJ_a(x_t)\ \,
\mathop{\longrightarrow}\limits_{t\to\infty}\,\ \int\CJ_a(x)\,
\nu_\epsilon(dx)\,,
\qqq
where $\,\nu_\epsilon\,$ is the invariant measure of the stationary
process with constant $\,\epsilon\,$ and using the
time-translation invariance of the right hand side, \,we obtain
from Eq.\,(\ref{togk}) the relation
\qq
\boxed{\,\int\limits_{-\infty}^t\mathbb E_{\nu}\,\CJ_b(x_s)\,\CJ_a(x_t)
\,ds\,=\,
\partial_{\epsilon^b}|_{\epsilon=0}\,\int\CJ^a(x)\,\,\nu_\epsilon(dx)\,}\,.
\label{GrKu}
\qqq
This is the {\,\bf Green-Kubo formula\,} \cite{G54,K57} \,that permits 
to extract the linear-regime 
change of the stationary expectation of observables $\,\CJ_a(x)\,$
under a perturbation of the dynamics from their unperturbed 
dynamical correlation function. 
\vskip 0.1cm

If the unperturbed process is time-reversible,
i.e. if $\,L=\ee^{\hspace{0.02cm}\varphi}L^\dagger\ee^{-\varphi}$,
\,then 
\qq
\mathbb E_{\nu}\,\CJ^b(x_s)\,\CJ^a(x_t)\,
=\,\mathbb E_{\nu}\,\CJ_a(x_{-t})\,\CJ_b(x_{-s})\,=\,
\mathbb E_{\nu}\,\CJ^a(x_s)\,\CJ^b(x_t)\,,
\label{A}
\qqq
and we may infer from the Eq.\,(\ref{GrKu}) the Onsager reciprocity relations
\qq
\boxed{\,\partial_{\epsilon^b}|_{\epsilon=0}\,\int\CJ^a(x)
\,\,\nu_\epsilon(dx)\,=\,
\partial_{\epsilon^a}|_{\epsilon=0}\,\int\CJ^b(x)\,\,\nu_\epsilon(dx)\,}\,.
\label{Orecip}
\qqq
The Green-Kubo formula itself may be rewritten in this case in 
the symmetrized form
\qq
\frac{_1}{^2}\int\limits_{-\infty}^\infty
\mathbb E_{\nu}\,\CJ^b(x_s)\,\CJ^a(x_t)\,ds\,=\,
\partial_{\epsilon^b}|_{\epsilon=0}\,\int\CJ^a(x)\,\nu_\epsilon(dx)\,.
\label{B}
\qqq
The last three relations also hold if the unperturbed process is 
time-reversible only relative to an involution $\,x\to x^*\,$ but 
the observables $\,\CJ^a\,$ are all either even or odd under it: 
$\,\CJ^a(x^*)=\pm\CJ^a(x)\,$ with the same sign for all $\,a$.
\vskip 0.4cm

\noindent{\bf Example 7.} \,Above, we considered for simplicity
only perturbations of the drift term in the diffusions, but similar 
strategy may be applied to perturbations involving also the pure 
diffusion part of the dynamics. For concreteness, let us 
consider the anharmonic chain (\ref{chain}) of Example 5 
in Sec.\,\ref{subsec:genfr} with $\,M_i^{-1}=0\,$ for $\,i\not=0,L\,$ 
and with $\,\beta_0=\beta-\frac{1}{2}\epsilon\,$ and 
$\,\beta_L=\beta+\frac{1}{2}\epsilon\,$ so that the functional 
(\ref{CWb}) corresponding to the piecewise constant interpolation of 
$\,\beta_i\,$ is equal to
\qq
\CW_\tau(\bm x)\,=\,\epsilon\int\limits_0^\tau j_{(i-1,i)}(x_t)\,
dt\,.
\qqq
Expanding the Jarzynski equality (\ref{5}) to the second order in 
$\,\epsilon\,$ and proceeding as before, one arrives at the identity
\qq
\int\limits j_{(i-1,i)}(x)\m\,\nu(dx)\ =\ 0\,,
\qqq
where $\,\nu(dx)\,$ is the Gibbs measure at inverse temperature $\,\beta\,$
for the chain, \,and at the Green-Kubo relation
\qq
\frac{_1}{^2}\int\limits_{-\infty}^\infty\mathbb E_\nu\m\,j_{(i-1,i)}(x_s)
\m\,j_{(i-1,i)}(x_t)\m\,ds\ \,
=\,\ \partial_\epsilon|_{\epsilon=0}\int\m\,
j_{(i-1,i)}(x)\,\nu_\epsilon(dx)\,,
\qqq
where $\,\nu_\epsilon(dx)\,$ is the invariant nonequilibrium measure 
for the perturbed boundary temperatures \,(the equilibrium underdamped 
$\,\epsilon=0\,$ dynamics is time-reversible under the involution 
that reverses the sign of momenta and the heat flux
$\,j_{(i-1,i)}\,$ is odd under it). \,One of the outstanding open problems 
of mathematical physics is the control of the large $\,L\,$
behavior of the thermal conductivity
\qq
\kappa(L,\beta)\,=\,L\hspace{0.03cm}\beta^2\,
\partial_\epsilon|_{\epsilon=0}\int\m\,j_{(i-1,i)}(x)\,\nu_\epsilon(dx)
\qqq
giving the proportionality constant between the heat flux and the 
(infinitesimally small) temperature gradient imposed at the boundary. 
In particular, one would like to establish the conjectured Fourier law 
which (in a weak form) states that the limit 
$\,\lim\limits_{L\to\infty}\,\kappa(L,\beta)\,$ exists and is strictly 
positive \cite{BLR00,BK07}.

\nsection{Large deviations and stationary fluctuation relations}

\noindent In the presence of a small parameter $\,\epsilon$, \,a family 
$\,\mu_\epsilon(dX)\,$ of measures may exhibit
a {\,\bf large deviations regime\,} in which it takes an exponential 
form, with the inverse of the small parameter as the prefactor in
front of a negative exponent. This is often formulated 
as the existence of a {\,\bf rate function} $\,\CI(X)\,$ such that
\qq
-\,\inf\limits_{X\in A^o}\,I(X)\,\leq\,\liminf\limits_{\epsilon\to0}\,
\epsilon\m\ln\mu_\epsilon(A)\,\leq\,\limsup\limits_{\epsilon\to0}\,
\epsilon\m\ln\mu_\epsilon(A)
\,\leq\,
-\,\mathop{\rm inf}\limits_{X\in\bar A}\,I(X)\,,
\qqq
where $\,A^o\,$ is the interior and $\,\bar A\,$ the closure of set $\,A$. 
\,In less formal, terms, this may be stated as the property
\qq
\mu_\epsilon(dX)\ \ \mathop{\sim}\limits_{\epsilon\to0}\ \ \ 
\ee^{-\frac{1}{\epsilon}I(X)}\,\lambda(dX) 
\qqq
or, \,if $\,\mu_\epsilon(dX)\,=\,\rho_\epsilon(X)\,\lambda(dX)$, 
\,as the existence in a sufficiently weak sense of the limit
\qq
\lim\limits_{\epsilon\to0}\ \epsilon\ln{\rho_\epsilon(X)}\ =\ -\,I(X)\,.
\label{convv}
\qqq
History of the large deviations theory is long as it originates in 
works of the founding fathers of statistical mechanics in the nineteenth
century. On the probability theory side, it goes back to contributions of
the Swedish mathematician  Harald Cram\'er from the thirties of the
last century. In application to stochastic processes, small parameters 
may have different origin. One possibility is a small noise in the stochastic
differential equations (e.g. low temperature in the Langevin equations).
This is the domain of application of Freidlin-Wentzell theory of large 
deviations \cite{FW84}. We shall encounter it below on a formal level
for diffusions in a functional space. Another possibility, developed
first by Donsker-Varadhan \cite{DV75}, is the long-time asymptotics 
of the solutions of stochastic equations, see also the textbooks 
\cite{DS89,DZ98}. We shall need its version that, to my knowledge, was 
not explicitly considered in mathematical texts but appeared in the papers 
of physicists \cite{CCM07,MNW08}. For an introduction to the subject
of large deviations, \,see also ref.\,\cite{T11}.

\subsection{Large deviations at long times}

\noindent For a stationary diffusion Markov process $\,(x_t)\,$ solving 
the stationary version of Eq.\,(\ref{gendif}), define the {\,\bf empirical 
density\,} and {\,\bf empirical current\,} by the formulae
\qq
\rho_\tau(x)\,=\,\tau^{-1}\int\limits_0^\tau\delta(x-x_t)\,dt\,,
\qquad
j_\tau(x)\,=\,\tau^{-1}\int\limits_0^\tau\delta(x-x_t)\circ dx_t\,,
\label{emprjT}
\qqq
where, as before, ``$\circ$'' signifies the Stratonovich convention. 
Assuming the ergodicity of the process, when $\,\tau\to\infty$, 
$\,\rho_\tau\,$ converges (in a weak sense) to the density $\,\rho\,$ 
of the invariant measure  $\,\nu(dx)=\rho(x)\,\lambda(dx)$, \,and 
$\,j_\tau\,$ converges to the probability current $\,j\,$ given by
\qq
j(x)\,=\,\big(\widehat X_0(x)\,-\,\CD(x)\nabla\big)\rho(x)
\,\equiv\,j_\rho(x)\,.
\label{js0}
\qqq
see Eq.\,(\ref{jt0}), \,which is conserved: 
\qq
\nabla\cdot j(x)\,=\,0\,.
\qqq 
We would like to inquire about the asymptotics of that convergence.
The answer is provided by the large deviations form of the joint
distribution function of the empirical density $\,\rho_\tau\,$ and 
current $\,j_\tau$:
\qq
\mathbb E_\nu\,\delta[\varrho-\rho_\tau]\,\,
\delta[\jmath-j_\tau]\ \,\mathop{\sim}\limits_{\tau\to\infty}\,\ 
\ee^{-\tau \CI[\varrho,\jmath]}\,,
\qqq
with the rate function
\qq
\CI[\varrho,\jmath]\,=\,\begin{cases}\,\infty\hspace{6.31cm}{\rm if}\qquad
\nabla\cdot\jmath\not=0\,,\cr\cr
\frac{_1}{^4}\int\big[\big(\jmath-j_\varrho\big)\cdot(\varrho
\hspace{0.02cm}\CD)^{-1}
\big(\jmath-j_\varrho\big)\big](x)\,\lambda(dx)\qquad{\rm if}
\qquad\nabla\cdot\jmath=0\,,
\end{cases}\label{DVrj}
\qqq
where $\,j_\varrho\,$ is given by Eqs.\,(\ref{js0})
with $\,\rho\,$ replaced by $\,\varrho$. \,The large-deviations
asymtotics for empirical densities or empirical currents only is
then obtained by the ``contraction principle'':
\qq
\mathbb E_\nu\,\,\delta[\varrho-\rho_\tau]\ \,
\mathop{\sim}\limits_{\tau\to\infty}\ 
\ee^{-\tau \CI[\varrho]}\,,\qquad
\mathbb E_\nu\,\,\delta[\jmath-j_\tau]\ \,
\mathop{\sim}\limits_{\tau\to\infty}\ 
\ee^{-\tau \CI[\jmath]}
\qqq
with
\qq
\CI[\varrho]\,=\,\mathop{\rm min}\limits_{\jmath}\,\CI[\varrho,\jmath]\,,\qquad
\CI[\jmath]\,=\,\mathop{\rm min}\limits_{\varrho}\,\CI[\varrho,\jmath]
\qqq
in a slightly abusive notation. The first minimum may 
be rewritten (why?) in terms of a maximum over Lagrange multipliers 
$\,f(x)$:
\qq
\CI[\varrho]&=&-\,\mathop{\rm min}\limits_{f}\,\int(\nabla f)(x)\cdot
\big[\varrho\hspace{0.02cm}\CD(\nabla f)+j_\varrho\big](x)\,\lambda(dx)\cr
&=&-\,\frac{_1}{^4}\int\big[(\nabla\cdot j_\varrho)
(\nabla\cdot\varrho\hspace{0.02cm}
\CD\nabla)^{-1}(\nabla\cdot j_\varrho)\big](x)\,\lambda(dx)\,.
\label{I(rho)}
\qqq
Note that $\,\CI[\varrho]\,$ is non-negative and attains its vanishing minimum
on the density $\,\rho\,$ of the invariant measure such that 
$\,\nabla\cdot j_\rho=0$. \,The first line of Eq.\,(\ref{I(rho)})
may be also rewritten as
\qq
\CI[\varrho]&=&-\,\mathop{\rm min}\limits_{f}\,\int\big[(\nabla f)\cdot
\varrho\hspace{0.02cm}\CD(\nabla f)+f\,L^\dagger\varrho\big](x)\,\lambda(dx)\cr
&=&-\,\mathop{\rm min}\limits_{u>0}\,\int\big(u^{-1}Lu\big)(x)
\,\varrho(x)\,\lambda(dx)\,,
\qqq
where the last minimum is over positive functions $\,u(x)=\ee^{f(x)}$.
\,In the last form, the formula for the rate function $\,\CI[\varrho]\,$ 
holds for general continuous-time stationary Markov processes \cite{DV75}.
\vskip 0.1cm

As for the rate function $\,\CI(\jmath)$, \,note that $\,\CI[\varrho,\jmath]\,$
is a local nonlinear functional of the density $\,\varrho>0\,$ constrained 
to be normalized, quadratic in its first derivatives, leading to the
$2^{\rm nd}$-order differential equation for the extrema. Even in one dimension
where the latter equation reduces to an ODE and $\,\nabla\cdot\jmath=0\,$ means 
that $\,\jmath={\rm const}.$, \, the minimization over $\,\varrho\,$
cannot be explicitly solved analytically. Nevertheless, for weak noise, 
one may resort to 
semiclassical instanton-gas type expansions which go back to 
ideas of Kramers from the 40's of the last century and to Freidlin-Wentzell 
large deviations theory, and which are still being actively developed, 
see e.g. \cite{E06,CCM09}.
\vskip 0.4cm

\noindent{\bf Example 8.} \ For the one-dimensional diffusion 
defined by Eqs.\,(\ref{poc}), (\ref{mu}) and (\ref{mf}), the drift
has two zeros, one unstable at $\,\theta_u\approx 0.214\,rad\,$ 
and one stable at $\,\theta_s=2\pi-\theta_u$. \,One obtains 
in this case the instantonic expression \cite{CCM09} 
\qq
\CI(\jmath)\,\approx\,-\,\jmath\,\ln{\frac{\sqrt{\jmath^2+4\kappa_+\kappa_-}
-\jmath}{2\kappa_-}}\,-\,\sqrt{\jmath^2+4\kappa_+\kappa_-} 
\,+\,\kappa_++\kappa_-
\label{ldc}
\qqq
for $\kappa_+=A_0\exp[D^{-1}\,A_{-}]$, $\,\kappa_- 
=A_0\exp[-D^{-1}A_{+}]\,$ where $\,A_-\,$ is the (tiny negative) area 
under the negative part of the drift graph, $\,A_+\,$ is the one under 
the positive part of the graph, see the left plot in FIG.\,8, 
and $\,A_0=M\sqrt{|(\partial_\theta^2U)(\theta_s)(\partial_\theta^2U)
(\theta_u)|}/(2\pi)$. \,The most 
probable value of $\,j_\tau\,$ for small $\,D\,$ is $j=\kappa_+-\kappa_-$. 
\,The same large deviation function (\ref{ldc}) may be obtained from a jump
process with plus or minus jumps occurring with rates $\,\kappa_\pm\,$
by looking at the statistics of the large sums of jumps \cite{LS99}.
Those jumps correspond in the diffusion to the tunneling to the right 
and to the left through the barriers separating the stable and unstable 
points of the drift.

\begin{figure}[H]
\begin{center}
\leavevmode
\vskip -0.2cm
\hspace*{1.4cm}
{%
      \begin{minipage}{0.4\textwidth}\hspace*{-0.4cm}
        \includegraphics[width=6.1cm,height=4cm,angle=0]{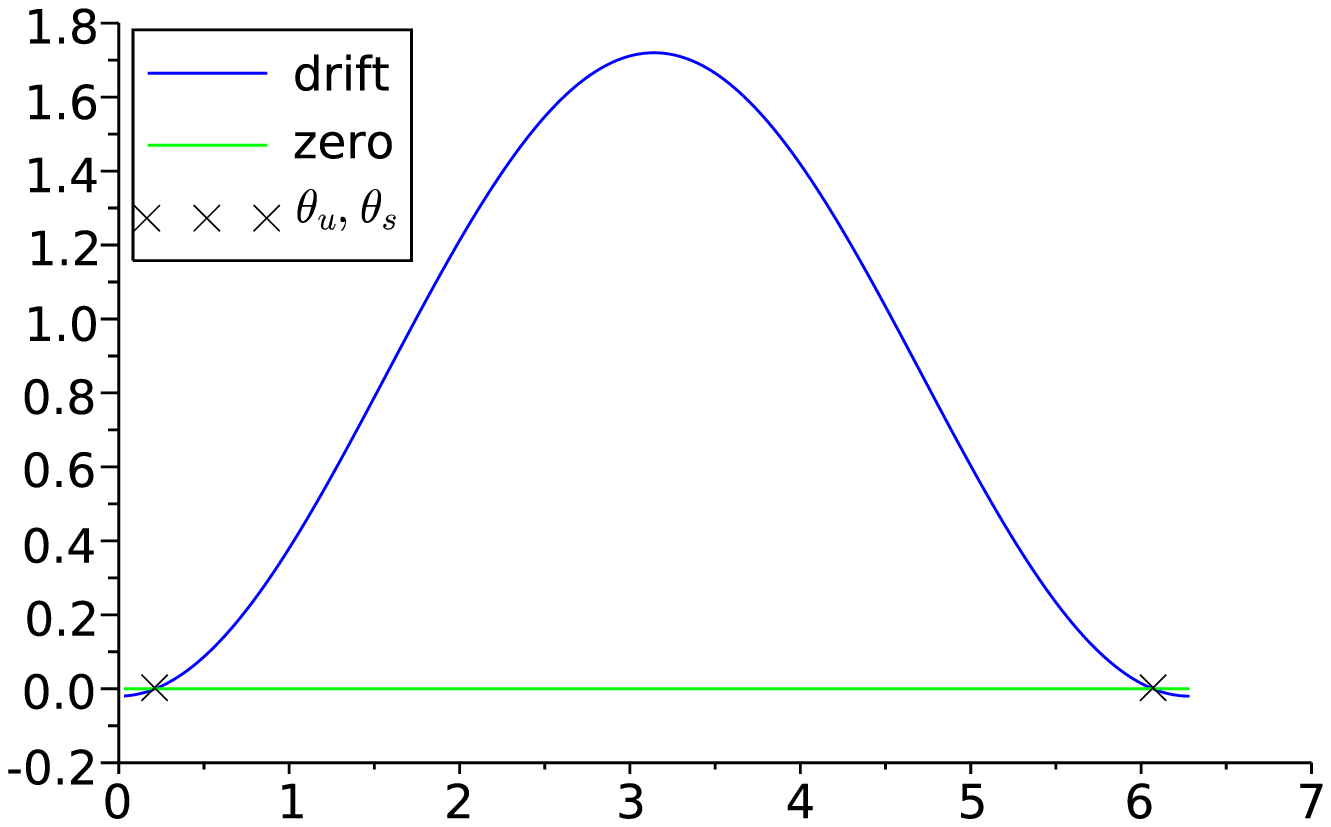}\\
        \vspace{-0.73cm} \strut%
        \end{minipage}}
    \hspace*{0.8cm}
{%
      \begin{minipage}{0.4\textwidth}\hspace*{-1.3cm}
        \includegraphics[width=6cm,height=4cm]{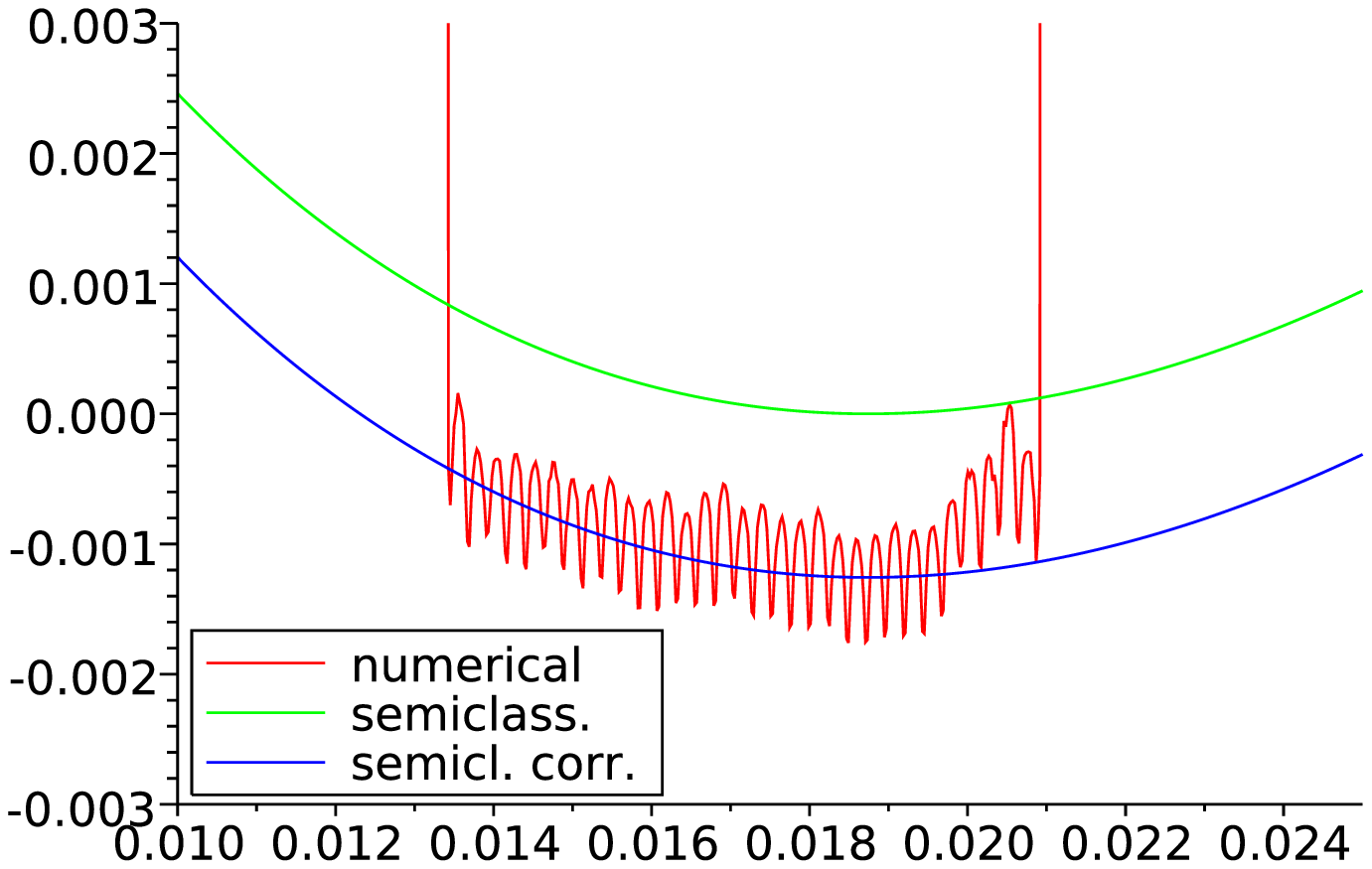}\\
        \vspace{-0.7cm} \strut
        \end{minipage}}
    \hspace*{0.63cm}
\hspace*{0.1cm}
\end{center}
\vskip -0.5cm
\caption{Drift in Eq.\,(\ref{poc}) (left) and the comparison of numerical
and theoretical rate functions for the empirical current (right)}
\end{figure}

\noindent The large deviations statistics of the empirical current $\,j_\tau$, 
\,that is $\,\theta$-independent in this regime, may be extracted from 
the one of its spatial mean
\qq
\frac{_1}{^{2\pi}}\int\limits_0^{2\pi}j_\tau(\theta)\,\lambda(d\theta)\,=\,
\frac{_1}{^{2\pi\tau}}\int\limits_0^\tau d\theta_t
\qqq
The numerical simulation of the minus logarithm of the distribution 
function of the latter divided by $\,\tau=4167\,s\,$ is shown on the right 
plot in FIG.\,8. Note its oscillatory character (with the period $1/\tau$). 
Its plot averaged over the oscillations compares well
on the interval with sufficient number of events with the semiclassical
formula (\ref{ldc}) shifted by 
$\,-(2\tau)^{-1}\ln\big(\frac{\partial^2_{\jmath}\CI(\kappa_+-\kappa_-)}
{2\pi}\big)\,$ to include the one-loop correction. I do not know whether 
the convergence
(\ref{convv}) (for $\,X=\jmath\,$ and $\,\epsilon=\tau^{-1}$) \,holds here
pointwise or only after smearing with test functions.

\subsection{Gallavotti-Cohen type fluctuation relation}

\noindent Defining the time-reversed diffusion process  
one of the ways described in Sec.\,\ref{subsec:genfr} and using
relation (\ref{comp}), we obtain the identity
\qq
\mathbb E_\nu\,\,\ee^{-\CW}\,
\CF(\rho_\tau,j_\tau)\,=\,
\mathbb E'_{\nu'}\,\,\CF(\rho^*_\tau,-j^*_\tau)\,,
\label{idd}
\qqq
where, by definition, 
\qq
\rho^*_\tau[\bm x]\,=\,\rho_\tau[\bm x^*]\,,\qquad 
j^*_\tau[\bm x]\,=\,-\,j_\tau[\bm x^*]
\qqq
(the minus sign in the transformation of the current 
comes from the change of the sign of the time derivative of the process 
under the time reversal). \,An easy calculation shows that in 
their dependence on points in space, $\,\rho^*_\tau\,$
and $\,j^*_\tau\,$ are related to empirical density 
$\,\rho_\tau\,$ and empirical current $\,j_\tau\,$ of Eqs.\,(\ref{emprjT})
by the geometric transformation rule for densities and currents: 
\qq
\varrho^*(x^*)\,\frac{\partial(x^*)}{\partial(x)}\,=\,\varrho(x)\,,
\qquad \jmath^*(x^*)\,\frac{\partial(x^*)}{\partial(x)}\,=\,
\frac{\partial x^*}{\partial x}\,\jmath(x)\,,
\label{rhoj}
\qqq
with $\,\frac{\partial x^*}{\partial x}\,$ denoting the Jacobi matrix
and $\,\frac{\partial(x^*)}{\partial(x)}\,$ the Jacobian of the involution
$\,x\mapsto x^*$. \,On the other hand, 
\,by Eq.\,(\ref{gCWT}) and (\ref{gCQ}),
\qq
\CW\,=\,-\ln\rho'(x_\tau^*)\,+\,\tau\,{\omega}[\rho_\tau,j_\tau]\,
+\,\ln\rho_0(x_0)\,,
\label{CWTw}
\qqq
where
\qq
{\omega}[\varrho,\jmath]\,=\,\int\Big[\widehat X^{+}_{0}\cdot
\CD^{-1}\big(\jmath-
X^{-}_{0}\varrho\big)\,-\,(\nabla\cdot X^{-}_{0})\,\varrho
\Big](x)\,\lambda(dx)
\label{omega}
\qqq
so that
\qq
\int\limits_0^\tau\CJ_t\,dt\,=\,\tau\,{\omega}[\rho_\tau,j_\tau]\,.
\label{CQint}
\qqq
Comparing the large $\,\tau\,$ asymptotics on both sides of identity
(\ref{idd}) we infer the identity
\qq
\boxed{\,\CI[\varrho,\jmath]\,+\,{\omega}[\varrho,\jmath]\,=\,\CI'[\varrho^*,
-\jmath^*]\,}
\label{ldI}
\qqq
where $\,\varrho^*\,$ and $\,\jmath^*\,$ are defined by the relations 
(\ref{rhoj}). This is the stationary fluctuation
relation for the rate functions describing the large deviations of empirical
density and current for the direct and time-reversed diffusion process. 
A simple exercise using Eqs.\,(\ref{DVrj}) and the identity
\qq
j'^*_{\varrho^*}\,=\,j_{\varrho}-2X^-_0\varrho\,=\,
-\,j_{\varrho}+2\widehat X^+_0\varrho-2\CD\nabla\varrho\,,
\label{j'j}
\qqq
where $\,j'_\rho\,$ defined by Eq.\,(\ref{js0}) but for the time-reversed
process, permits to verify Eq.\,(\ref{ldI}) directly.
The probability distribution of the quantity $\,\omega[\rho_\tau,j_\tau]\,$ 
given by Eq.\,(\ref{CQint}), representing the rate of the entropy production
in environment in the units of $\,k_B$, \,has also a large deviations 
regime with the rate function given by the contraction
\qq
\CI(\varpi)\ =\,\mathop{\rm min}\limits_{{\omega}[\varrho,\jmath]=\varpi}\,
\CI[\varrho,\jmath]\,.
\qqq
From Eq.\,(\ref{ldI}), using also the relation
\qq
{\omega}'[\rho^*,-\jmath^*]=-\,{\omega}[\varrho,\jmath]\,,
\qqq 
where $\,\omega'\,$ corresponds to the time-reversed process, 
\,a consequence of the
second equality in (\ref{gCWT}), \,we obtain immediately the fluctuation 
relation
\qq
\boxed{\,\CI(\varpi)\,+\,\varpi\,=\,\CI'(-\varpi)\,}\,.
\label{GC}
\qqq The latter identity holds, in particular, for the time reversal
with $\,\widehat X^{+}_0=0\,$ considered in the case ${\bf(d)}$ of
Sec.\,\ref{subsec:genfr}. In this instance,
\qq
{\omega}[\rho_\tau,j_\tau]\,=\,
-\int(\nabla\cdot X_0^-)(x)\,\rho_\tau(x)\,
\lambda(dx)\,=\,-\,\tau^{-1}\int\limits_0^\tau(\nabla\cdot X_0^-)(x_t)\,dt 
\,\equiv\,{\omega}[\rho_\tau]\,,
\qqq
which reduces to the phase-space contraction rate in the deterministic
case with $\,X_\alpha\equiv0\,$ and $\,X^-_0=X_0$. \,The stationary
fluctuation relation (\ref{GC}) for uniformly hyperbolic deterministic
dynamical systems (in the case when the time-reversed dynamics
coincides with the direct one) was proven in \cite{GC95a,GC95b}) as
the {\,\bf Fluctuation Theorem}.  \,The existence of large deviations 
regime for $\,{\omega}[\rho_\tau]\,$ representing the phase-space 
contraction rate followed in that case from the thermodynamical 
formalism for such dynamical systems so that the Fluctuation Theorem of 
Gallavotti-Cohen is not a direct consequence of the relation (\ref{GC}) 
for the stochastic diffusions. In the latter case, we could obtain 
(\ref{GC}) from the transient fluctuation relation (\ref{idd}) holding 
for the stationary dynamics on any time interval, whereas there is no 
such relation for the general stationary deterministic systems that 
typically have singular invariant measures. The transient Evans-Searles 
relation for such systems employs the non-invariant smooth initial measures 
and a non-trivial work using the thermodynamical formalism would be needed 
to show that they lead for long times to the Gallavotti-Cohen relation. 
\vskip 0.1cm

The fluctuation relation (\ref{GC}) with $\,\CI'=\CI\,$ should also hold 
for the large-deviations rate function of the cumulated heat flux 
$\,\CW_{\tau}\,$ given by Eqs.\,(\ref{CWbetai}), (\ref{CWa}) or (\ref{CWb}) 
in the non-equilibrium stationary state of the anharmonic chain 
with Hamiltonian dynamics in the interior that we discussed in Example 5 
in Sec.\,\ref{subsec:genfr}), see \cite{RT02} for a proof of this fact
for a closely related model.

\subsection{Large deviations for replicated diffusions}

\noindent The final part of these lectures is based on a joint work
in progress with F. Bouchet et C. Nardini \cite{BGN}. The first half 
concerning the large deviations for independent replicated systems
is rather well known, but we present it in the spirit of the 
{\,\bf macroscopic fluctuation theory\,} developed for the dynamics 
of  boundary driven lattice gases in a series of papers of the Rome 
group, see e.g. \cite{BDG06}. The second half that develops the macroscopic 
fluctuation theory for a non-equilibrium system of replicated diffusions 
with a mean-field interaction seems original.
\vskip 0.1cm
 
Let us consider $\,N\,$ independent copies $\,(x^n_t),\ n=1,\dots,N$, 
\,of identical diffusions satisfying stochastic equation (\ref{gendif}).
For such replicated system, we may define the dynamical empirical density 
and empirical current by the formulae 
\qq
\bm\rho_N(t,x)\,=\,\frac{1}{N}\sum\limits_{n=1}^N\delta(x-x^n_t)\,,\qquad
\bm j_N(t,x)\,=\,\frac{1}{N}\sum\limits_{n=1}^N\delta(x-x^n_t)
\circ\frac{dx^n_t}{dt}\,.\label{dempdc}
\qqq
Here and below, we use bold letters for quantities that depend on
time and (phese-)space coordinates. We shall also employ the notation 
$\,\rho_{Nt}(x)\,$ and $\,j_{Nt}(x)\,$ whenever we consider only the 
$\,x$-dependence for fixed $\,t$. \,Note the continuity equation
\qq
\partial_t\bm\rho_N\,+\,\nabla\cdot\bm j_N\,=\,0\,.
\label{ceq}
\qqq
Let $\,\varPhi[\varrho]\,$ be a functional of (possibly
distributional) densities of the cylindrical form:
\qq
\varPhi[\varrho]\,=\,f\Big(\int h_1(x_1)\,\rho(x_1)\,\lambda(dx_1),\cdots,
\int h_k(x_k)\,\varrho(x_k)\,\lambda(dx_k)\Big)\,.
\qqq
In its $t$-dependence, random variable $\,\varPhi(\rho_{Nt})\,$ satisfies the
stochastic equation
\qq
&&d\varPhi[\rho_{Nt}]\ =\ \int\frac{\delta\varPhi[\rho_{Nt}]}{\delta\varrho(x)}
\Big(-\frac{_1}{^N}\sum\limits_n(\nabla\delta)(x-x^n_t)\big(X_{0t}(x^n_t)\,dt
+X_{\alpha t}(x^n_t)\circ dW_\alpha^n(t)\big)\Big)\,\lambda(dx)\cr
&&=\ \int\Big(\rho_{Nt}\,X_{0t}\nabla\frac{\delta\varPhi[\rho_{Nt}]}
{\delta\varrho}
\Big)(x)\,\lambda(dx)\,dt\,+\,\frac{_1}{^N}\sum\limits_n
X_\alpha(x^n_t)\Big(\nabla\frac{\delta\varPhi[\rho_{Nt}]}
{\delta\varrho}\Big)(x^n_t)\circ dW_\alpha^n(t)
\qqq
which implies that
\qq
&&\frac{d}{dt}\ \un{\mathbb E}_{\un{\nu}}\ \varPhi[\rho_{Nt}]\ =\ 
\un{\mathbb E}_{\un{\nu}}\,\bigg[\int\Big(j_{\rho_{N\hspace{-0.02cm}t}}\nabla
\frac{\delta\varPhi[\rho_{Nt}]}{\delta\varrho}\Big)(x)\,\lambda(dx)\cr
&&+\ 
\frac{_1}{^N}\int\rho_{Nt}(x)\,\CD^{ij}_t(x)\,\delta(x-y)\,
\nabla_{x^i}\nabla_{y^j}\frac{\delta^2\varPhi[\rho_{Nt}]}{\delta\varrho(x)\,
\delta\varrho(y)}\,\lambda(dx)\lambda(dy)\  \equiv\ 
\un{\mathbb E}_{\un{\nu}}\ \big(\CL_{Nt}\varPhi\big)[\rho_{Nt}]\bigg]\,,\qquad
\label{CLN}
\qqq
where $\,\un{\mathbb E}_{\un{\nu}}\,$ denotes the expectation value
over the replicated processes and $\,j_{\rho_t}\,$ is given by
Eq.\,(\ref{jt0}).  
Note that the $2^{\rm nd}$-order term in the functional operator
$\,\CL_{Nt}\,$ is proportional to $\,\frac{1}{N}$. \,Formally, this is 
the same equation as the one for the expectation of the diffusion process 
in the space of densities solving the stochastic PDE
\qq
\partial_t\bm\rho\,+\,\nabla\cdot\bm j_{\bm\rho,\bm\xi}\,=\,0
\label{nFP}
\qqq
where
\qq
\bm j_{\bm\rho,\bm\xi}\,=\,\bm j_{\bm\rho}\,+\,
(2N^{-1}\hspace{-0.05cm}\bm\rho\bm\CD)^{1/2}\bm\xi
\label{jrhoxi}
\qqq
with the space-time white noise $\,\bm\xi$,
\qq
\mathbb E\,\,\bm\xi^i(t,x)\,=\,0\,,\qquad \mathbb E\,\,
\bm\xi^i(t,x)\,\bm\xi^j(s,y)\,=\,
\delta^{ij}\,\delta(t-s)\,\delta(x-y)\,.
\qqq
Compare Eq.\,(\ref{nFP}) to the continuity equation (\ref{ceq}).
In the limit $\,N\to\infty$, \,Eq.\,(\ref{nFP}) reduces to the
standard Fokker-Plank equation 
\qq
\partial_t\bm\rho+\nabla\cdot\bm j_{\bm\rho}\,=\,0\,,
\label{FPbis}
\qqq
for the instantaneous probability densities of a 
single copy of the process, see Eqs.\,(\ref{FP}) and (\ref{jt0}).

\subsection{Hamilton-Jacobi equation and Sanov Theorem}

\noindent The probability distributions of the empirical densities 
$\,\rho_{Nt}\,$ evolve by the adjoint operator $\,\CL_{Nt}^*\,$ and their 
hypothetical densities by the formal adjoint $\,\CL_{Nt}^\dagger\,$
built with the use of the rule 
\qq
\Big(\frac{\delta}{\delta\varrho}\Big)^\dagger\,=\,-\,\frac{\delta}
{\delta\varrho}\,.
\qqq
In particular, assuming that those densities have the large-deviation
form $\,\ee^{-N\CF_t[\varrho]}\,$, \,we obtain in the leading order
the {\,\bf Hamilton-Jacobi equation\,} for $\,\CF_t$:
\qq
\boxed{\,\partial_t\CF_t[\varrho]\,+\,\int
\bigg(\Big(\nabla\frac{\delta\CF_t[\varrho]}
{\delta\varrho}\Big)\cdot\varrho\hspace{0.02cm}\CD_t
\Big(\nabla\frac{\delta\CF_t[\varrho]}{\delta\varrho}\Big)\,
+\,j_{\varrho\hspace{0.02cm}t}
\cdot\nabla\frac{\delta\CF_t[\varrho]}{\delta\varrho}\bigg)(x)\,
\lambda(dx)\,=\,0\,}
\label{dHJ}
\qqq
We shall call $\,\CF_t[\rho]\,$ the free energy of the replicated system. 
Eq.\,(\ref{dHJ}) is solved by the relative entropy functional
in the units of $\,k_B$
\qq
\boxed{\,\CF_t[\varrho]\,=\,\int\Big(\varrho\,
\ln\big(\frac{\varrho}{\rho_t}\big)\Big)(x)\,\lambda(dx)\,=\,
k_B^{-1}S[\varrho\Vert\rho_t]\,}
\label{dSanov}
\qqq
where $\,\rho_t\,$ solves the Fokker-Planck equation (\ref{FPbis}).
\vskip 0.4cm

\noindent{\bf Theorem} \,(Dynamical version of the Sanov Theorem).
\vskip 0.1cm

\noindent The solution (\ref{dSanov}) of the Hamilton-Jacobi equation
(\ref{dHJ}) describes the time evolution 
of the rate function for the large deviations of the 
distribution of empirical density $\,\rho_{Nt}\,$ if 
the initial points $\,x^n_0\,$ of the replicated processes
are distributed (independently) with the probability density 
$\,\rho_0$.
\vskip 0.4cm

\noindent{\bf Corollary.} \ In the particular case of the replicated 
stationary diffusion process,
the distribution of the empirical densities $\,\rho_{Nt}\,$ stays 
time independent and for large $\,N\,$ it takes the large deviations form 
with the rate function 
\qq
\boxed{\,\CF[\varrho]\,=\,k_B^{-1}S[\varrho\Vert\rho]\,}
\label{sSanov}
\qqq
where $\,\rho\,$ is the density of the invariant measure of the process.
$\,\CF[\varrho]\,$ solves the stationary Hamilton-Jacobi equation
\qq
\boxed{\,\int
\bigg(\Big(\nabla\frac{\delta\CF[\varrho]}
{\delta\varrho}\Big)\cdot\varrho\hspace{0.02cm}\CD
\Big(\nabla\frac{\delta\CF[\varrho]}{\delta\varrho}\Big)\,+\,j_{\varrho}
\cdot\nabla\frac{\delta\CF[\varrho]}
{\delta\varrho}\bigg)(x)\,\lambda(dx)\,=\,0\,}\,.
\label{sHJ}
\qqq
\vskip 0.4cm

Introducing the stationary current velocity in the space of densities 
by the formula
\qq
\CV[\varrho]\,=\,-\nabla\cdot\Big(j_\varrho\,+\,\varrho\hspace{0.02cm}\CD
\nabla\frac{\delta\CF[\varrho]}{\delta\varrho}\Big)\,,
\label{CV}
\qqq
the stationary Hamiltonian-Jacobi equation 
(\ref{sHJ}) may be rewritten as the orthogonality condition
\qq
\int\frac{\delta\CF[\varrho]}{\delta\varrho(x)}\,\CV[\varrho](x)
\,\lambda(dx)\,=\,0\,.
\label{orth}
\qqq
For the time reversal corresponding of case ${\bf(a)}$ 
in Sec.\,\ref{subsec:genfr}, \,one obtains from Eq.\,(\ref{j'j})
the relation
\qq
j'^*_{\varrho^*}\,=\,-\,j_{\varrho}-
2\varrho\CD\nabla\varphi-2\CD\nabla\varrho
\qqq
which may be rewritten in the form
\qq
\boxed{\,j'^*_{\varrho^*}\,=\,-\,j_{\varrho}\,-\,
2\hspace{0.02cm}\rho\hspace{0.02cm}\CD\nabla
\frac{\delta\CF[\varrho]}{\delta\varrho}\,}\,.
\label{j'rho}
\qqq
A comparison with Eq.\,(\ref{CV}) yields the relations:
\qq
\CV[\varrho]\,=\,\nabla\cdot\Big(j'^*_{\varrho^*}\,+\,\varrho\hspace{0.02cm}\CD
\nabla\frac{\delta\CF[\varrho]}{\delta\varrho}\Big)
\label{CV1}
\qqq
and
\qq
\CV[\varrho]\,=\,\frac{_1}{^2}\,\big(\nabla\cdot j'^*_{\varrho^*}
-\nabla\cdot j_{\varrho}\big)\,.
\label{CV2}
\qqq
\vskip 0.4cm

\noindent{\bf Example 9.} \ For the diffusion on a circle (\ref{poc}),
\qq
\CF[\varrho]\,=\,\int\limits_0^{2\pi}
\varrho(\theta)\big(\ln{\varrho(\theta)}+\varphi(\theta)\big)\,
\lambda(d\theta)
\label{Sanovcirc}
\qqq
with $\,\varphi(\theta)\,$ given by 
Eq.\,(\ref{invm}). \,In particular, \,in the equilibrium case 
with $\,F=0$,
\qq
\CF[\varrho]\,=\,\int\limits_0^{2\pi}
\varrho(\theta)\Big(\ln{\varrho(\theta)}\,
+\,\beta\hspace{0.02cm}U(\theta)\Big)\,\lambda(d\theta)\ +\ {\rm const.}
\label{CFeq}
\qqq
so that $\,\CF\,$ is equal in this instance to $\,\beta\,\,$ times 
the free energy of the gas of noninteracting particles in the thermal 
equilibrium at inverse temperature $\,\beta$. \,Quantity $\,\varrho\,$ 
is the density of the gas and $\,U\,$ is the external potential. 
\,For the Langevin equation (\ref{poc}) (with any $F$),
\qq
j_\varrho(\theta)\,=\,M\big(F-(\partial_\theta U)(\theta)\big)\hspace{0.02cm}
\varrho(\theta)-D\hspace{0.03cm}\partial_\theta\varrho(\theta)\,,
\qqq
and for the time reversed one of Eq.\,(\ref{invF}),
\qq   
j'^*_{\varrho^*}(\theta)\,=\,j'_\varrho(\theta)
\,=\,\big(M(F-(\partial_\theta U)(\theta))-2v(\theta)\big)
\hspace{0.02cm}\varrho(\theta)-D\hspace{0.03cm}\partial_\theta\varrho(\theta)\,,
\qqq
where the current velocity $\,v(\theta)\,$ is given by Eq.\,(\ref{v0}).
\,In this case
\qq
\CV[\varrho](\theta)\,=\,-\partial_\theta\big(\varrho\,v\big)(\theta)\,.
\label{CVF}
\qqq

\subsection{Dynamical large deviations for the replicated process} 

\noindent One can show, at least formally, that the joint distribution
of the dynamical empirical density $\,\bm\rho_N\,$ and current $\,\bm j_N\,$
exhibits for large $\,N\,$ the large-deviations regime with the rate function
\qq
\CI[\bm\varrho,\bm\jmath]\,=\,\begin{cases}
\infty\hspace{6.91cm}{\rm if}\qquad
\partial_t\bm\varrho+\nabla\cdot\bm\jmath\not=0\,,\cr\cr
\frac{_1}{^4}\int\big[(\bm\jmath-\bm j_{\bm\varrho}
\cdot(\bm\varrho\bm\CD)^{-1}(\bm\jmath-\bm j_{\bm\varrho})\big](t,x)\,dt
\lambda(dx)
\qquad{\rm if}
\qquad\partial_t\bm\varrho+\nabla\cdot\bm\jmath=0\,.
\end{cases}\label{DLDrj}
\qqq
Note the similarities and the differences with the long-time
rate function \,(\ref{DVrj}). \,In the formal argument using functional
integrals, we shall replace
$\,\bm\rho_N\,$ and $\,\bm j_N\,$ that satisfy Eq.\,(\ref{ceq})
by $\,\bm\rho\,$ and $\,\bm j_{\bm\rho,\bm\xi}\,$ connected 
by Eq.\,(\ref{nFP}). Thus
\qq
&&\mathbb E\ \varPsi[\bm\rho_N,\bm j_N]\,=\,\mathbb E\int
\varPsi[\bm\varrho,\bm j_{\bm\varrho,\bm\xi}]\ 
\delta[\partial_t\bm\varrho+\nabla\cdot\bm j_{\bm\varrho,\bm\xi}]\,
\det\hspace{-0.05cm}\Big(\frac{\delta(\partial_t\bm\varrho
+\nabla\cdot\bm j_{\bm\varrho,\bm\xi})}{\delta\bm\varrho}\Big)\,D\bm\varrho\cr
&&=\ \mathbb E\int
\varPsi[\bm\varrho,\bm\jmath]\,\,\ee^{\,i\int\bm a\cdot(\bm\jmath-
j_{\bm\varrho,\bm\xi})}\,\delta[\partial_t\bm\varrho+\nabla\cdot\bm\jmath]\,
\,D\bm a\,D\bm\varrho
\,D\bm\jmath\,,
\qqq
where we dropped the determinant which will not contribute 
to the large deviations (and is equal to $1$ if properly regularized).
Averaging over the white noise $\,\bm\xi$, \,we obtain:
\qq
\mathbb E\ \varPsi[\bm\rho_N,\bm j_N]
&=&\int\varPsi[\bm\varrho,\bm\jmath]\ \ee^{\,i\int\bm a\cdot(\bm\jmath-
\bm j_{\bm\varrho})-\frac{1}{N}\int\bm a\cdot\bm\varrho\bm\CD\,\bm a}\,
\delta[\partial_t\bm\varrho+\nabla\cdot\bm j]\,\,D\bm a
\,D\bm\varrho\,D\bm\jmath\cr
&=&\int\varPsi[\bm\varrho,\bm\jmath]\ \ee^{-\frac{N}{4}\int(\bm\jmath
-\bm j_{\bm\varrho})\cdot(\bm\varrho\bm\CD)^{-1}(\bm\jmath
-\bm j_{\bm\varrho})}\,\delta[\partial_t\bm\varrho+\nabla\cdot\bm j]\,
\,D\bm\varrho\,D\bm\jmath\,.
\qqq
From the last functional integral expression, we read off the large 
deviations rate function (\ref{DLDrj}) (a similar functional-integration 
argument may be used to obtain formula (\ref{DVrj})).
\vskip 0.1cm

The rate functions for the dynamical large deviations of the empirical 
density $\,\bm\rho_N\,$ alone or for the empirical current $\,\bm\jmath_N\,$
alone are given by the contraction:
\qq
\CI[\bm\varrho]&=&\mathop{\min}\limits_{\bm\jmath}\ \CI[\bm\varrho,\bm\jmath]
\,=\,\frac{_1}{^4}\int\Big[\big(\partial_t\bm\varrho
+\nabla\cdot\bm j_{\bm\varrho}\big)\big(-\nabla\cdot\bm\varrho
\hspace{0.02cm}\bm\CD\hspace{0.03cm}\nabla\big)^{-1}
\big(\partial_t\bm\varrho+\nabla\cdot\bm j_{\bm\varrho}\big)\Big](t,x)\,
dt\lambda(dx)\,,\qquad\label{CIcontr1}\\
\CI[\bm\jmath]&=&\mathop{\min}\limits_{\bm\varrho}\ \CI[\bm\varrho,\bm\jmath]\,,
\label{CIcontr2}
\qqq
with no closed expression in the latter case where an appropriate initial 
conditions for $\,\bm\varrho\,$ should be specified. As for the first
formula, it follows by a formal application of the Freidlin-Wentzell theory
to the diffusion (\ref{nFP}) with noise (\ref{jrhoxi}) and it was
rigorously established in \cite{DG87}.
\vskip 0.1cm

We shall denote by $\,\CI_{A}[\bm\varrho,\bm\jmath]\,$ the rate functions
given by Eq.\,(\ref{DLDrj}) with the time-integral restricted to
the interval $A$. \,The functionals $\,\CI_{[0,\tau]}(\bm\varrho,\bm\jmath)\,$
and $\,\CI'_{[0,\tau]}[\bm\varrho,\bm\jmath]\,$ for the replicated direct 
and time reversed process, the latter obtained with the use of one of the 
rules of Sec.\,\ref{subsec:genfr}, satisfy the stationary fluctuation relation
\qq
\boxed{\,\CI_{[0,\tau]}[\bm\varrho,\bm\jmath]\,+\,
{\omega}_{[0,\tau]}[\bm\varrho,\bm\jmath]\,
=\,\CI'_{[0,\tau]}[\bm\varrho^*,-\bm\jmath^*]\,}
\label{ldIbm}
\qqq
where 
\qq
\bm\varrho^*(t,x)\,=\,\bm\varrho(t^*,x^*)\,
\frac{\partial(x^*)}{\partial(x)}\,,
\qquad \bm\jmath^*(t,x)\,=\,\frac{\partial x}{\partial x^*}\,
\jmath(t^*,x^*)\,\frac{\partial(x^*)}{\partial(x)}\,,
\label{rhojbm}
\qqq
and
\qq
{\omega}_{[0,\tau]}[\bm\varrho,\bm\jmath]&=&\int\limits_0^\tau dt\int
\Big[{\widehat X}^{+}_{0t}\cdot\CD_t^{-1}\big(\jmath_t-
X^{-}_{0t}\hspace{0.02cm}\varrho_t\big)\,
-\,(\nabla\cdot X^{-}_{0t})\hspace{0.02cm}\varrho_t
\Big](x)\,\lambda(dx)\cr
&-&\int\varrho_t(x)\,\ln{\varrho_t}(x)\,
\lambda(dx)\,\bigg|^\tau_0\,,
\label{omegabm}
\qqq
compare to relations (\ref{ldI}), (\ref{rhoj}) and (\ref{omega}).
These identities follow in a straightforward way from the relations
\qq
\bm j'^*_{\bm\varrho^*}\,=\,\bm j_{\bm\varrho}-2\bm X^-_0\bm\varrho\,=\,
-\bm j_{\bm\varrho}+2\widehat{\bm X}^+_0\bm\varrho-2\bm\CD\nabla\bm\varrho
\qqq
that generalize Eqs.\,(\ref{j'j}). In the particular case of the 
stationary process and the time reversal corresponding of case ${\bf(a)}$ 
in Sec.\,\ref{subsec:genfr},
\qq
{\omega}_{[0,\tau]}[\bm\varrho,\bm\jmath]\,=\,\CF[\varrho_0]-\CF[\varrho_\tau]\,.
\label{omegabms}
\qqq
By contraction, we infer then from relation (\ref{rhojbm}) the identity
\qq
\boxed{\,\CI_{[0,\tau]}[\bm\varrho]\,+\,
\CF[\varrho_0]-\CF[\varrho_\tau]
=\,\CI'_{[0,\tau]}[\bm\varrho^*]\,}
\label{ldIrbm}
\qqq
Let $\,\varrho_0=\rho$, \,where $\,\rho=\ee^{-\varphi}\,$ is the invariant 
density of the single process, so that $\,\CF[\varrho_0]=0\,$ follows. 
\,Take $\,\tau\to\infty$. Then the minimum of the right hand side over 
$\,\bm\varrho^*\,$ with $\,\varrho_0\,$ and $\,\varrho_\infty\,$ fixed
is realized by the trajectory $\,\bm\varrho'\,$ solving the 
reversed process Fokker-Planck equation
\qq
\partial_t\bm\varrho'+\nabla\cdot\bm j'_{\bm\varrho'}\,=\,0
\qqq
that relaxes from $\,(\varrho_\infty)^*\,$ to the invariant density
$\,\rho^*\,$ and $\,\CI'_{[0,\infty[}(\bm\varrho')\,$ vanishes 
for such a trajectory.
This is an expression of the generalized {\,\bf Onsager-Machlup principle}
\cite{BDG06}: \,the most probable trajectory that describes the creation 
of the spontaneous fluctuation $\,\varrho_\infty\,$ from the vacuum
configuration $\,\rho\,$ is the time reversal of the trajectory
that describes the the most probable relaxation of the spontaneous 
fluctuation $\,\varrho_\infty^*\,$ to the vacuum $\,\rho^*\,$ in 
the time-reversed dynamics. Taking the minima on the both hand sides
of Eq.\,(\ref{ldIrbm}), we obtain the identity
\qq
\boxed{\,\CF[\varrho]\,\ =\hspace{-0.2cm}\mathop{\rm min}
\limits_{\bm\varrho\atop\varrho_0=\rho,\ \varrho_\infty=\varrho}
\hspace{-0.25cm}\CI_{[0,\infty[}[\bm\varrho]\ \,=\,\ 
\CI_{[0,\infty[}[\bm\varrho'^*]\,}\,.
\label{OM}
\qqq  
that connects the rate functions for the large deviations of  
the invariant distribution and for the dynamical large deviations
of the empirical density $\,\bm\varrho_N$.
   
\subsection{Replicated diffusions with mean-field coupling}

\noindent One may perturb the $\,N\,$ replicated 
diffusions (\ref{gendif}) by introducing a {\,\bf mean-field\,} type coupling
between the replicated processes by a pair force $\,\frac{1}{N}Y_t(x_n,x_m)
=-\frac{1}{N}Y(x_m,x_n)$, \,obtaining a coupled system
of stochastic equations
\qq
dx\,=\,X_{0t}(x)\hspace{0.03cm}dt\,+\,\frac{_1}{^N}
\sum\limits_{m=1}^NY_t(x_n,x_m)\,dt\,+
\,X_{\alpha t}(x)\circ dW_\alpha(t)\,.
\label{gendifN}
\qqq
One may still define the empirical dynamical densities 
$\,\bm\rho_N\,$ and currents $\,\bm j_N\,$ by Eqs.\,(\ref{dempdc}).
The discussion concerning the large deviations of $\,\bm\rho_N\,$
and $\,\bm\jmath_N\,$ for the replicated diffusions carries over 
to the interacting case after a modification of the formula 
Eq.\,(\ref{jt0}) for the current $\,j_{\rho_t}$ which becomes
\qq
j_{\rho_t}(x)\,=\,
\big(\rho_t\hspace{0.01cm}(\widehat X_{0t}\,+\,Y_t)\,
-\,\CD_t\nabla\rho_t\big)(x)\,,
\label{modj}
\qqq
picking up an additional term involving the effective mean field force
\qq
(Y_t\rho_t)(x)\,\equiv\,\int Y_t(x,y)\,\rho_t(y)\,\lambda(dy)\,.
\qqq
After this modification, one still obtains the formal stochastic 
equation (\ref{nFP}) reducing in the limit $\,N\to\infty\,$ to 
Eq.\,(\ref{FPbis}). \,The latter becomes now a {\,\bf nonlinear 
Fokker-Planck equation\,} due to the presence of a quadratic term 
in $\,\rho_t\,$ in the expression for $\,j_{\rho_t}$.
\,The Hamilton-Jacobi equations 
(\ref{dHJ}) and (\ref{sHJ}) have still the same form but the Sanov
solutions (\ref{dSanov}) and (\ref{sSanov}) are no more valid.
Finding, in particular,  the right solution of the free energy 
$\,\CF[\varrho]\,$ in the stationary case is a mayor problem, see below, 
except of the special instance of equilibrium 
dynamics. The dynamical large deviations rate functions 
$\,\CI[\bm\varrho,\bm\jmath])$, $\,\CI[\bm\varrho]\,$ and 
$\,\CI[\bm\jmath]\,$ are still given by Eqs.\,(\ref{DLDrj}) and
(\ref{CIcontr1}), (\ref{CIcontr2}). In the stationary case, 
if one defines the
time reversed process as corresponding to the formal stochastic
equation (\ref{nFP}) with $\,\bm j_{\bm\rho}\,$ in relation 
(\ref{jrhoxi}) replaced by $\,\bm j'_{\bm\rho}\,$ given by 
Eq.\,(\ref{j'rho}) and $\,\bm\CD'\,$ defined as before
then the stationary fluctuation relation (\ref{ldIbm}) still
holds for $\,\omega[\bm\varrho,\bm\jmath]\,$ given by 
Eq.\,(\ref{omegabms}) implying the Onsager-Machlup-type
relations (\ref{ldIrbm}) and (\ref{OM}).

\subsection{Perturbative solutions for free energy $\,\CF[\varrho]$}

\noindent In the stationary case, we may search for the solution 
the Hamilton-Jacobi 
equation (\ref{sHJ}) for the nonequilibrium free energy functional 
$\,\CF[\varrho]\,$ in the form of a formal power series 
\qq
\CF[\varrho]\,=\,\sum\limits_{k=0}^\infty\CF_k(\varrho)\,,
\label{CFn}
\qqq
in the interaction force $\,Y\,$ treated as a perturbation,
where the term $\,\CF_k\,$ is of order $\,k\,$ in $\,Y\,$ and where
\qq
\CF_0[\varrho]\,=\,\int
\Big(\varrho\,\ln{\big(\frac{\varrho}{\rho_0}\big)}\Big)(x)\,\lambda(dx)
\label{CF0}
\qqq
is the Sanov solution (\ref{sSanov}) for $\,Y=0\,$ with $\,\rho_0(x)\,$ 
standing for the density of the invariant measure of the diffusion 
(\ref{gendif}). \,Inserting expansion (\ref{CFn}) into the  
Hamilton-Jacobi equation (\ref{sHJ}) and gathering terms of the same 
order in $\,V$, \,we obtain the relations
\qq
&&\int\varrho(x)\,L'\,\frac{\delta\CF_k[\varrho]}{\delta\varrho(x)}\,
\lambda(dx)\cr
&=&\int\varrho(x)\bigg[(Y\varrho)\cdot
\nabla\frac{\CF_{k-1}
[\varrho]}{\delta\varrho}\,+\,\sum\limits_{l=1}^{n-1}
\Big(\nabla\frac{\CF_l[\varrho]}{\delta\varrho}\Big)
\cdot\CD\Big(\nabla\frac{\CF_{k-l}[\varrho]}{\delta\varrho}\Big)
\bigg](x)\,\lambda(dx)\,,\qquad
\label{algebr}
\qqq
where
\qq
L'\,=\,-\bigg(L\,+\,2\Big(\nabla\frac{\CF_0[\varrho]}
{\delta\varrho}\Big)\cdot\CD\nabla\bigg)\,=\,\rho_0^{-1}L^\dagger\rho_0\,=\,-\widehat X_0\cdot\nabla+\big(\nabla
+2(\nabla\ln{\rho_0})\big)\cdot\CD\nabla
\qqq
is the backward generator of the single uncoupled process time-reversed 
according to the rules of case ${\bf(a)}$ in Sec.\,\ref{subsec:genfr}
(with $x^*=x$) and $\,L\,$ is the backward generator of the single uncoupled 
direct process. \,One can find a solution of these equations
assuming that, for $k\geq 1$, $\,\CF_k[\rho]\,$ is a polynomial 
of degree $\,k+1\,$ in $\,\varrho$:
\qq
\CF_k[\varrho]\,=\,\frac{_1}{^{(k+1)!}}\int\cdots
\int\phi_k(x_0,\dots,x_{k})
\,\varrho(x_0)\cdots\varrho(x_{k})\,\lambda(dx_0)\cdots
\lambda(dx_{k})
\label{CFpert}
\qqq
with $\,\phi_k\,$ symmetric in its arguments. Then Eqs.\,(\ref{algebr})
reduce at order $\,k=1\,$ to the equation
\qq
\big(L'(x_0)-L'(x_1)\big)\phi_1(x_0,x_1)\,=\,-\rho_0(x_0)^{-1}
\rho_0(x_1)^{-1}(\nabla_{x_0}-\nabla_{x_1})\cdot
\big[\rho_0(x_0)\rho_0(x_1)Y(x,y)\big]
\label{eqfp1}
\qqq
for kernel $\,\phi_1$. \,For higher orders, one obtains
iterative equations for the kernels $\,\phi_k(x_0,\dots,x_k)\,$
whose solution may be written in terms of a sum over tree diagrams. 
\vskip 0.2cm

A different perturbative scheme consists of developing $\,\CF[\varrho]\,$
around the (stable) stationary solution $\,\rho_{st}\,$ of the nonlinear
Fokker-Planck equation:
\qq
\CF[\varrho]\,=\,\sum\limits_{k=1}^\infty\widetilde\CF_k[\widetilde\varrho]\,,
\qqq
where $\,\widetilde\varrho=\varrho-\rho_{st}\,$ and
\qq
\widetilde\CF_k[\widetilde\varrho]\                          
=\ \frac{_1}{^{(k+1)!}}\int\widetilde\phi_k(x_0,\dots,x_n)\,                    
\widetilde\varrho(x_0)\cdots\widetilde\varrho(x_k)\,\lambda(dx_0)\cdots 
\lambda(dx_k)
\qqq  
with $\,\widetilde\phi_k\,$ symmetric in its arguments fixed by assuming that
\qq
\int\widetilde\phi_k(x_0,\dots,x_k)\,\lambda(dx_0)\,=\,0\,.
\qqq
Substitution into the stationary Hamilton-Jacobi equation (\ref{sHJ})
gives for $\,k>1\,$ the recursion
\qq
&&\int\widetilde\varrho\,\Phi R\hspace{0.02cm}\Phi^{-1}
\frac{\delta\widetilde\CF_k[\widetilde\varrho]}{\delta\widetilde\varrho}
\ =\,\int
\widetilde\varrho\bigg[\big(Y*\widetilde\varrho)
\cdot
\nabla\frac{\delta\widetilde\CF_{k-1}[\widetilde\varrho]}
{\delta\widetilde\varrho}\cr\cr
&&+\,\sum\limits_{l=1}^{k-1}
\Big(\nabla\frac{\delta\widetilde\CF_{l}[\varrho]}
{\delta\varrho}\Big)\cdot D
\Big(\nabla\frac{\delta\widetilde\CF_{k-l}[\varrho]}
{\delta\varrho}\Big)\bigg]\ +\ \sum\limits_{l=2}^{k-1}\int\Big(
\nabla\frac{\delta\widetilde\CF_{l}[\widetilde\varrho]}
{\delta\widetilde\varrho}\Big)
\cdot\rho_{st}D
\Big(\nabla\frac{\delta\widetilde\CF_{k+1
-l}[\widetilde\varrho]}{\delta\widetilde
\varrho}\Big),
\qqq
where $\,R\,$ is the linearization of the nonlinear Fokker-Planck operator 
around $\,\rho_{st}\,$ and
\qq\big(\Phi\widetilde\varrho\big)(x)\,                                       
=\,\int{\widetilde\phi}_{1}(x,y)\,\widetilde\varrho(y)\,
\lambda(dy)
\qqq
solves the operator equation
\qq
R\hspace{0.02cm}\Phi^{-1}+\hspace{0.03cm}\Phi^{-1}\hspace{-0.04cm}        
R^\dagger\,=\,2                                                                 
\nabla\cdot\rho D\hspace{0.02cm}\nabla\,.
\qqq
that comes from the stochastic Lyapunov equation and determines
$\,\widetilde\CF^1[\widetilde\varrho]$. \,Kernels $\,{\widetilde\phi}_{k}
\,$ of $\,\widetilde\CF^k[\widetilde\varrho]\,$ for $\,k\geq2\,$ may again
be iteratively calculated from the above recursion and represented in terms 
of a sum over tree diagrams.
\vskip 0.2cm

This way the replicated diffusions coupled in a mean-field way seem 
to be among a few non-equilibrium systems, along with some special
models of one-dimensional lattice gases, see \cite{BDG06}, where 
nonequilibrium free energy $\,\CF[\varrho]\,$ may be controlled 
analytically at least to some extent.

\subsection{Static large deviations of currents}
\label{subsec:currfl}

\noindent Considering again the stationary regime of replicated diffusions
with a mean-field coupling, \,one may ask what 
is the probability
distribution to see a dynamical empirical-current fluctuation 
with prescribed temporal mean 
\qq
\jmath(x)\,=\,\frac{_1}{^\tau}\,\int\limits_0^\tau\bm\jmath_N(t,x)\,dt\,.
\qqq
The large deviations regime of that distribution in the limit 
$\,\tau\to\infty\,$ will be governed by
the contracted rate function
\qq
\CJ[\jmath]\ =\ \lim\limits_{\tau\to\infty}
\mathop{\rm min}\limits_{\bm\jmath\atop\frac{1}{\tau}
\int\limits_0^\tau\bm\jmath(t,\cdot)\,dt=\jmath}\hspace{-0.2cm}
\CJ[\bm\jmath]\ \ =\ \ \lim\limits_{\tau\to\infty}
\hspace{-0.1cm}\mathop{\rm min}\limits_{\bm\varrho,\ \bm\jmath\atop\frac{1}{\tau}
\int\limits_0^\tau\bm\jmath(t,\cdot)\,dt=\jmath}\hspace{-0.2cm}
\CI[\bm\varrho,\bm\jmath]\,.
\label{ldfmc}
\qqq  
The minimum of $\,\CI[\jmath]\,$ defined this way is equal to zero and
is attained on $\,\jmath=j_{\rho_{st}}\,$ and corresponds to time-independent
configurations $\,\bm\varrho=\rho_{st}\,$ and $\,\bm\jmath=j_{\rho_{st}}$.
\,For $\,\jmath\,$ close to $\,j_{\rho_{st}}$, \,the minimum on the right
hand side will still be attained on time independent configurations
$\,\bm\varrho\,$ and $\,\bm\jmath$, \,i.e.
\qq
\CI[\jmath]\ =\ \begin{cases}\,\infty\hspace{6.76cm}
{\rm if\quad}\nabla\cdot\jmath\not=0\,,\cr
\,\mathop{\rm min}\limits_{\varrho}\ \frac{_1}{^4}
\int\big[(\jmath-j_\varrho)\cdot(\varrho\CD)^{-1}(\jmath-j_\varrho)(x)\,
\lambda(dx)\hspace{0.6cm}{\rm otherwise}\,,
\end{cases}
\label{ldfmcs}
\qqq
see Eq.\,(\ref{DLDrj}), where for $\,\nabla\cdot\jmath=0\,$
the minimizing $\,\varrho\,$ satisfies the equation
\qq
\int g(x)\cdot\frac{\delta j_\varrho(x)}{\delta\varrho(y)}\,\lambda(dx)
\,+\,(g\cdot\CD g)(y)
\,=\,\lambda\ =\ {\rm const.}\qquad{\rm for}\qquad 
g=(2\CD\varrho)^{-1}(j-j_\varrho)\,.
\qqq 
It was observed in 
\cite{BD05} and \cite{BDG05}, see also \cite{BDL08}, that in some examples 
of lattice gasses the minimum on the right hand side of (\ref{ldfmc}) 
is not always
realized on time-independent configurations, leading to the phenomena
of dynamical phase transitions in systems with preimposed 
temporal mean of current fluctuations. Here, we shall concentrate, however, 
on values of $\,\jmath\,$ sufficiently close to $\,j_{\rho_{st}}\,$ so that
(\ref{ldfmcs}) holds. \,In particular, 
\,for $\,\jmath=j_{\rho_0}+\delta\jmath\,$ with $\,\nabla\cdot\delta\jmath=0$,
\,we have $\,\varrho=\rho_0+\delta\varrho\,$ and $\,g=\delta g\,$ so that
\qq
\CI(j_{\rho_0}+\delta\jmath)\,=\,\int(\delta g\cdot\CD\,
\delta g)(x)\,\rho_0(x)
\,\lambda(dx)\ +\ o((\delta\jmath)^2)\,,
\label{tqor}
\qqq
where, to the linear order,
\qq
\int\delta g(x)\cdot\frac{\delta j_{\rho_0}(x)}{\delta\varrho(y)}\,dx\,=\,
\delta\lambda\ =\ {\rm const.}
\label{eqfdg}
\qqq
If we denote by $\,S\,$ the linear operator that acts on functions 
$\,\delta\varrho\,$ with vanishing integral assigning to them vector
field by the formula
\qq
(S\delta\varrho)(x)\,=\,\int\frac{\delta j_{\rho_0}(x)}{\delta\varrho(y)}\,
\delta\varrho(y)\,\lambda(dy)
\qqq
then Eq.\,(\ref{eqfdg}) for $\,\delta g\,$ may be rewritten as the condition
\qq
S^\dagger\delta g\,=\,0
\label{1g}
\qqq
whereas the extremal variation $\,\delta\varrho\,$ has to satisfy  
the linear equation
\qq
\delta j-S\delta\varrho\,=\,2\CD\rho_0\delta g\,.
\label{1rho}
\qqq
It is easy to see that the last two equations fix $\,\delta g\,$
uniquely. The above calculation fixes the covariance of the fluctuations
$\,\delta\jmath\,$ of the time average empirical current on the 
central-limits scale $\,\frac{1}{\sqrt{N}}\,$ as the corresponding
covariance operator $\,C\,$ is given by the formula
\qq
\CI(j_{\rho_0}+\delta\jmath)\,=\,\frac{_1}{^2}\int\delta\jmath(x)\cdot
(C^{-1}\delta\jmath)(x)\,\lambda(dx)\ +\ o((\delta\jmath)^2)\,.
\qqq
\vskip 0.2cm

In the $\,N\to\infty\,$ limit the phase transitions in an autonomous 
system of
replicated diffusions with mean-field coupling correspond to
changes in the form of attractors of the nonlinear Fokker-Planck
dynamics: the stable fixed points or stable
periodic solutions in the simplest cases. In particular, the second 
order transitions 
correspond to bifurcations in that dynamics where eigenvalues
of linearized Fokker-Planck operator $\,R\,$ cross the imaginary
axis. We shall describe an example of such a systems exhibiting
a rich phase diagram in the next subsection. Here, let us only
remark that at the transitions corresponding to bifurcations where 
an eigenvalue of $\,R\,$ itself crosses zero (e.g. at saddle-node
bifurcations), the covariance of the fluctuations
of the time averaged empirical current on
the central-limit scale diverges in the directions
proportional to $\,\delta\jmath=S\delta\rho_0$, \,where
$\,\delta\rho_0\,$ is the zero mode of the linearized Fokker-Planck operator
$\,R=-\nabla\cdot S$.
\,Indeed, for such $\,\delta\jmath$, \,Eqs.\,(\ref{1g}) and (\ref{1rho})
are satisfied for $\,\delta g=0\,$ and $\,\delta\varrho=\delta\rho_0\,$
so that the right hand side of (\ref{tqor}) vanishes to the second order, 
implying the divergence of $\,\int\delta\jmath(x)(C\delta\jmath)(x)\,
\lambda(dx)$. \,Note that such a divergence is specific to non-equilibrium
transitions because at equilibrium transitions relation
$\,R\hspace{0.02cm}\delta\varrho_0=0\,$ implies that 
$\,S\hspace{0.01cm}\delta\varrho_0=0$. 
\,For other non-equilibrium transitions corresponding to bifurcations 
where an eigenvalue of $\,R\,$ crosses the imaginary axis at a non-zero value,
as in the case of Hopf bifurcations, similar phhenomeon 
occurs but in time averages of the current fluctuations multiplied
by a time-periodic function \cite{BGN}.

\subsection{Diffusions on a circle with mean field coupling}

\noindent Let us look more closely at the case when the original diffusion 
process is given by the stationary overdamped Langevin equation 
(\ref{poc}) on a circle and where we take the time-independent coupling force 
$\,Y(\theta,\vartheta)=-M\partial_\theta V(\theta-\vartheta)\,$ for
a symmetric potential $\,V(\theta)=V(-\theta)$,
\,arriving at a system of stochastic equations
\qq
d\theta^n\,=\,M\Big(F-\partial_{\theta^n}\big(U(\theta^n)-
\frac{_1}{^N}\sum\limits_{m=1}^NV(\theta^n-\theta^m)\big)\Big)\hspace{0.02cm}dt\,
+\,\sqrt{2D}\,dW^n(t)\,.
\label{pocN}
\qqq
For $\,F=0$, \,equations (\ref{pocN}) describe an equilibrium dynamics 
with the invariant measure given by the Gibbs state
\qq
\nu(d(\theta^n))\,=\,Z_N^{-1}\,\,\ee^{-\beta\big(\sum\limits_{n=1}^NU
(\theta^n)\,+\frac{1}{2N}\hspace{-0.1cm}\sum\limits_{n,m=1}^N\hspace{-0.1cm}
V(\theta^n-\theta^m)\big)}\prod\limits_{n=1}^N\lambda(d\theta^n)\,.
\label{gibbsN}
\qqq
In this case, the large deviations rate function for the 
stationary distribution of empirical density $\,\rho_{Nt}(\theta)\,$ is
\qq
\CF[\varrho]\,=\,\int\limits_0^{2\pi}
\varrho(\theta)\Big(\ln{\varrho(\theta)}\,
+\,\beta\hspace{0.02cm}\big(U(\theta)+\frac{_1}{^2}(V*\varrho)(\theta)\big)
\Big)\,\lambda(d\theta)\ +\ {\rm const.}
\label{CFeqV}
\qqq
generalizing Eq.\,(\ref{CFeq}). \,It is a solution of the
stationary Hamilton-Jacobi equation (\ref{sHJ}) with
\qq
j_\varrho=\,\rho\big(F-\partial_\theta(U+V*\rho)\big)-D\hspace{0.02cm}
\partial_\theta\rho
\qqq
with $\,F=0\,$ and it is of the form (\ref{CFpert}) with 
\qq
\phi_1(\theta_0,\theta_1)\,=\,\beta\hspace{0.02cm}V(\theta_0-\theta_1)
\label{tosol1}
\qqq
and with $\,\phi_k=0\,$ for $\,k\geq 2$. \,It is easy to see that this, 
indeed, provides a solution of the recursion (\ref{algebr}).
Note that expression (\ref{CFeqV}) is equal to $\,\beta\,$ times
free energy of the gas of interacting 
particles in the thermal equilibrium at inverse temperature $\,\beta$.
\vskip 0.2cm
 
For $\,F\not=0$, \,stochastic equation (\ref{pocN}) describes 
a nonequilibrium $\,N$-particle dynamics, generalizing the single particle 
one (\ref{poc}). In this case, the perturbative solution
(\ref{CFn}) and (\ref{CFpert}) starts with the term
\qq
\phi_1(\theta_0,\theta_1)&=& 
\beta\hspace{0.02cm}V(\theta_0-\theta_1)
\,+\,\beta\,(L'(\theta_0)+L'(\theta_1))^{-1}\Big(\big(v_0(\theta_0)
-v_0(\theta_1)\big)\,\partial_{\theta_0}V(\theta_0,\theta_1)\Big),
\label{phi1}
\qqq
where $\,v_0(x)\,$ is the current velocity (\ref{currv0}) 
of the stationary diffusion (\ref{gendif}), see (\ref{eqfp1}).
For $\,F=0\,$ this reduces to solution (\ref{tosol1}) since $\,v_0=0\,$
in that case. The higher order kernels $\,\phi_k\,$  are non-trivial 
for $\,F\not=0$.
\vskip 0.4cm

\noindent{\bf Example 10 (Active rotators model of \cite{SK86}).} \ Consider 
the case when  
$\,U(\theta)\,=\,-h\cos(\theta)\,$ and $\,V(\theta)=
J(1-\cos(\theta))\,$ with $\,J>0$.  \,Such a system was studied in 
\cite{SK86} as a model of  ``active rotators'' closely related to the original
Kuramoto's model \cite{K75} of synchronization phenomena. It has been
a subject of rich mathematical literature, see e.g.\,\cite{GPP12}, and
it is also related to the models of depinning transition in disordered elastic
media, see e.g.\,\cite{M05}. \,In this example, one may 
also view the angles $\,\theta^n\,$ as 
describing planar spin vectors 
$\,\vec{\sigma}_n=(\cos{\theta^n},\sin{\theta^n})$, \,with mean-field
ferromagnetic interaction, in a planar magnetic field with magnitude 
$\,h\,$ that rotates with angular velocity $\,MF\,$, \,provided 
that we describe the spin angles in the frame co-rotating 
with the magnetic field.
For general $\,F$,  
\,the $N=\infty$ dynamics is described by
the nonlinear Fokker-Planck equation (\ref{FPbis}) whose stationary
solutions have the form
\qq
\rho(\theta)\,=\,Z^{-1}\,\,\ee^{\,\beta\big(F\theta+(h+x_1)\cos{\vartheta}+
x_2\sin{\theta}\big)}\hspace{-0.1cm}
\int\limits_\theta^{\theta+2\pi}\hspace{-0.1cm}
\ee^{-\beta\big(F\vartheta+(h+x_1)\cos{\vartheta}+
x_2\sin{\vartheta}\big)}\lambda(d\vartheta)\,,
\qqq
compare to Eq.\,(\ref{invm}), where the coefficients
\qq
\frac{x_1}{J}\,=\,\int\limits_0^{2\pi}\cos{\vartheta}\,
\rho(\vartheta)\,\lambda(d\vartheta)
\,,\qquad
\frac{x_2}{J}\,=\,\int\limits_0^{2\pi}\sin{\vartheta}\,\rho(\vartheta)
\,\lambda(d\vartheta)\,.
\label{x1x2}
\qqq
have to be found self-consistently. \,The linearly stable solution
solutions are then selected by the analysis of the spectrum of
the linearization $\,R\,$ of the nonlinear Fokker-Plank operator. 
\,For $\,F=0$, \,integrals in Eqs.\,(\ref{x1x2}) are expressible
by Bessel functions, leading to the equations
\qq
\frac{x_1}{J}\,=\,\frac{(x_1+h)}{x(h)}\,
\frac{I_1(\beta\hspace{0.02cm}x(h))}
{I_0(\beta\hspace{0.02cm}x(h))}\,,
\qquad
\frac{x_2}{J}\,=\,\frac{x_2}{x(h)}\,\frac{I_1(\beta\hspace{0.02cm}x(h))}
{I_0(\beta\hspace{0.02cm}x(h))}\,,
\label{seq}
\qqq
where $\,x(h)=\sqrt{(x_1+h)^2+x_2^2}$.

\begin{figure}[H]
\begin{center}
\leavevmode
\vskip 0.1cm
\hspace*{-0.3cm}
        \includegraphics[width=7cm,height=5cm]{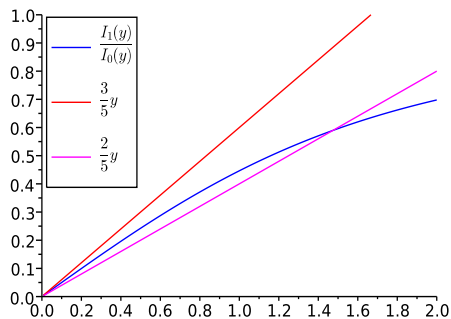}\\
        \caption{Ratio of the Bessel functions and the curves 
$\frac{y}{\beta J}$ for $\beta J=\frac{5}{3}<2$ and $\beta J=\frac{5}{2}>2$}
\end{center}
\end{figure}

\noindent For $\,h=+0$, \,these equations are solved by $\,x_1=0=x_2\,$
but the corresponding stationary solution of the nonlinear Fokker-Planck
equation becomes unstable for $\,\beta J>2\,$ where another stable solution
with $\,x_1>0\,$ and $\,x_2=0$ (accompanied by the one reflected in
the $\,x_2=0\,$ axes corresponding to $\,h=-0$) \,appears by a 
supercritical pitchfork bifurcation, see FIG.\,9. 
\,For $\,h\not=0\,$ there is a single solution of Eqs.\,(\ref{seq}) 
with $\,{\rm sgn}(h)\,x_1>0\,$ and $\,x_2=0$. 
\,These solutions describe the $2^{\rm nd}$ order phase transition
in the $\,N\to\infty\,$ limit of the Gibbs state
(\ref{gibbsN}) for $\,h=0\,$ from the disordered
stationary phase with $\,\mathbb E\,\vec{\sigma}_n=0\,$ for 
$\,\beta J\leq 2\,$ to the mixture of ordered ones with 
$\,\mathbb E\,\vec{\sigma}_n\not=0\,$ for $\,\beta J>2\,$ \cite{SFN72}. 
A pure ordered state is selected by turning on an infinitesimal 
magnetic field $\,h=+0\,$ and the other pure states are obtained
by a simultaneous rotation of all spins. \,The sharp transition disappears 
for (non-infinitesimal) $\,h\not=0$. 
\vskip 0.2cm

For $\,F>0\,$ and $\,h=+0$, there is no stable stationary solution
for $\beta J>2$ but a stable periodic solution of the nonlinear 
Fokker-Planck equation 
\qq
\rho(t,\theta)=\rho_{F=0}(\theta-MFt)\,,
\qqq
where $\,\rho_{F=0}\,$ is the stable solution for $\,F=0$, \,appears 
by a Hopf bifurcation: the phase transition becomes dynamical.
This type of transition persists for $\,h>0\,$ sufficiently small
and some intermediate $\,\beta$, \,with a stable stationary solution
occurring for smaller $\,\beta\,$ and stable periodic solution for 
higher $\,\beta$. \,If we increase $\,h\,$ then the stable periodic 
solution turns back to a stable stationary solution for a sufficiently 
high $\,h$. \,The corresponding phase diagram obtained by numerical
simulations in refs.\,\cite{SK86} and \cite{SSK88} is sketched in FIG.\,10.

\begin{figure}[H]
\begin{center}
\leavevmode
\vskip -1cm
\hspace*{-0.1cm}
        \includegraphics[width=9cm,height=5.9cm]{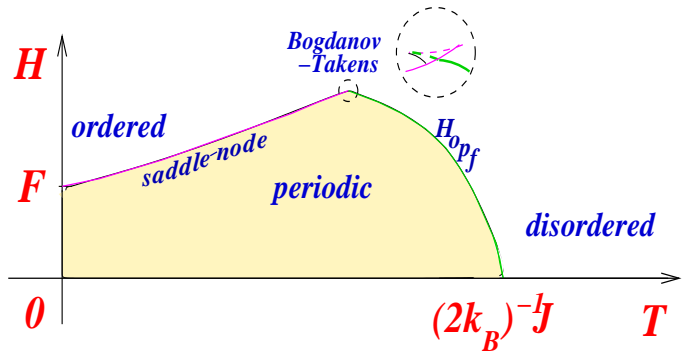}\\
        \caption{Phase diagram of the active rotator model of
        \cite{SK86,SSK88}}
\end{center}
\end{figure}

\noindent Note the presence of a $2^{nd}$ order transition corresponding 
to a saddle-node bifurcation, to which the analysis of 
Sec.\,\ref{subsec:currfl} applies directly, as well as of the one
corresponding to the Hopf bifurcation, with the crossing of the two
involving a more complicated small transition region with a Bogdanov-Takens
bifurcation \cite{SSK88}. More results on the large-deviations for 
empirical density and empirical current in the active rotators model,
a prototypical example of a nonequilibrium system with mean-field 
interactions, will be described in ref.\,\cite{BGN}. 
\vskip 0.5cm

\noindent{\bf Acknowledgements.} \ The author thanks 
the Mathematics Department of Helsinki University,
and, in particular, Prof. A. Kupiainen, for the invitation to give 
a series of lectures on fluctuation relations and for the hospitality 
during author's stay in Helsinki. The final work on these notes was
done during author's visit to KITPC in Beijing in July 2013. 
Support of the ANR project ANR-11-BS01-015-02 ``STOSYMAP'' is also 
acknowledged.

\end{document}